\documentclass[12pt]{article}

\usepackage{graphics,epsfig,subfigure}

\newcommand{\bea}{\begin{eqnarray}}
\newcommand{\eea}{\end{eqnarray}}
\newcommand{\beq}{\begin{equation}}
\newcommand{\eeq}{\end{equation}}

\newcommand{\gev}{{\rm GeV}}

\newcommand{\msb}{\overline{\rm{MS}}}

\def\slash#1{\mbox{$\not\!\! #1$}}

\newcommand{\beqn}{\begin{eqnarray}}   
\newcommand{\eeqn}{\end{eqnarray}}

\def\dfrac#1#2{{\displaystyle {#1 \over #2}}}

\def\simge{\mathrel{\rlap{\raise 0.511ex \hbox{$>$}}{\lower 0.511ex
 \hbox{$\sim$}}}}
\def\simle{\mathrel{\rlap{\raise 0.511ex \hbox{$<$}}{\lower 0.511ex
 \hbox{$\sim$}}}}
\def\slash#1{\setbox0=\hbox{$#1$}\dimen0=\wd0 \setbox1=\hbox{/} \dimen1=\wd1
 \ifdim\dimen0>\dimen1 \rlap{\hbox to \dimen0{\hfil/\hfil}} #1
 \else \rlap{\hbox to \dimen1{\hfil$#1$\hfil}} / \fi}

\def\cyp{a}
\def\rmi{b}
\def\rmii{c}
\def\infntv{d}
\def\zeut{e}
\def\rmiii{f}
\def\infntre{g}

\begin{document}
\begin{titlepage}
{ \vspace{-0.5cm} 
\normalsize\hfill 
\parbox{120mm}{DESY 10-037,~RM3-TH/10-6,~ROME1/1468,~ROM2F/2010/06}}
\vspace{0.5cm}
\begin{center}
\begin{Large}
\textbf{Non-perturbative renormalization of quark bilinear \\ operators
with $N_f=2$ (tmQCD) Wilson fermions \\ and the tree-level improved gauge
action} \\
\end{Large}
\end{center}

\begin{figure}[h]
  \begin{center}
    \includegraphics[draft=false]{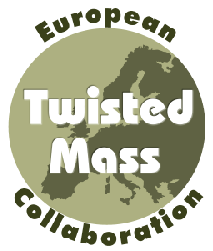}
  \end{center}
\end{figure}

\baselineskip 20pt plus 2pt minus 2pt
\begin{center}
  \textbf{
    M.~Constantinou$^{(\cyp)}$,
    P.~Dimopoulos$^{(\rmi)}$,
    R.~Frezzotti$^{(\rmii,\infntv)}$,
    G.~Herdoiza$^{(\zeut)}$
    K.~Jansen$^{(\zeut)}$
    V.~Lubicz$^{(\rmiii,\infntre)}$,
    H.~Panagopoulos$^{(\cyp)}$,
    G.C.~Rossi$^{(\rmii,\infntv)}$,\\
    S.~Simula$^{(\infntre)}$,
    F.~Stylianou$^{(\cyp)}$,
    A.~Vladikas$^{(\infntv)}$}
\end{center}

\begin{center}
  \begin{footnotesize}
    \noindent

$^{(\cyp)}$ Department of Physics, University of Cyprus, P.O.Box 20537, Nicosia
CY-1678, Cyprus
\vspace{0.2cm}

$^{(\rmi)}$ Dip. di Fisica, Universit\`a di Roma ``La Sapienza'',
    and INFN, Sezione di Roma, \\P.le A.~Moro 2, I-00185 Rome, Italy\\
\vspace{0.2cm}

$^{(\rmii)}$ Dip. di Fisica, Universit{\`a} di Roma ``Tor Vergata",\\ 
Via della Ricerca Scientifica 1, I-00133 Rome, Italy
\vspace{0.2cm}

$^{(\infntv)}$ INFN, Sez. di Roma ``Tor Vergata",
c/o Dip. di Fisica, Universit{\`a} di Roma ``Tor Vergata",\\ 
Via della Ricerca Scientifica 1, I-00133 Rome, Italy
\vspace{0.2cm}

$^{(\zeut)}$ NIC, DESY, Platanenallee 6, D-15738 Zeuthen, Germany 
\vspace{0.2cm}

$^{(\rmiii)}$ Dip. di Fisica, Universit{\`a} di Roma Tre, Via della Vasca Navale
84, I-00146 Rome, Italy
\vspace{0.2cm}

$^{(\infntre)}$ INFN, Sez. di Roma III, Via della Vasca Navale 84, I-00146 Rome,
Italy
\vspace{0.2cm}

\end{footnotesize}
\end{center}

\vspace{0.5cm}

\begin{abstract}{
We present results for the renormalization constants of bilinear quark operators
obtained by using the tree-level Symanzik improved gauge action and the $N_f=2$
twisted mass fermion action at maximal twist, which guarantees automatic ${\cal
O}(a)$--improvement. Our results are also relevant for the corresponding
standard (un-twisted) Wilson fermionic action since the two actions only differ,
in the massless limit, by a chiral rotation of the quark fields. The
scale-independent renormalization constants $Z_V$, $Z_A$ and the ratio $Z_P/Z_S$
have been computed using the RI-MOM approach, as well as other alternative
methods. For $Z_A$ and $Z_P/Z_S$, the latter are based on both standard twisted
mass and Osterwalder-Seiler fermions, while for $Z_V$ a Ward Identity has been
used. The quark field renormalization constant $Z_q$ and the scale dependent
renormalization constants $Z_S$, $Z_P$ and $Z_T$ are determined in the RI-MOM
scheme. Leading discretization effects of ${\cal O}(g^2a^2)$, evaluated in
one-loop perturbation theory, are explicitly subtracted from the RI-MOM
estimates.
}
\end{abstract}

\end{titlepage}


\section{Introduction}

Non-perturbative renormalization is an essential ingredient of lattice QCD
calculations which aim at a percent level accuracy. In this paper, we present
calculation methods and results for the renormalization constants (RCs) of
bilinear quark operators for the lattice action used by the ETM Collaboration.
The simulations with $N_f=2$ dynamical flavours employ the tree-level Symanzik
improved gauge action and the twisted mass fermionic action at maximal twist. In
the chiral limit, the latter is related to the standard Wilson fermionic action
by a chiral transformation, under which quark composite operators behave in a
definite and simple way. Therefore the RCs we compute here can be also employed
to renormalize bilinear quark operator matrix elements computed with the same
glue action but standard (un-twisted) Wilson quarks.  

For the computation of the bilinear operator RCs we have used the RI-MOM method
\cite{rimom}. For the calculation of the scale independent RCs, $Z_V$, $Z_A$ and
the ratio $Z_P/Z_S$, a set of alternative methods have also been used. In
particular, we have determined $Z_V$ by imposing the validity of the axial Ward
identity whereas for $Z_A$ and $Z_P/Z_S$ we have employed a new method which
makes a combined use of standard twisted mass (tm) and Osterwalder-Seiler (OS)
formalism. A comparison of the results obtained with different approaches
provides an estimate of the size of the systematic errors affecting the
calculation of the RCs at fixed lattice bare coupling. Needless to say, in
renormalized quantities the discretization errors coming from both RCs and bare
quantities will be removed by continuum extrapolation, as illustrated with an
example in sect. 3.3. 

Throughout the paper we follow the standard practice of labelling the RCs
according to the notation adopted in the un-twisted Wilson case. Thus, for
instance, $Z_V$ and $Z_A$ denote the RCs of the local vector and axial-vector
currents of the standard Wilson action even though, in the maximal twist case,
they renormalize in the physical basis the local (charged) axial-vector and
polar-vector currents respectively. We refer to Table~\ref{table_dict} for the
explicit presentation of the renormalization pattern.

We have computed the RCs at three values of the gauge coupling, namely
$\beta=3.80$, $3.90$ and $4.05$, which correspond to inverse lattice spacing
$a^{-1} \simeq 2.0$, $2.3$ and $2.9$ GeV. Preliminary results of this work have
been presented in ref.~\cite{Dimopoulos:2007fn}. Further details concerning the
ETMC simulations can be found in refs.~\cite{plb}-\cite{Baron:2009wt}.

The paper's contents are the following. In Section \ref{sec:SI} we present the
novel approach of computing the scale independent RCs $Z_A$ and $Z_P/Z_S$ and
the Ward identity determination of $Z_V$. In Section \ref{sec:rimom} we discuss
the RI-MOM approach applied to tm quarks and provide the details of the
numerical analysis. A collection of our final results for the RCs can be found
in Tables~\ref{tab:final1}, \ref{tab:final2} and \ref{tab:final3}. In
Tables~\ref{tab:final1} and \ref{tab:final2} the predictions of one loop boosted
perturbation theory~\cite{pt,Og2a2} for the various RCs are also given. Finally
in the Appendix we present the proof of the automatic ${\cal O}(a)$--improvement
of the RI-MOM RCs  calculated with maximally twisted quarks.

An RI-MOM calculation of the RCs of bilinear quark operators similar to the one
described in the present paper but for the standard Wilson's plaquette gauge
action and non-perturbatively improved Wilson (clover) fermions has been
recently presented in ref.~\cite{Gockeler:2010yr}.

\section{A novel approach to the calculation of the scale independent RCs
\label{sec:SI}}

In this section we present a calculation of the scale independent RCs, namely
$Z_V$, $Z_A$ and $Z_P/Z_S$. The evaluation of $Z_V$ is based on the PCAC Ward
identity, which in the quenched approximation has led to very precise results
(see e.g.\ refs.~\cite{xlf_1,becirevic}). On the other hand, $Z_A$ and $Z_P/Z_S$
are computed from two-point correlators only with a new method, which is based
on the simultaneous use of two regularizations of the valence quark action. One
is the standard tm action at maximal twist, while the other is the OS
variant~\cite{OS}. In the so called physical basis these actions can be written
in the form:
\beq
\label{action}
S_{val} = a^4
\sum_{x} \sum_{q=u,d} \bar q(x) \left(\gamma \tilde{\nabla} - i\gamma_5 ~r_q~
W_{cr} + \mu_q \right) q(x) ~ ,
\eeq
with $ W_{cr} = -\frac{a}{2} \sum_{\mu} \nabla_{\mu}^{*} \nabla_{\mu} + M_{cr}$.
The Wilson parameters are $r_u=-r_d=1$ for the standard tm case and $r_u=r_d=1$
for OS quarks. In the sea sector we have two degenerate quarks, regularized in
the standard tm framework.

Let us illustrate the general idea of the calculation of the scale independent
RCs, $Z_A$ and $Z_P/Z_S$. It is based on the observation that the axial
transformations of the quark fields from the physical basis to the so called
twisted basis (primed fields),
\beq
\label{chirot}
q=\exp\left( i \gamma_5 r_q \pi/4 \right) q' ~~~~~\mbox{and}~~~~~ 
\bar q = \bar q' \exp\left( i \gamma_5 r_q \pi/4 \right) ~,
\eeq
transform the tm and OS actions of Eq.~(\ref{action}) into an action with the
Wilson term in the standard form (i.e. having no $\gamma_5$ and $r_u=r_d=1$) and
a mass term which takes the form $\left(i \mu_u \bar u' \gamma_5 u' - i \mu_d
\bar d' \gamma_5 d' \right)$ in the tm case and $\left(i \mu_u \bar u' \gamma_5
u' + i \mu_d \bar d' \gamma_5 d' \right)$ in the OS case. This also implies, in
turn, that the RCs for the operators in the twisted basis, being defined in the
chiral limit, are the same for the Wilson, tm and OS cases. Consider, now, a
non-singlet quark  bilinear operator, $O_{\Gamma} = \bar u \Gamma d$, defined in
terms of the fields of the physical basis. Under the axial transformations of
Eq.~(\ref{chirot}) this operator transforms into an operator in the twisted
basis, denoted as $O_{\tilde{\Gamma}}$ and $O_{\hat{\Gamma}}$ for the tm and OS
case respectively. In general, the two operators are not of the same form.
However their renormalized matrix elements between given physical states, say
$Z_{O_{\tilde{\Gamma}}} \langle \alpha \vert O_{\tilde{\Gamma}} \vert \beta
\rangle^{tm}$ and $Z_{O_{\hat{\Gamma}}} \langle \alpha \vert O_{\hat{\Gamma}}
\vert \beta \rangle^{OS}$ are estimates of the same physical matrix element
$\langle \alpha \vert (O_{\Gamma})_R \vert \beta \rangle$ up to ${\cal O}(a^2)$
errors:
\beq
\label{renorm}
 \langle \alpha \vert (O_{\Gamma})_R \vert \beta \rangle =
 Z_{O_{\tilde{\Gamma}}} \langle \alpha \vert O_{\tilde{\Gamma}} \vert \beta
\rangle^{tm} + {\cal O}(a^2) =
 Z_{O_{\hat{\Gamma}}} \langle \alpha \vert O_{\hat{\Gamma}} \vert \beta
\rangle^{OS} + {\cal O}(a^2) ~~. 
\eeq
RCs are named, as anticipated, after the twisted basis, in which the Wilson term
has its standard form. The operator renormalization pattern in the physical and
twisted bases is given in Table~\ref{table_dict} for both OS and tm formulations
at maximal twist. The primed operators refer to the twisted basis while the
unprimed ones to the physical basis. Notice that the condition of maximal twist
ensures that cut-off effects are of order ${\cal O}(a^2)$~\cite{FR}.

\begin{table}[t]
\begin{center}
\renewcommand{\arraystretch}{1.2}
\begin{tabular}{cccccccc}
\hline \hline
               OS case &&&&&&&                tm case    \\
$(A_R)_{\mu, ud} = Z_A A_{\mu, ud} =  Z_A A_{\mu, ud}^{'}$ &&&&&&& 
$~~~(A_R)_{\mu, ud} =  Z_V A_{\mu, ud} =  -i Z_V V_{\mu, ud}^{'}$ \\
$(V_R)_{\mu, ud}  = Z_V V_{\mu, ud} =  Z_V V_{\mu, ud}^{'}$  &&&&&&&
$~~~~(V_R)_{\mu, ud} = Z_A V_{\mu, ud} =  -i Z_A A_{\mu, ud}^{'}$ \\
$(P_R)_{ud}      = Z_S P_{ud}      ~~~=  i Z_S S_{ud}^{'}$     &&&&&&&
\hspace{-0.3cm}$(P_R)_{ud}     ~~= Z_P P_{ud}     ~~= Z_P P_{ud}^{'}$\\
$(T_R)_{\mu\nu, ud}  = Z_T T_{\mu\nu, ud} =  i Z_T \tilde T_{\mu\nu, ud}^{'}$
&&&&&&&
$~~(T_R)_{\mu\nu, ud}  = Z_T T_{\mu\nu, ud} =  Z_T T_{\mu\nu, ud}^{'}$\\
\hline \hline
\end{tabular}
\renewcommand{\arraystretch}{1.0}
\end{center}
\caption{\sl Renormalization pattern of the bilinear quark operators for the OS
and tm case at maximal twist. The unprimed operators refer to the physical basis
while the primed ones to the twisted basis. The symbols $P_{ud}$ and $\tilde
T_{\mu\nu,ud}$ indicate the operators $\bar u \gamma_5 d$ and $\bar u
\sigma_{\mu\nu} \gamma_5 d$.}
\label{table_dict}
\end{table}

The main point of Eq.~(\ref{renorm}) is that the tm and OS determinations of the
same continuum matrix element are equal up to discretization effects. Our
proposal for the determination of  $Z_P/Z_S$ and $Z_A$, described in the
following subsections, is based on this observation.

\subsection{Calculation of $Z_P/Z_S$}
The method is based on comparing the  amplitude  $g_{\pi} = \langle 0|P|\pi
\rangle$, computed both in tm and OS regularizations. We start by considering,
in the physical basis, the correlation function
\beq
\label{CPP_def}
C_{PP}(t) \equiv - \sum_{\bf x} 
\langle\bar{u}\gamma_5 d (x) ~ \bar{d}\gamma_5 u
(0)\rangle \,\, ,
\eeq
which at large time separations behaves like
\beq
\label{CPP}
 C_{PP}(t) \simeq \frac{|g_{\pi}|^2}{2m_{\pi}} [\exp(-m_{\pi} t) + 
\exp(-m_{\pi} (T-t))]
\eeq
From Table~\ref{table_dict} we see that in the twisted basis this corresponds to
$C_{S^{'}S^{'}}(t)$ (for OS regularization) and $C_{P^{'}P^{'}}(t)$ (for tm
regularization):
\beq 
\label{C_PP}
[C_{PP}(t)]^{cont} \,\, = \,\, Z_P^2 \, [C_{P^\prime P^\prime}(t)]^{tm} + {\cal
O}(a^2) \,\, = \,\,  Z_S^2 \, [C_{S^\prime S^\prime}(t)]^{OS} + {\cal O}(a^2)
\,\,\,\, ,
\eeq
which implies
\beq 
\label{g}
[g_{\pi}]^{cont} \,\,   = \,\, Z_P \, [g_{\pi}]^{tm} + {\cal O}(a^2)
     \,\, = \,\, Z_S \, [g_{\pi}]^{OS} + {\cal O}(a^2) \,\,\,\, .
\eeq
The ratio $Z_P/Z_S$ is extracted from the above equation in the chiral limit. As
our simulations are performed at finite quark masses, an extrapolation of our
$Z_P/Z_S$-estimators to zero quark mass is necessary.

\subsection{Calculation of $Z_A$}
In order to compute $Z_A$, we consider the calculation of the charged
pseudoscalar meson decay constant $f_{\pi}$, in both OS and tm regularizations.
We start with the tm regularization. From the axial Ward identity in the
physical basis, we have the well-known result for the pseudoscalar meson decay
constant~\cite{frezzotti}:
\beq
\label{fpi_tm}
[f_{\pi}]^{cont} \,\, = \,\, [f_\pi ]^{tm} + {\cal O}(a^2)  \,\, = \,\, 
\Bigg [ \dfrac{ 2 \mu_q g_\pi}{m_{\pi}^2} \Bigg ]^{tm} + {\cal O}(a^2)
\,\, .
\eeq
Note that in this case no RC is needed \cite{frezzotti}. Thus the pion decay
constant can be extracted from the large time asymptotic behaviour of
$[C_{P^\prime P^\prime}(t)]^{tm}$.

In the OS formulation we consider, besides the correlator $C_{PP}$ of
Eq.~(\ref{CPP_def}), also $C_{AP}$, which is defined by
\beq
\label{CAP_def}
C_{AP}(t) \equiv \sum_{\bf x} \langle \bar{u}\gamma_0\gamma_5 d (x)~
\bar{d}\gamma_5
u (0)\rangle ~.
\eeq
Its large time asymptotic behaviour is:
\beq
\label{CAP}
C_{AP}(t) \simeq \frac{\xi_{AP}}{2m_{\pi}} [\exp(-m_{\pi} t) - \exp(-m_{\pi}
(T-t))] \,\, ,
\eeq
with
\beq
\label{xiAP}
\xi_{AP} \,\, = \,\, \langle 0 \vert A_0 \vert \pi \rangle \langle \pi \vert P
\vert 0 \rangle \,\, = \,\, 
[ f_\pi m_\pi i g_\pi ]^{OS} \,\, .
\eeq
The last equation may be solved for $f_\pi$, which is thus obtained from the
correlation functions of Eqs.~(\ref{CPP}) and (\ref{CAP}).  With the aid of
Table~\ref{table_dict}, this solution is expressed in terms of OS-regularized
quantities in the twisted basis (note $\xi_{AP} = i\xi_{A'S'}$)  as follows:
\beq
\label{fpi_OS}
[f_{\pi}]^{cont} \,\, = \,\, Z_A [f_{\pi}]^{OS} + {\cal O}(a^2)  \,\, =
\,\, Z_A \frac{\xi_{A^\prime S^\prime}}{[g_{\pi}]^{OS} [m_{\pi}]^{OS}} + 
{\cal O}(a^2)  \,\, .
\eeq

Combining Eqs.~(\ref{fpi_tm}) and (\ref{fpi_OS}), we obtain
\beq 
\label{fp}
[f_{\pi}]^{cont} \,\, = \,\, [f_\pi]^{tm} +{\cal O}(a^2) \,\,
 = \,\, Z_A [f_\pi]^{OS} + {\cal O}(a^2) \,\,\, ,
\eeq
from which an estimate of $Z_A$ can be extracted. Again, results obtained at
finite quark masses need to be extrapolated to the chiral limit. In
ref.~\cite{dsv} the discretization effects affecting the quenched pseudoscalar
decay constant in the tm and OS regularizations have been studied in detail. The
tm and OS results were compatible within one standard deviation for three values
of the lattice spacing, indicating the smallness of ${\cal O}(a^2)$ cut-off
effects.

\subsection{Calculation of $Z_V$}
The determination of the renormalization constant $Z_V$ can be done using {\em
solely} tm quarks. It is based on the comparison of the point-like axial current
$A_{\mu,ud}$, which renormalizes with $Z_V$ (see Table~\ref{table_dict}), with
the exactly conserved one-point split current $[V_\mu^\prime]^{\rm 1PS}_{ud} =
[A_\mu]^{\rm 1PS}_{ud}$. The four-divergence of the latter is exactly equal to
the sum of the valence quark mass values, $(\mu_1+\mu_2)$, times the
corresponding pseudoscalar density. Therefore, we extract $Z_V$ by solving the
equation:
\beq
Z_V \,\tilde \partial_0\, [C_{V^\prime P^\prime}(t)]^{tm} = (\mu_1+\mu_2)\, 
[C_{P^\prime P^\prime}(t)]^{tm}
\eeq
where $\tilde \partial_0$ is the symmetric lattice time derivative, and
extrapolating results to the chiral limit.

\subsection{Results}

\begin{figure}[t]
\begin{center}
\subfigure[]{\includegraphics[scale=0.28,angle=-90]{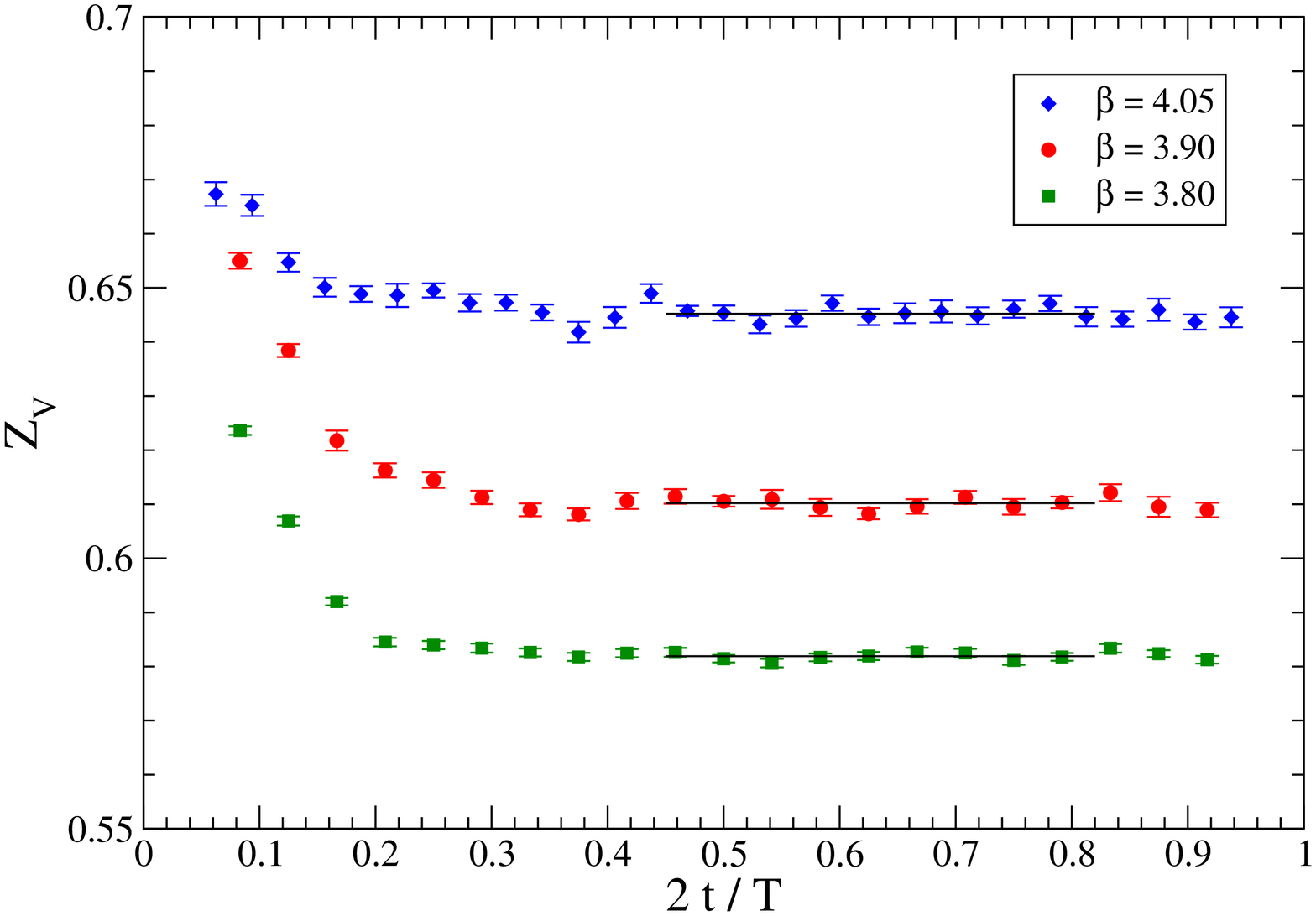}}
\subfigure[]{\includegraphics[scale=0.28,angle=-90]{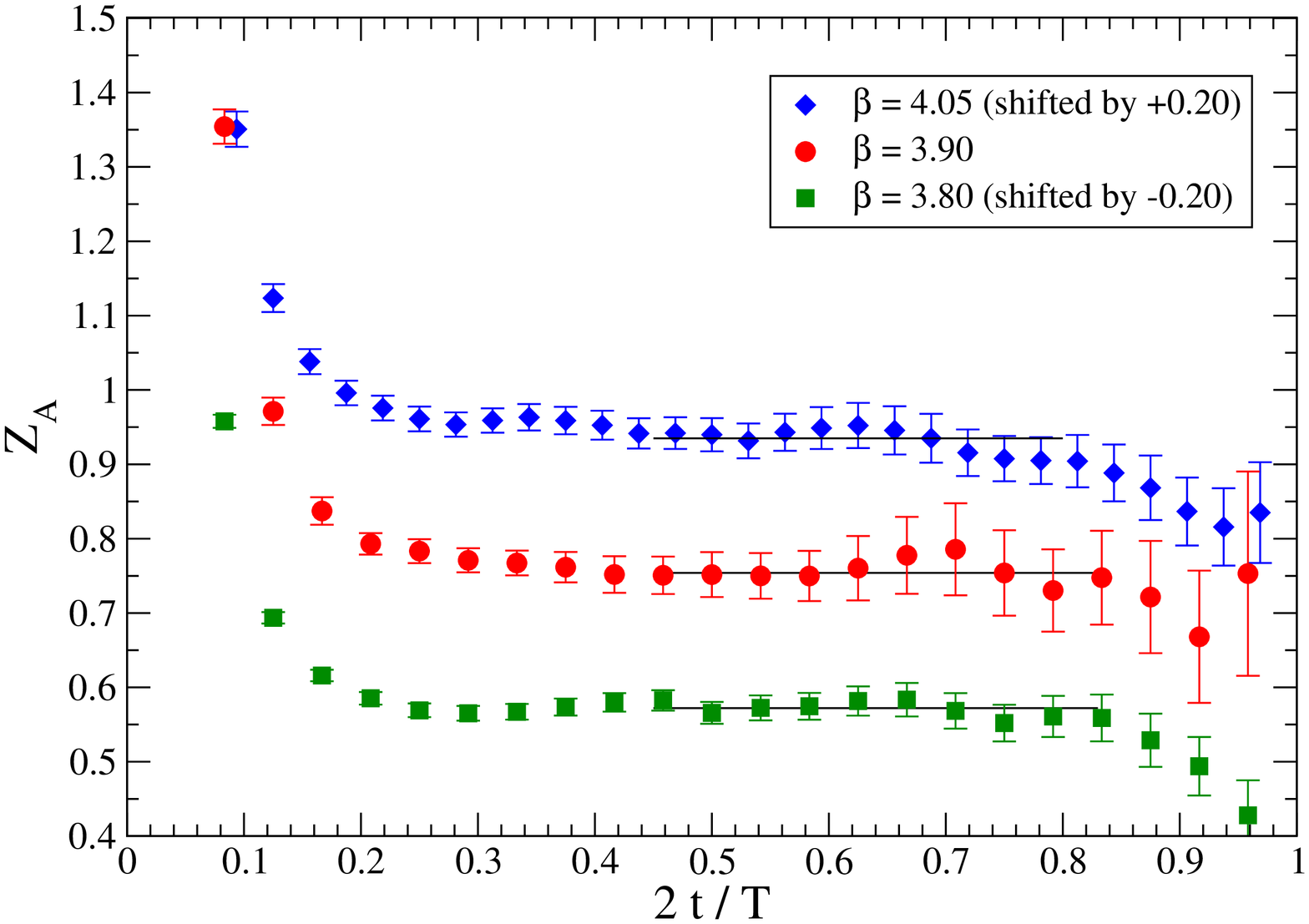}}
\subfigure[]{\includegraphics[scale=0.28,angle=-90]{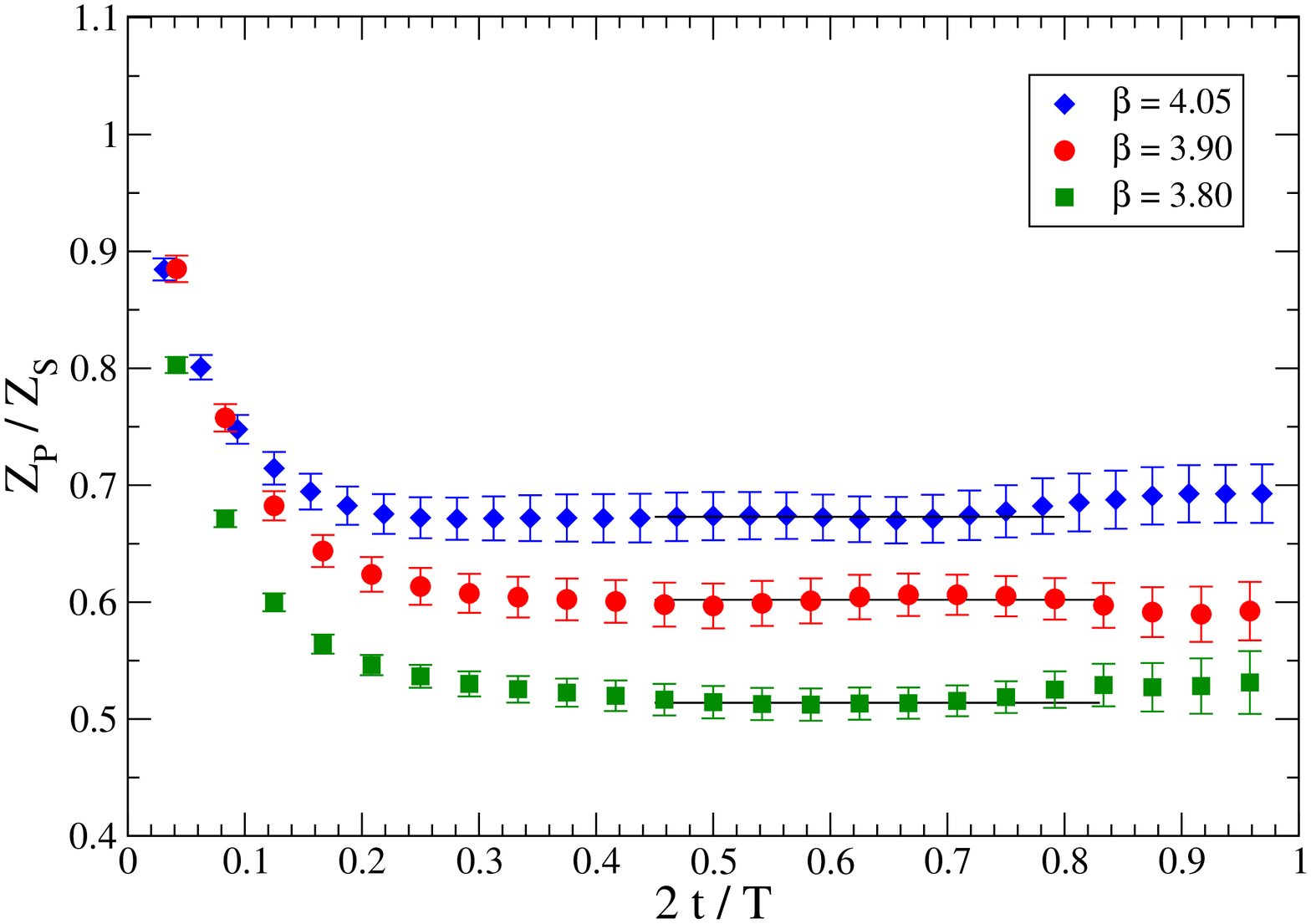}}
\end{center}
\vspace{-0.5cm}
\caption{\sl The plateaux quality for $Z_V$ (a), $Z_A$ (b) and $Z_P/Z_S$ (c) for
three gauge couplings and masses $a\mu_{sea}^{min}= a\mu_{1}= a\mu_{2}$. From
Table~\ref{simuldetails} we have $a\mu_{sea}^{min}(\beta=3.80) =0.0080$,
$a\mu_{sea}^{min}(\beta=3.90)=0.0040$ and $a\mu_{sea}^{min} (\beta=4.05)=
0.0030$. Note that in Fig.(b) the data have been shifted for clarity.}
\label{plateaux}
\end{figure}

In Table~\ref{simuldetails} we give various details of our $N_f=2$, partially
quenched simulations. The smallest sea quark mass corresponds to a pion of about
300 MeV for the two larger $\beta$ values, while for the smallest coupling the
lightest pion weights approximately 400 MeV. The highest sea quark mass is
around half the strange quark mass. For the inversions in the valence sector we
have made use of the stochastic method (one--end trick of ref.~\cite{chris}) in
order to increase the signal to noise ratio. Propagator sources have been placed
at randomly located timeslices. This turned out to be an optimal way to further
reduce the autocorrelation time. Our correlation functions are computed for all
combinations of valence quark masses appearing in Table~\ref{simuldetails}. In
order to deal with effectively independent measurements of RC-estimators, we
have selected gauge field configurations separated by 20 HMC (length 1/2)
trajectories.  Within each ensemble of such gauge configurations statistical
errors are evaluated using a jackknife procedure. For results involving an
extrapolation in the sea quark mass and/or combined fits at several lattice
couplings, statistical errors have been estimated with a bootstrap procedure for
a 1000 bootstrap events. Typical plots on the data quality of $Z_V$, $Z_A$ and
$Z_P/Z_S$ are shown in Fig.~\ref{plateaux}; in this example we show the case
$a\mu_{sea}^{min}=a\mu_{1}=a\mu_{2}$ for the three values of the gauge coupling.
\begin{table}[t]
\begin{center}
\renewcommand{\arraystretch}{1.5}
\begin{tabular}{|c|c c c l c|} \cline{2-6}
\multicolumn{1}{c|}{} & $\beta$ & $L^3 \times T$ & $a \mu_{sea}$ &
\hspace{3cm} $a \mu_{val}$ & $N_{conf}$ \\ \hline
\hline 
$A_2$ & 3.80  & $24^3 \times 48$
          & 0.0080 & \{0.0080, 0.0110, 0.0165, 0.0200\} & 400/240 \\
$A_3$ &&  & 0.0110 & \{0.0080, 0.0110, 0.0165, 0.0200\} & 400/240 \\
$A_4$ &&  & 0.0165 & \{0.0080, 0.0110, 0.0165, 0.0200\} & 400/240 \\ \hline
\hline 
$B_1$ & 3.90  & $24^3 \times 48$
          & 0.0040 & \{0.0040, 0.0064, 0.0085, 0.0100, 0.0150\} & 240/240 \\
$B_2$ &&  & 0.0064 & \{0.0040, 0.0064, 0.0085, 0.0100, 0.0150\} & 240/240 \\
$B_3$ &&  & 0.0085 & \{0.0040, 0.0064, 0.0085, 0.0100, 0.0150\} & 240/240 \\
$B_4$ &&  & 0.0100 & \{0.0040, 0.0064, 0.0085, 0.0100, 0.0150\} & 240/240 \\
\hline
\hline 
$C_1$ & 4.05 & $32^3 \times 64$
          & 0.0030 & \{0.0030, 0.0060, 0.0080, 0.0120\} & 130/240 \\
$C_2$ &&  & 0.0060 & \{0.0030, 0.0060, 0.0080, 0.0120\} & 130/160 \\
$C_3$ &&  & 0.0080 & \{0.0030, 0.0060, 0.0080, 0.0120\} & 130/160 \\
\hline 
\end{tabular}
\renewcommand{\arraystretch}{1.0}
\end{center}
\caption{\sl Details of the simulations performed for computing RCs. The number
$N_{conf}$ of gauge configurations we analysed is given for both the case of the
scale independent RCs discussed in this section (first figure) and the RI-MOM
analysis of Sect.~3 (second figure).}
\label{simuldetails}
\end{table}

\begin{figure}[p]
\begin{center}
\subfigure[]{\includegraphics[scale=0.28,angle=-90]{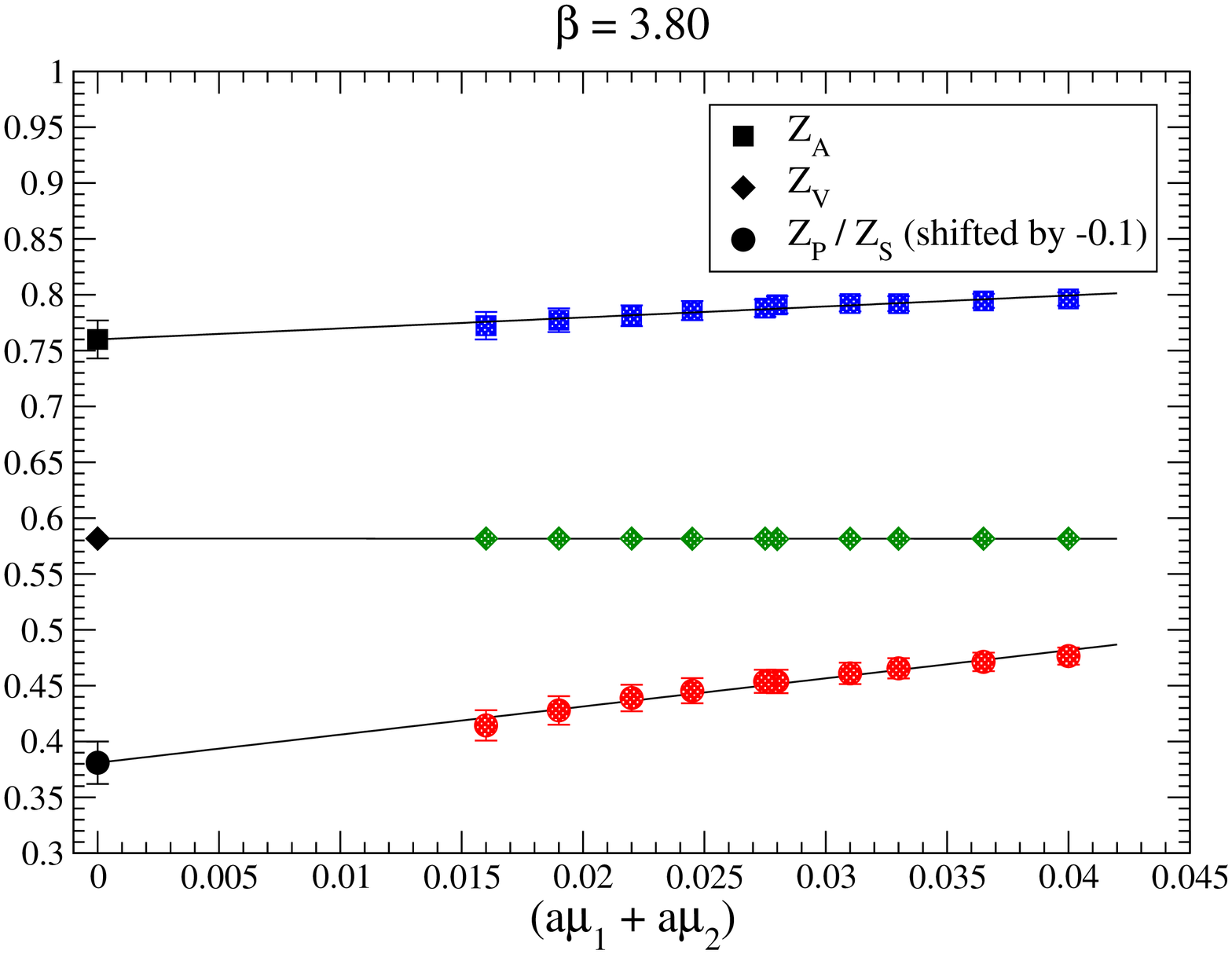}}
\subfigure[]{\includegraphics[scale=0.28,angle=-90]{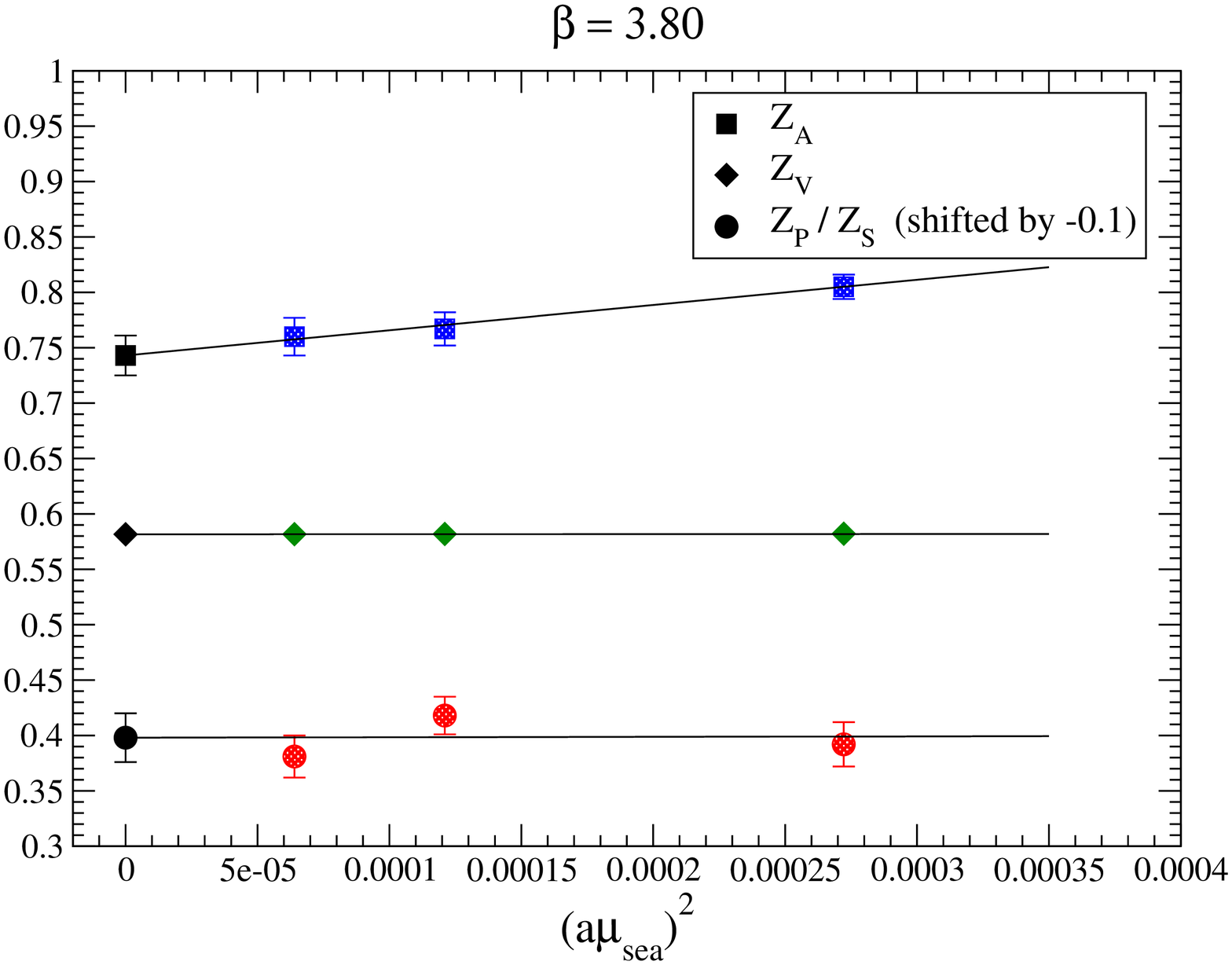}}
\subfigure[]{\includegraphics[scale=0.28,angle=-90]{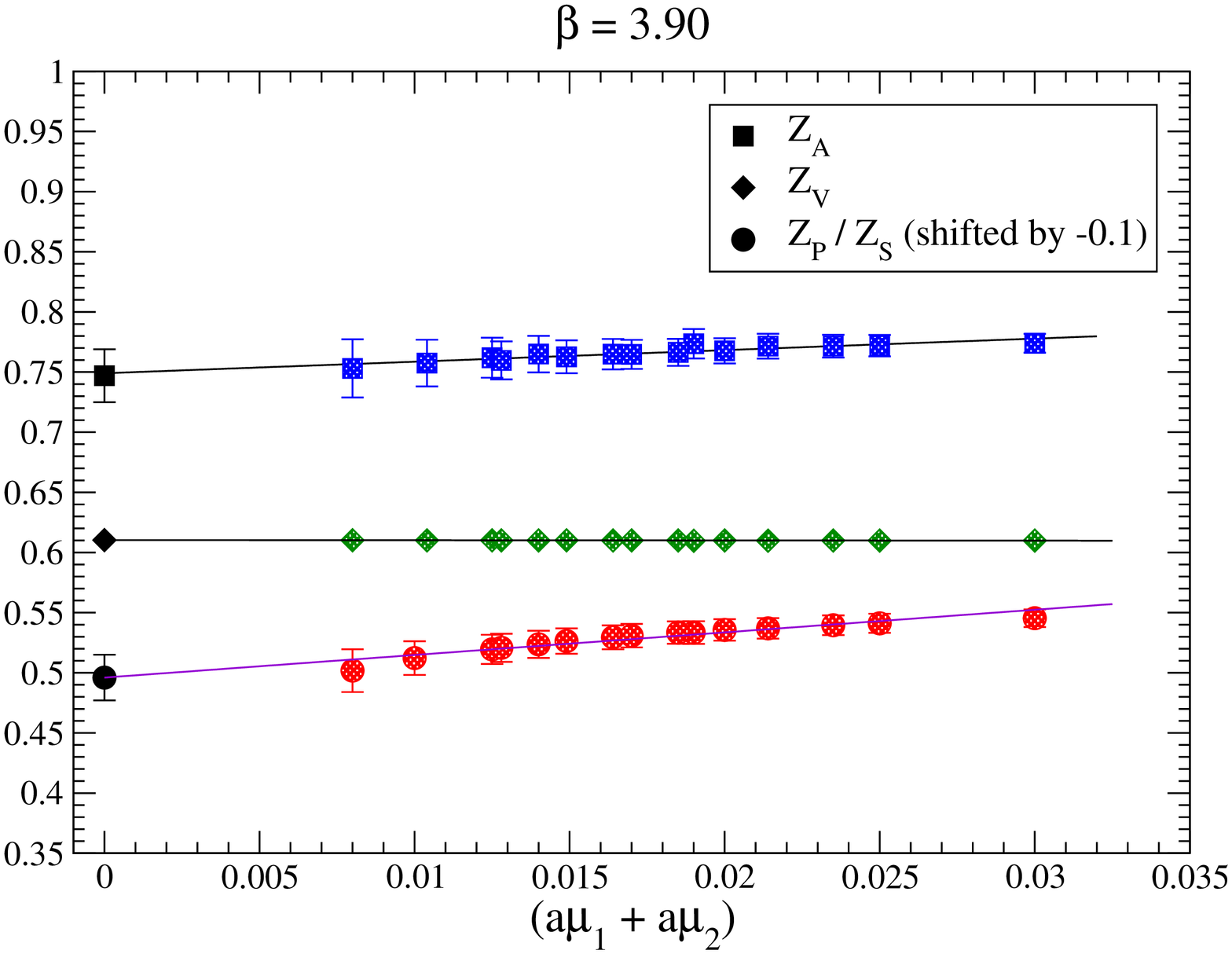}}
\subfigure[]{\includegraphics[scale=0.28,angle=-90]{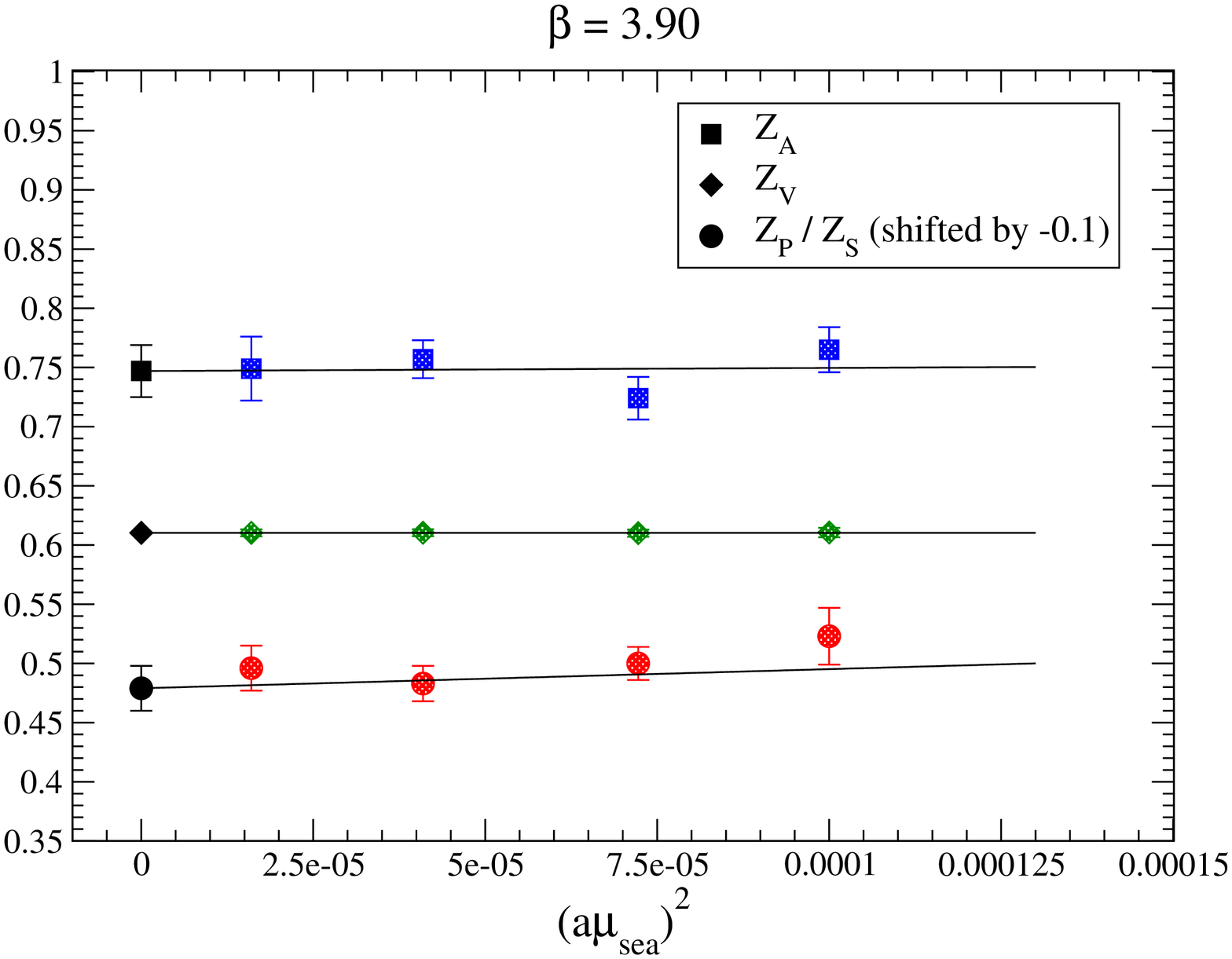}}
\subfigure[]{\includegraphics[scale=0.28,angle=-90]{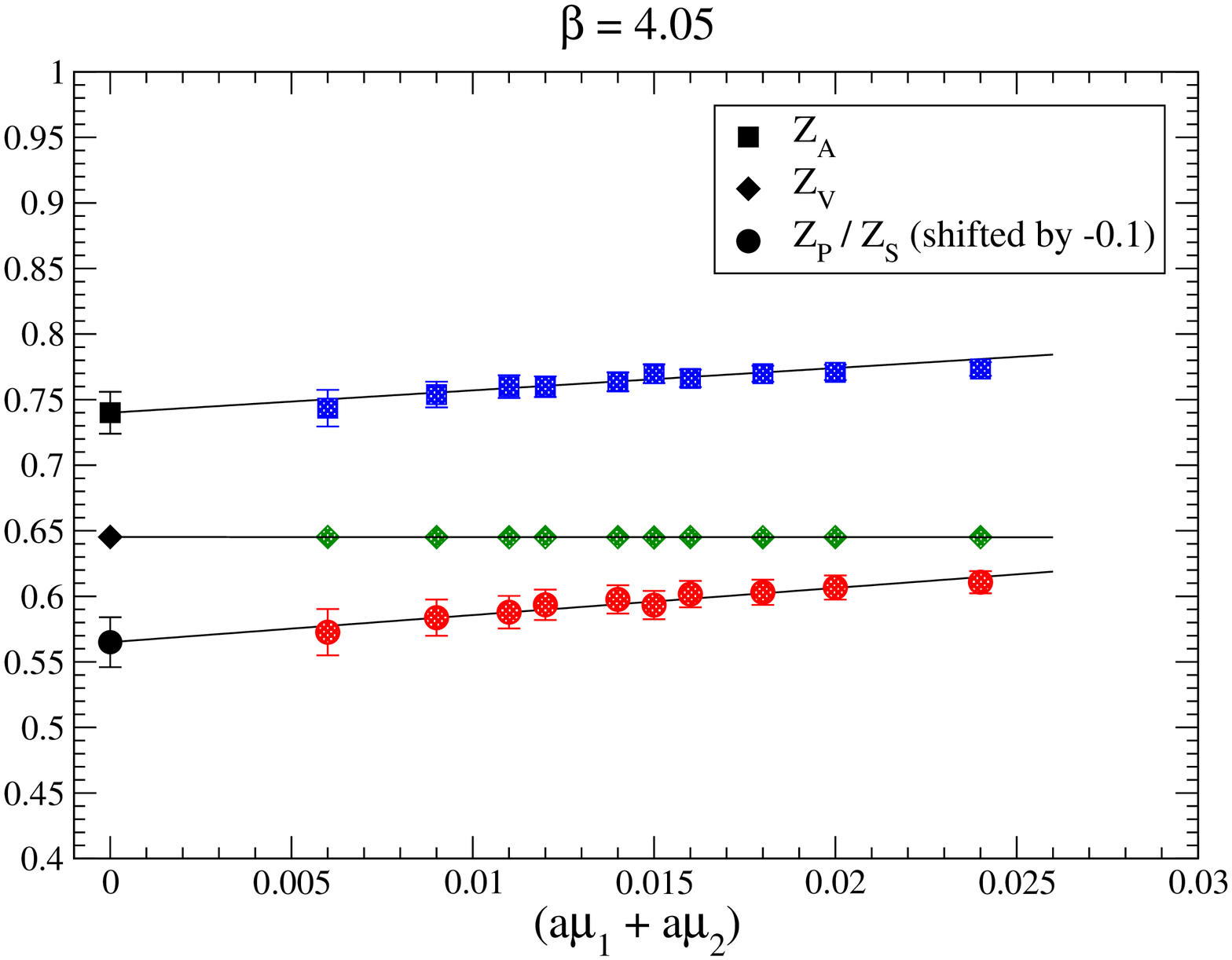}}
\subfigure[]{\includegraphics[scale=0.28,angle=-90]{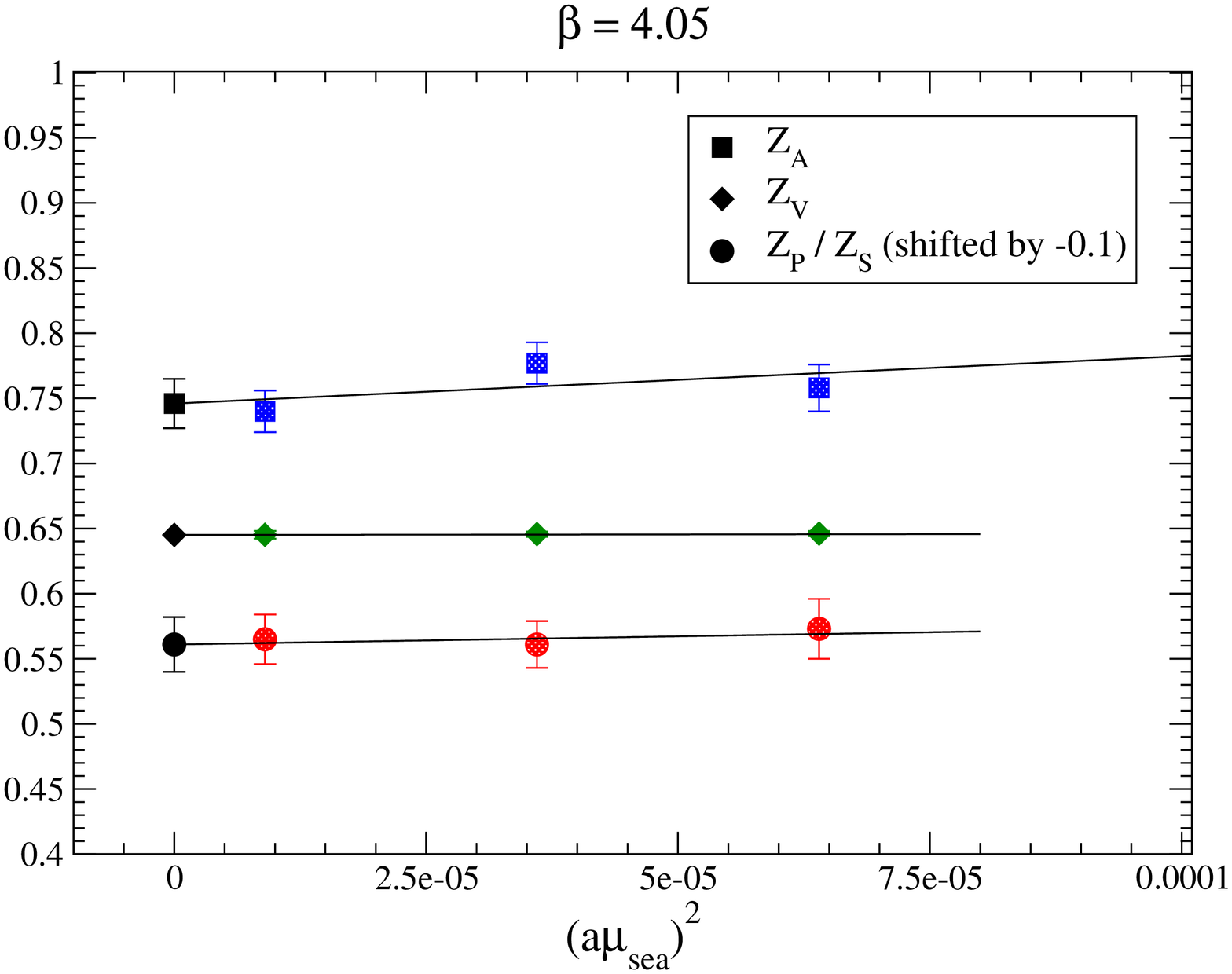}}
\end{center}
\vspace{-0.5cm}
\caption{\sl Left panel: valence chiral limit for the three scale independent
RCs at $\beta=3.80$, $a\mu_{sea}=0.0080$ (a), $\beta=3.90$, $a\mu_{sea}=0.0040$
(c) and $\beta=4.05$, $a\mu_{sea}=0.0030$ (e). Right panel: sea chiral limit
for the same constants at $\beta=3.80$ (b), $\beta=3.90$ (d) and $\beta=4.05$
(f). Note that the data for $Z_P/Z_S$ have been shifted for clarity.}
\label{extra}
\end{figure}

Various methods have been implemented in taking the chiral limit (in the sea and
valence quark sector). A first method consists in calculating the RCs at fixed
sea quark mass for three choices of valence quark pairs, namely with
$a\mu_{1}=a\mu_{2}$, or for all pairs of ($a\mu_{1}, a\mu_{2}$) or for
$(a\mu_{1}, a\mu_{2}) \ge a\mu_{sea}$ and taking the ``valence chiral limit"
using a linear fit in terms of the sum of the valence quark masses.
Subsequently, the RCs were quadratically extrapolated to the sea quark chiral
limit. Note that a quadratic dependence on $a\mu_{sea}$ is expected from the
form of the sea quark determinant, assuming that  lattice artefacts on the RCs
are not sensitive to  spontaneous chiral symmetry breaking. However we have
verified  that a linear fit in $\mu_{sea}$ leads to compatible results, albeit
with larger final errors. A second method consists in inverting the order of the
two chiral limits. A third method is simply the extraction of the RCs from the
subset of data satisfying $\mu_{valence}=\mu_{sea}$. This allows reaching the
chiral limit with a single, linear extrapolation in the quark mass. In most
cases the quality of the fits is  very good and the final results, obtained from
these different extrapolations, are compatible within one standard deviation. We
have opted to quote as final results those produced by the first method, for all
pairs of ($a\mu_{1}, a\mu_{2}$), followed by a linear fit in terms of
$a^2\mu_{sea}^2$. In Fig.~\ref{extra} (left panel) we show the ``valence chiral
limit" extrapolation of the three scale independent RCs, computed at the
lightest sea quark mass for each gauge coupling. In Fig.~\ref{extra} (right
panel) we present the chiral limit in the sea sector for all the RCs for each
gauge coupling.

Our final results are collected in Table~\ref{tab:final1} (see rows labelled as
``Alt. methods"). For comparison with the RI-MOM results and conclusions on the
precision of the methods used, see below. Here we point out that the $Z_V$ value
at $\beta=3.90$ compares nicely with the result of Ref.~\cite{Frezzotti:2008dr},
produced from 3- and 2-point correlation functions, in the context of a
calculation of the pion form factor.

\section{Renormalization constants in the RI-MOM scheme 
\label{sec:rimom}}
The non-perturbative determination of the RCs in the RI-MOM scheme~\cite{rimom}
is based on the numerical evaluation, in momentum space, of correlation
functions of the relevant operators between external quark and/or gluon states.
In this work we are interested in the case of the bilinear quark operators
$O_\Gamma = \bar u \Gamma d$, where $\Gamma = S,P,V,A,T$ stands for $I,
\gamma_5, \gamma_\mu, \gamma_\mu\gamma_5, \sigma_{\mu\nu}$ respectively. The
fields $u$ and $d$ of the operator $O_\Gamma$ belong to a twisted doublet of
quarks, regularized by the tm action of Eq.~(\ref{action}), with $r_u=-r_d=1$.

The relevant Green functions are the quark propagator
\beq
S_q(p) = a^4 \sum_x \, e^{-i p\cdot x} \, \langle q(x) \bar q(0) \rangle 
\qquad \qquad (q=u,d)\, ,
\eeq
the forward vertex function
\beq
{G}_\Gamma(p) = a^8 \sum_{x,y} \, e^{-i p\cdot (x-y)} \, \langle u(x) O_\Gamma(0)
\bar d(y) \rangle \, ,
\label{eq:gg}
\eeq
and the amputated Green function\footnote{The calculation of ${G}_\Gamma(p)$ in
Eq.~(\ref{eq:gg}) involves the propagator $S_d(0,y)$ which, in the tm approach,
equals $\gamma_5 S_u(y,0)^\dagger \gamma_5$. Similarly, in Eq.~(\ref{eq:ll}) the
propagator $S_d(p)$ stands for $\sum_y \, e^{i p\cdot y} S_d(0,y) = \gamma_5
S_u(p)^\dagger \gamma_5$.}
\beq
\Lambda_\Gamma(p) = S_u(p)^{-1} {G}_\Gamma(p) S_d(p)^{-1} \, .
\label{eq:ll}
\eeq
The RI-MOM renormalization scheme consists in imposing that the forward
amputated Green function, computed in the chiral limit in the Landau gauge and
at a given (large Euclidean) scale $p^2=\mu^2$, is equal to its tree-level
value. In practice, the renormalization condition is implemented by requiring in
the chiral limit that
\beq
Z_q^{-1} Z_\Gamma \, {\cal V}_{\tilde\Gamma} (p) \vert_{p^2=\mu^2} \equiv
Z_q^{-1} Z_\Gamma \, {\rm Tr} [\Lambda_{\tilde\Gamma} (p)\, P_{\tilde\Gamma}]\vert_{p^2=\mu^2}=1 \,\, ,
\qquad \tilde\Gamma = e^{-i\gamma_5\pi/4} \Gamma e^{i\gamma_5\pi/4} \,\, ,
\label{eq:rimom}
\eeq
where $P_{\tilde\Gamma}$ is a Dirac projector satisfying ${\rm Tr} \,
[\tilde\Gamma \, P_{\tilde\Gamma}] =1$.\footnote{The appearance of
$\tilde\Gamma$ in Eq.~(\ref{eq:rimom}) is a trivial consequence
of the fact that RCs, as we said, are named after the (here unphysical) twisted
quark basis while the operator $O_\Gamma = \bar u \Gamma d$ is expressed in the
physical quark basis.} The quark field RC $Z_q$, which enters
Eq.~(\ref{eq:rimom}), is obtained by imposing, in the chiral limit, the
condition\footnote{Strictly speaking, the renormalization condition of
Eq.~(\ref{eq:zqri}) defines the so called RI$^\prime$ scheme. In the original
RI-MOM scheme the quark field renormalization condition reads
\[
Z_q^{-1} \,\frac{-i}{48}\,{\rm Tr}\left[\gamma_\mu \frac{\partial S_q(p)^{-1}}
{\partial p_\mu}\right] _{p^2=\mu^2}=1 \,.
\]
The two schemes differ in the Landau gauge at the N$^2$LO. In this work, when
perturbative predictions are used, the difference between the two schemes has
been properly taken into account.}
\beq
Z_q^{-1} \, \Sigma_1(p) \vert_{p^2=\mu^2} \equiv
Z_q^{-1} \,\frac{-i}{12}\,{\rm Tr}\left[\frac{\slash p\, S_q(p)^{-1}}{p^2}
\right]_{p^2=\mu^2}=1 \, .
\label{eq:zqri}
\eeq
In order to minimize discretization effects, we select momenta with components
$p_\nu=(2\pi/L_\nu)\, n_\nu$ in the intervals 
\beq
n_\nu= \left\{
\begin{array}{l}
([0,2],[0,2],[0,2],[0,3]) \\ ([2,3],[2,3],[2,3],[4,7]) 
\end{array} \right.
~~, \quad \mathrm{for~} L=24 ~~~ (\beta=3.8 ~ \mathrm{and} ~ 3.9)
\eeq
and
\beq
n_\nu= \left\{
\begin{array}{l}
([0,2],[0,2],[0,2],[0,3]) \\ ([2,5],[2,5],[2,5],[4,9]) 
\end{array} \right.
~~, \quad \mathrm{for~} L=32 ~~~ (\beta=4.05)
\eeq
($L_\nu$ is the lattice size in the direction $\nu$). The time component of the
four-momentum is shifted by the constant $\Delta p_4=\pi/L_4$, in order to
account for the use of anti-periodic boundary conditions on the quark fields in
the time direction. Furthermore, we have only considered for the final RI-MOM
analysis the momenta satisfying
\beq
\Delta_4(p) \equiv \frac{\sum_\rho \tilde p^{\,4}_\rho}{\left(\sum_\rho \tilde
p^2_\rho\right)^{\,2}} < 0.28 \, ,
\eeq
where
\beq
\tilde p_\nu \equiv \frac{1}{a} \sin(a\, p_\nu)  \, ,
\eeq
which helps in minimizing the contribution of Lorentz non-invariant
discretization effects (cfr Eq.~(\ref{eq:v1loop})).

In order to improve the statistical accuracy of the RCs, we have averaged the
results obtained from the correlation functions of the operators $O_\Gamma =
\bar u \Gamma d$ and $O_\Gamma^\prime = \bar d \Gamma u$. Similarly, in the case
of the quark field RC, we have computed $Z_q$ by averaging the results obtained
for the up and down quark propagators.

The RCs, calculated in the chiral limit in the way described above, are
automatically improved at ${\cal O}(a)$. Actually in the maximal twist situation
${\cal O}(a)$ improvement holds for all external momenta $p$ and non-zero
(valence and sea) quark mass at the level of the basic quantities ${\rm Tr}
[\Lambda_\Gamma (p)\, P_\Gamma]$ and $ {\rm Tr}\left[(\slash p\,
S_q(p)^{-1})/p^2\right] $ entering in Eqs.~(\ref{eq:rimom})-(\ref{eq:zqri}). A
proof of this statement, which follows from an analysis based on the symmetries
of the tm action at maximal twist and the $O(4)$ symmetry of the underling
continuum theory, is given in the appendix.\footnote{With the standard Wilson or
Clover actions, the RCs  determined with the RI-MOM method are also ${\cal
O}(a)$-improved. The RC-estimators are however improved only at large momenta,
where spontaneous chiral symmetry breaking effects can be neglected~\cite{zeta}.
This is sufficient, as the RI-MOM scheme is defined precisely in the
high-momentum region.}

Details of the RI-MOM simulation are collected in Table~\ref{simuldetails}. The
lattice parameters are those used also in the determination of the scale
independent constants with the methods discussed in the previous section.
However, a different ensemble of independent gauge configurations has been
analyzed in each case. Moreover, the inversions in the valence sector for the
RI-MOM study have been performed without using the stochastic approach, but
with point-like sources randomly located on the lattice for each gauge
configuration.

\subsection{Analysis of the twisted mass quark propagator}
The necessary Green functions for the RI-MOM determination of RCs of bilinear
quark operators are the quark propagator $S_q(p)$ and the amputated vertex
$\Lambda_\Gamma(p)$, evaluated in momentum space in the Landau gauge. In this
section, we first illustrate the results on the tm quark propagator. 

At ${\cal O}(a)$, the spin-flavour structure of the lattice artefacts of the
quark propagator differs from that of the standard Wilson case, due to the
twisted  Wilson term in the action. With tm fermions, the explicit breaking of
parity at finite lattice spacing allows for the presence of a parity violating
contribution in the quark propagator. By neglecting O(4) symmetry violating
effects, which only appear at ${\cal O}(a^2)$, the inverse quark propagator can
be expressed in terms of three scalar form factors, as follows:
\beq
\label{eq:spm1}
S_q(p)^{-1} = i\, \slash p \, \Sigma_1(p^2) + \Sigma_2(p^2) - i \, \gamma_5 \,
\Sigma_3(p^2) \, .
\eeq
At large $p^2$, $\Sigma_1$ and $\Sigma_2$ are related to the quark field RC
$Z_q$ (see Eq.~(\ref{eq:zqri})), and to the renormalized quark mass $\hat\mu_q$
respectively~\cite{qprop}. The parity violating term proportional to $\Sigma_3$
represents an ${\cal O}(a)$ discretization effect, induced by the twisted Wilson
term in the action.\footnote{In the standard Wilson case, the analogous ${\cal
O}(a)$ artifact contributes to the form factor $\Sigma_2$.} At maximal twist,
the form factors $\Sigma_{1,2,3}$ are given at tree-level by
\beq
\label{eq:tlsigmai}
\Sigma_1(p^2) =1 \quad , \quad \Sigma_2(p^2) = \mu_q \quad , \quad \Sigma_3(p^2)
=\frac{a \,r_q}{2}\, p^2 \,.
\eeq

\begin{figure}[p]
\begin{center}
\vspace{-0.5cm}
\subfigure{\includegraphics[scale=0.35,angle=270]{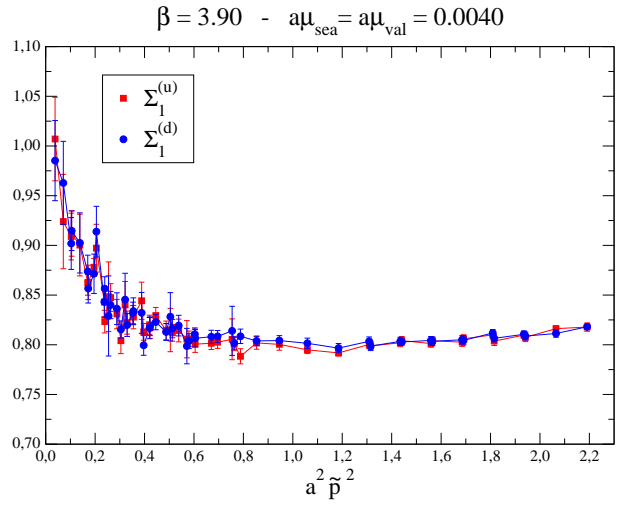}}
\subfigure{\includegraphics[scale=0.35,angle=270]{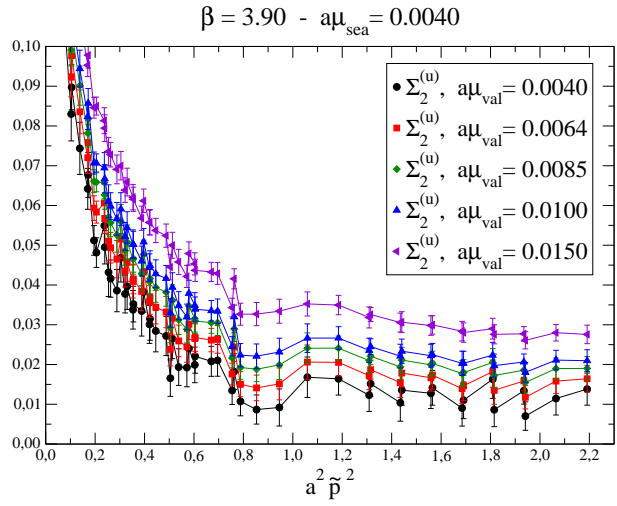}}
\subfigure{\includegraphics[scale=0.35,angle=270]{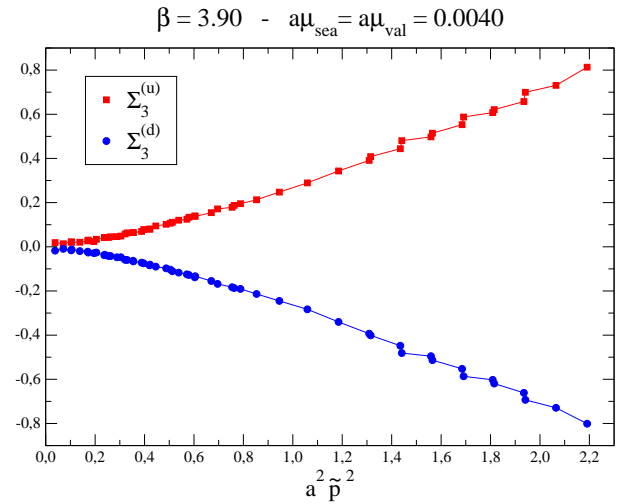}}
\end{center}
\vspace{-1.0cm}
\caption{\sl The form factors $\Sigma_1$ (top), $\Sigma_2$ (center) and
$\Sigma_3$ (bottom), computed at $\beta=3.90$ with $a \mu_{sea}=0.0040$, as a
function of $a^2 \tilde p^2$.}
\label{fig:sigma}
\end{figure}

In Fig.~\ref{fig:sigma}, the non-perturbatively determined form factors
$\Sigma_{1,2,3}$ are plotted against $a^2 \tilde p^2$, for $\beta=3.90$ and $a
\mu_{sea}=0.0040$.
The following observations are in order:
\begin{itemize} 
\item The form factor $\Sigma_1$ exhibits a nice plateau at large $p^2$, with a
tiny slope which is partly due to a NLO renormalization scale dependence (the
quark field anomalous dimension vanishes at LO in the Landau gauge) and partly
to ${\cal O} (a^2)$ discretization effects.  The quark field RC $Z_q$ is
obtained from the $\Sigma_1$ results at large $p^2$, extrapolated to the chiral
limit (see Eq.~(\ref{eq:zqri})).

\item The form factor $\Sigma_2$ also exhibits a plateau at large $p^2$, which
is expected to be proportional, up to the quark field RC, to the renormalized
valence quark mass $\hat\mu_q$ in the RI-MOM scheme. Indeed, at variance with
the cases of $\Sigma_1$ and $\Sigma_3$, a clear separation among the results
obtained for $\Sigma_2$ at different bare valence quark masses is visible in the
plot.

\item As expected, the form factor $\Sigma_3$, which represents a pure ${\cal O}
(a)$ discretization effect, has opposite sign for the up and down quark
propagators ($r_u=-r_d=1$). Its behaviour is quite close to the tree level
estimate of Eq.~(\ref{eq:tlsigmai}). We find $\Sigma_3 \simeq \pm \, c_q\,
ap^2/2$, with $c_q \simeq 0.8 \div 0.9$.
\end{itemize} 

\subsection{Renormalization constants}
The RCs in the RI-MOM scheme are determined by closely following the procedure
summarised in section~\ref{sec:rimom} above and described in detail in
ref.~\cite{zeta}. This procedure involves two main steps: the chiral limit
extrapolation of the RCs, at fixed coupling and renormalization scale, and the
study of the renormalization scale dependence.

\subsubsection{Chiral extrapolations}
RI-MOM is a mass independent renorma\-li\-za\-tion scheme. Since in practice
the RCs are obtained at non vanishing values of the quark masses, an
extrapolation of the results to the chiral limit, both in the valence and the
sea quark masses, must be performed.

The validity of the RI-MOM approach relies on the fact that, at large $p^2$,
Green functions are expected to depend smoothly on the quark masses, since their
non-perturbative contributions vanish asymptotically in that limit. However,
specific care must be taken in the study of the pseudoscalar Green function
${\cal V}_P$, since the leading $1/p^2$-suppressed contribution is divergent in
the chiral limit, due to the coupling with the Goldstone
boson~\cite{rimom,alain1,alain2}. Moreover, for tm fermions, the explicit
breaking of parity at finite lattice spacing induces a coupling of the Goldstone
boson also with the scalar operator. Though the latter is suppressed at ${\cal
O}(a^2)$, its effect may be not negligible at the couplings considered in the
present simulations.

In order to subtract the pseudoscalar mass dependence of the amputated vertex
functions ${\cal V}_P$ and ${\cal V}_S$, at each given $p^2$ we fit their 
values at different quark masses to the Ansatz
\beq
\label{eq:gpsfit}
{\cal V}_{P(S)}(p^2,\mu_1,\mu_2) = A_{P(S)}(p^2) + B_{P(S)}(p^2) \,
M_{PS}^2(\mu_1,\mu_2) + \frac{C_{P(S)}(p^2)}{M_{PS}^2(\mu_1,\mu_2)} \, ,
\eeq
where $M_{PS}(\mu_1,\mu_2)$ is the mass of the pseudoscalar meson composed by
valence quarks of mass $\mu_1$ and $\mu_2$ (with $r_1=-r_2$)~\footnote{In our
preliminary analysis presented in~\cite{Dimopoulos:2007fn}, the fitting
function used to subtract the Goldstone pole contribution was expressed in
terms of the sum of quark masses $\mu_1+\mu_2$ rather than the pseudoscalar
meson mass squared $M_{PS}(\mu_1,\mu_2)^2$, as in Eq.~(\ref{eq:gpsfit}).
Since the latter is simply proportional to $\mu_1+\mu_2$ at maximal twist,
the two fits lead to completely consistent results.}. The first term in
the fit, i.e. the function $A_{P(S)}(p^2)$, represents the Green function ${\cal
V}_{P(S)}$ in the (valence) chiral limit, from which the RC $Z_{P(S)}$ is
eventually extracted. The last term in Eq.~(\ref{eq:gpsfit}) accounts for the
presence of the Goldstone pole. Using the results of the fit, we can define the
subtracted correlators
\beq
\label{eq:gpssub}
{\cal V}_{P(S)}^{\rm sub}(p^2,\mu_1,\mu_2) = {\cal V}_{P(S)}(p^2,\mu_1,\mu_2) -
\frac{C_{P(S)}(p^2)}{M_{PS}^2(\mu_1,\mu_2)} \,.
\eeq

\begin{figure}[p]
\begin{center}
\vspace{-0.5cm}
\subfigure{\includegraphics[scale=0.33,angle=270]{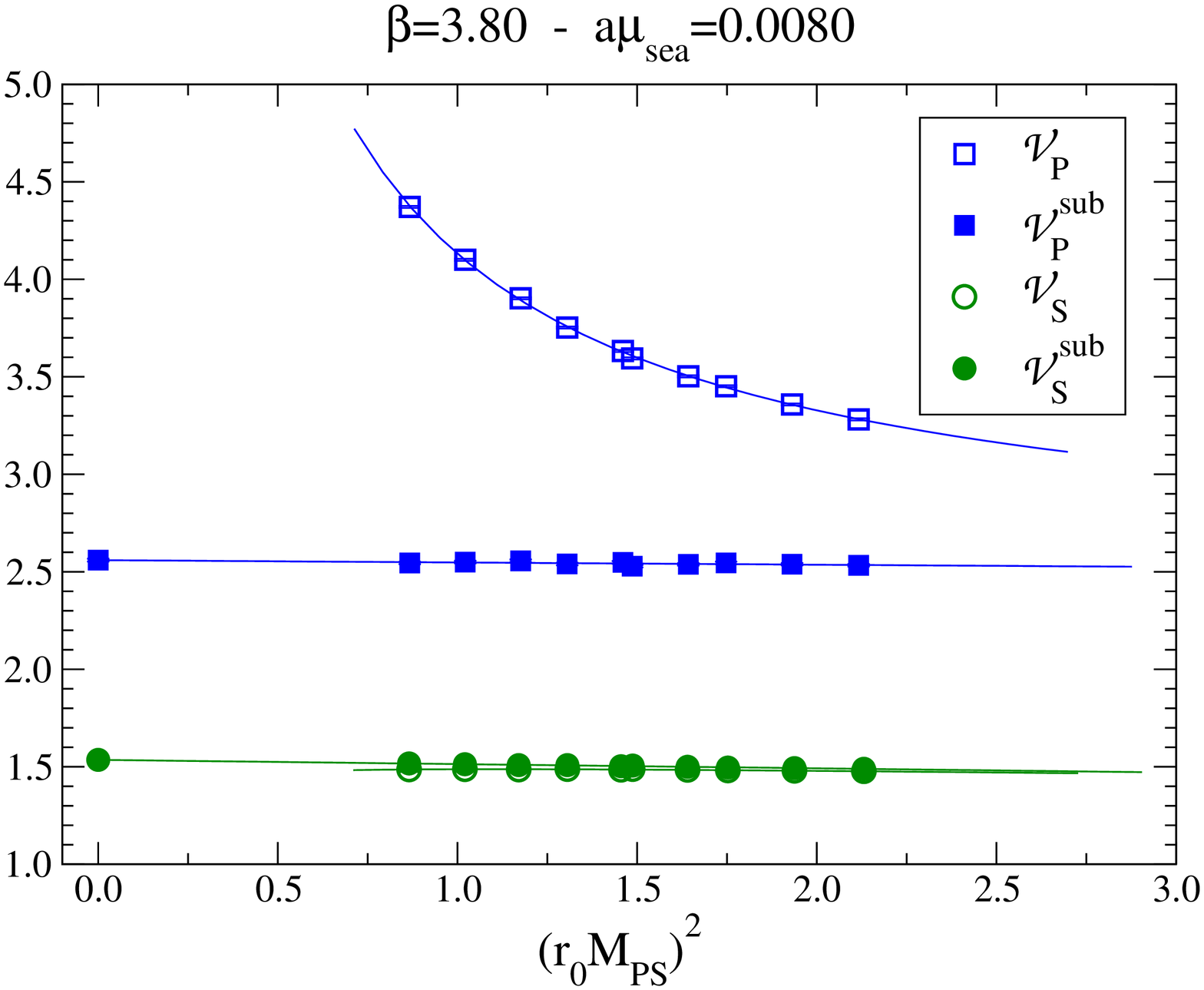}}
\subfigure{\includegraphics[scale=0.33,angle=270]{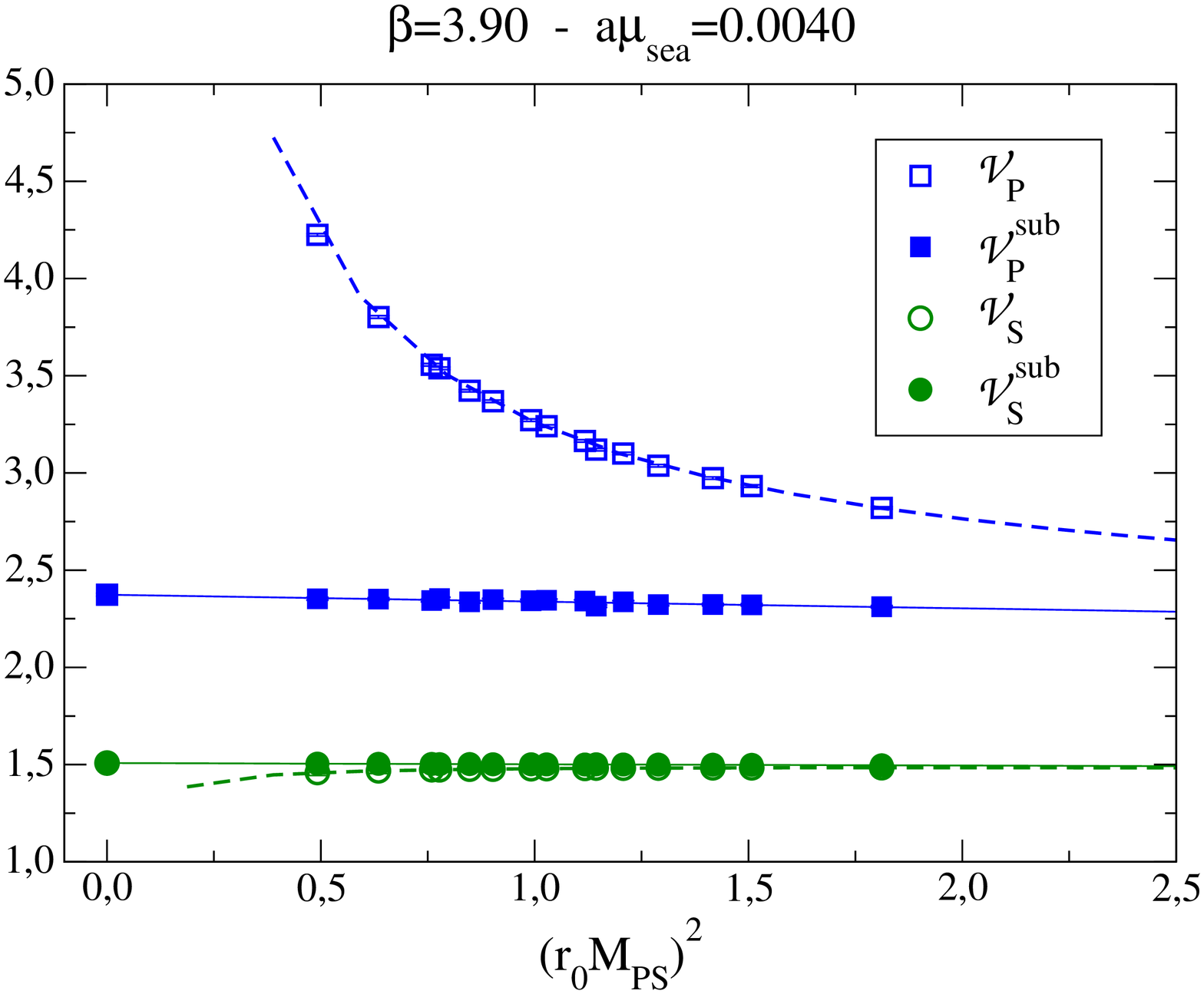}}
\subfigure{\includegraphics[scale=0.33,angle=270]{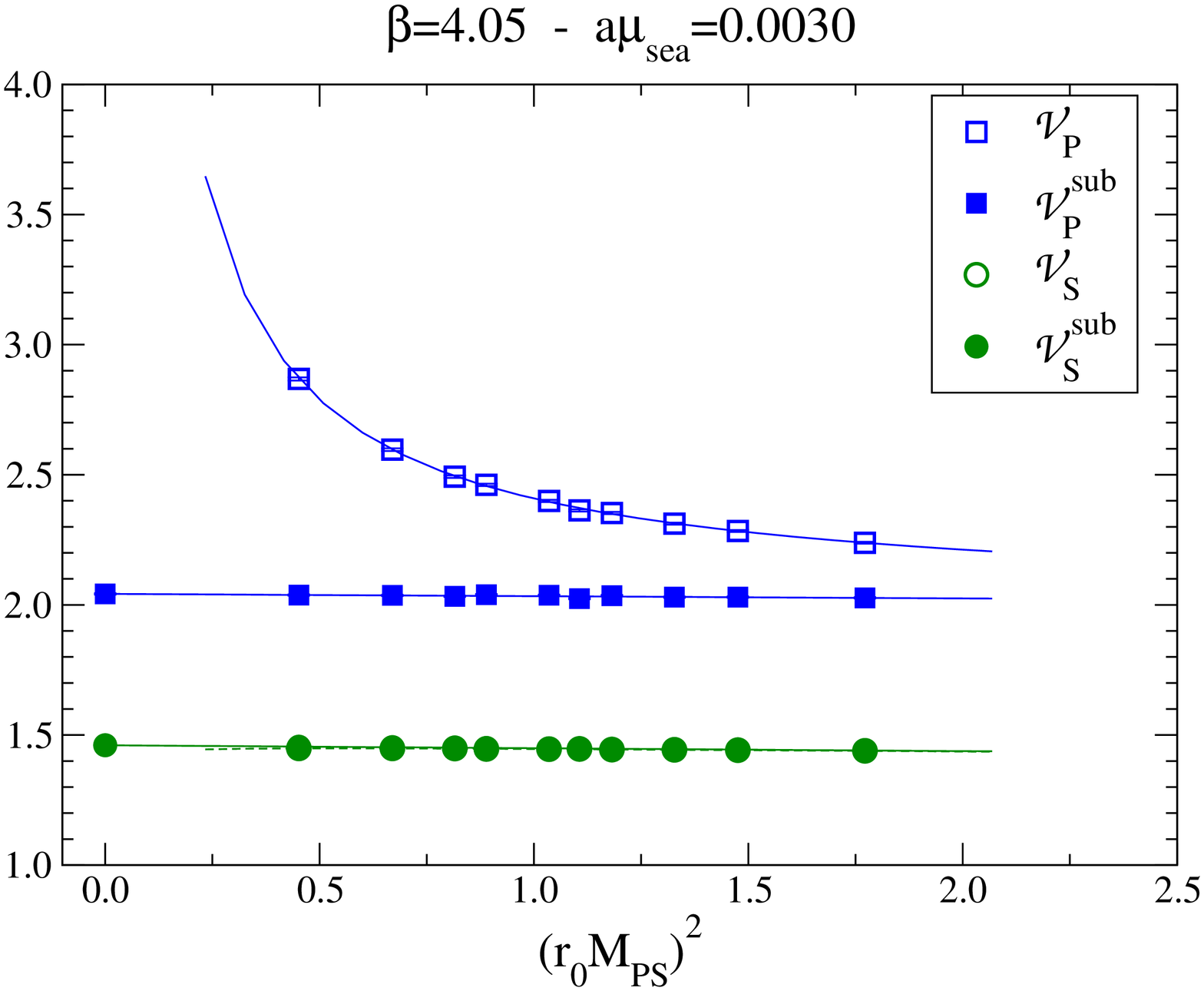}}
\end{center}
\vspace{-1.0cm}
\caption{\sl The amputated Green functions ${\cal V}_P(p^2,\mu_1,\mu_2)$ and
${\cal V}_S(p^2,\mu_1,\mu_2)$, evaluated at a scale $a^2 \tilde p^2\simeq 1$,
as a function of the pseudoscalar meson mass squared. The subtracted Green
functions are defined in Eq.~(\ref{eq:gpssub}). Dashed and solid lines
illustrate the results of the fit to Eq.~(\ref{eq:gpsfit}) (with $C=0$ for
${\cal V}_{P(S)}^{\rm sub}$).}
\label{fig:gps_pgb}
\end{figure}
The effect of the subtractions is illustrated in Fig.~\ref{fig:gps_pgb}. We find
that, while in the case of ${\cal V}_P$ the contribution of the Goldstone pole
is clearly visible, the subtraction is practically irrelevant in the case of
${\cal V}_S$, indicating a strong ${\cal O}(a^2)$ suppression of the parity
violating coupling. We have also verified that, the alternative
procedure~\cite{gv} for the Goldstone pole subtraction, based on the definition
\beq
{\cal V}_{P(S)}^{\rm sub}(p^2,\mu_1,\mu_2,\mu_3,\mu_4) = 
\frac{M_{PS}^2(\mu_1,\mu_2)\, {\cal V}_{P(S)}(p^2,\mu_1,\mu_2) - 
M_{PS}^2(\mu_3,\mu_4)\, {\cal V}_{P(S)}(p^2,\mu_3,\mu_4)}
{M_{PS}^2(\mu_1,\mu_2)-M_{PS}^2(\mu_3,\mu_4)} \, .
\label{eq:gv_pgb}
\eeq
leads to consistent results in the chiral limit, within our statistical errors.

\begin{figure}[p]
\begin{center}
\vspace{-0.5cm}
\subfigure{\includegraphics[scale=0.35,angle=270]{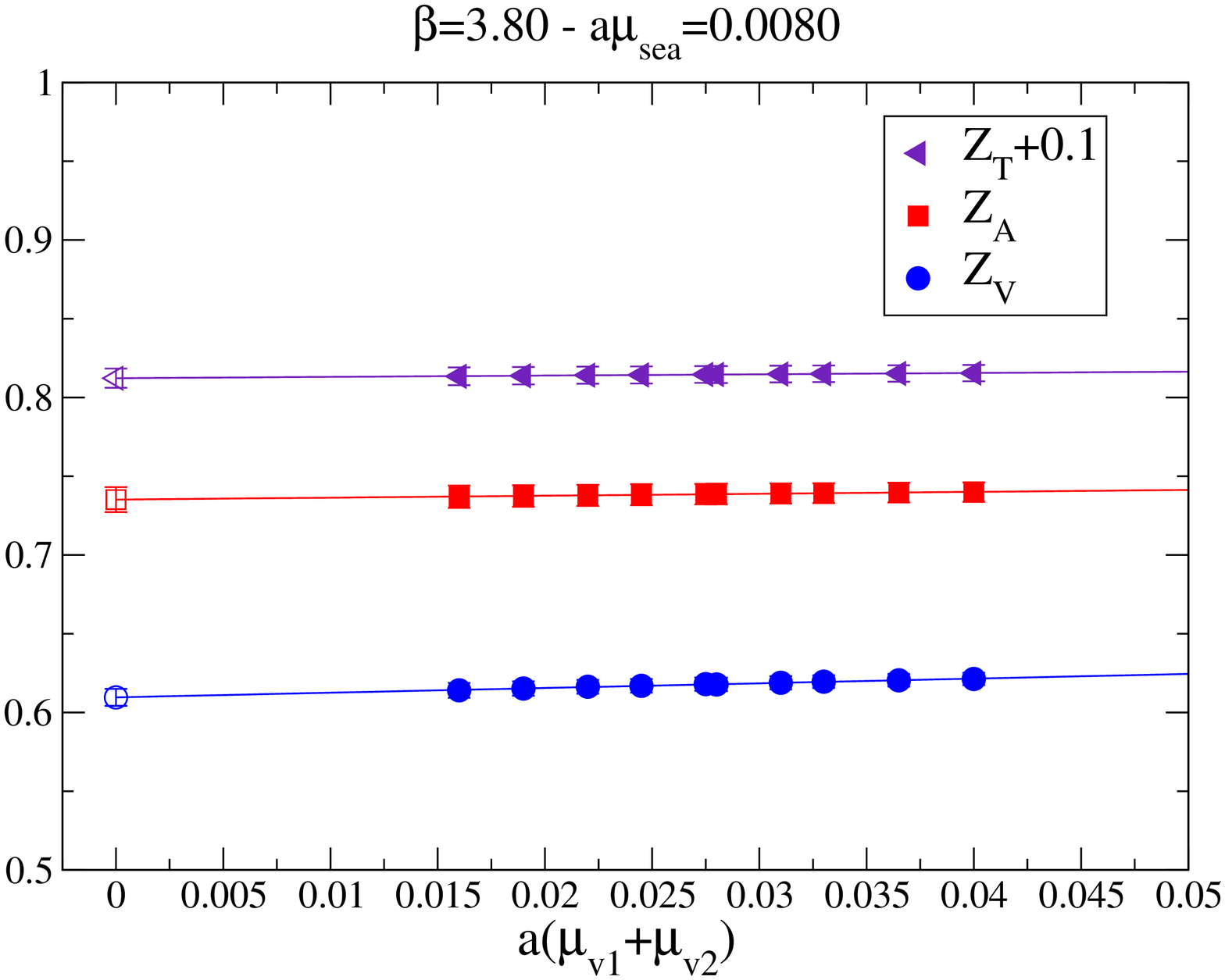}}
\subfigure{\includegraphics[scale=0.35,angle=270]{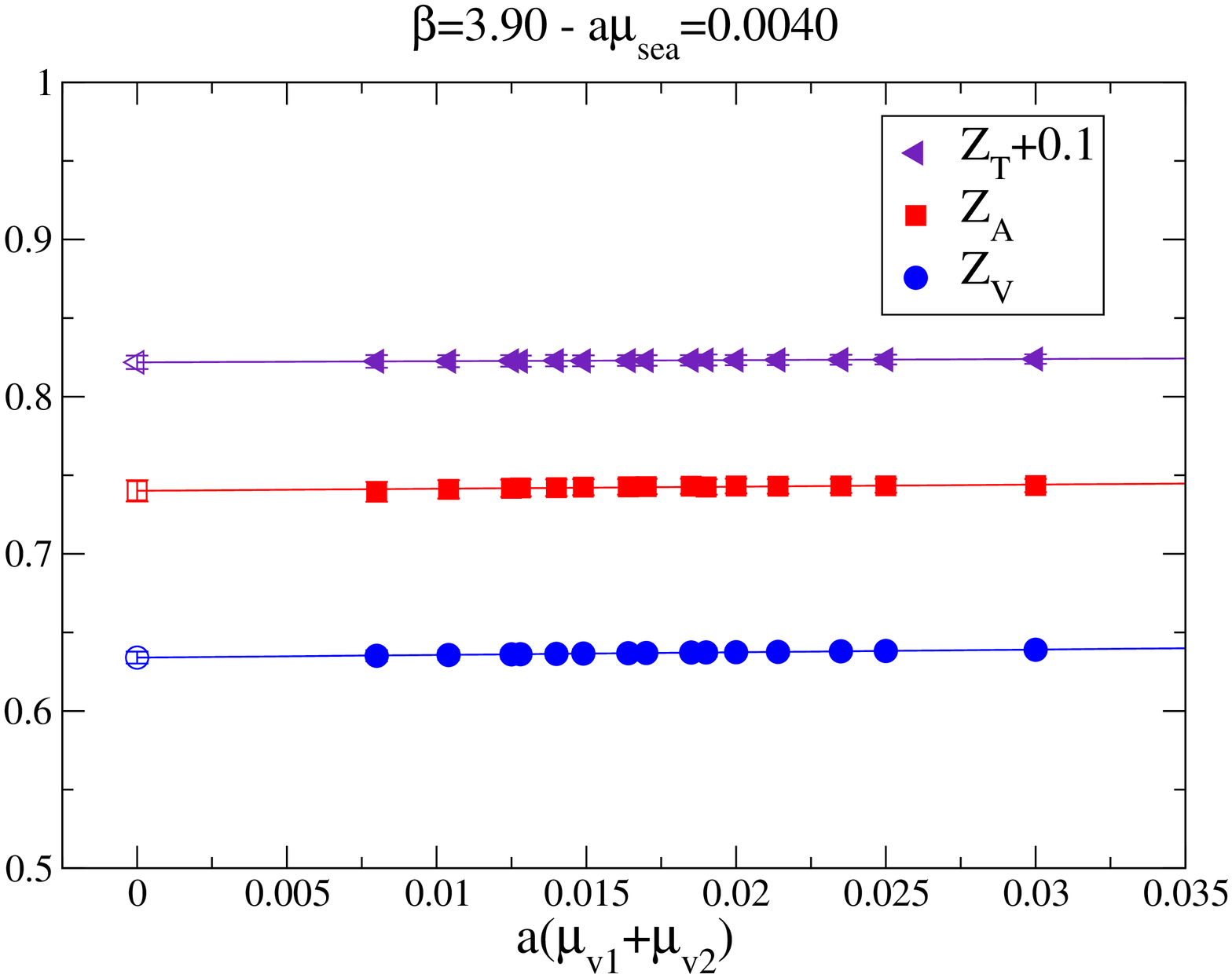}}
\subfigure{\includegraphics[scale=0.35,angle=270]{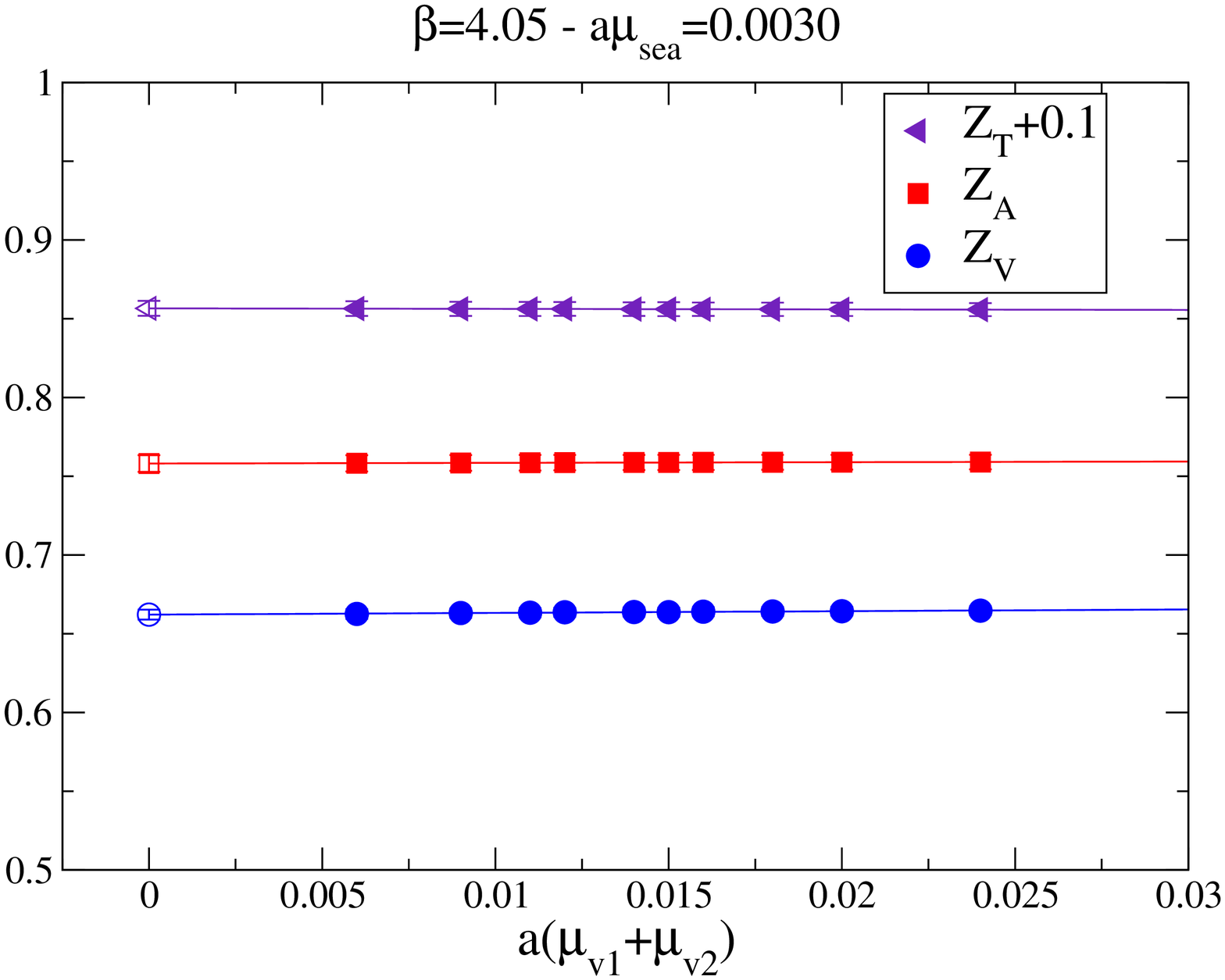}}
\end{center}
\vspace{-1.0cm}
\caption{{\sl Dependence of $Z_V$, $Z_A$ and $Z_T$ on the valence quark mass, at
$a^2 \tilde p^2\simeq 1$. The solid lines are linear extrapolations to the
(valence)
chiral limit.}}
\label{fig:zextrav}
\end{figure}

The dependence of $Z_V$, $Z_A$ and $Z_T$, which are not affected by the
Goldstone pole con\-ta\-mi\-na\-tion, on the valence quark masses, is shown in
Fig.~\ref{fig:zextrav},  for all three couplings. For illustration purposes, a
specific value of the sea quark mass has been chosen in each case. We see that
the valence quark mass dependence of the RCs is rather weak and well consistent
with a linear behaviour.

\begin{figure}[p]
\begin{center}
\vspace{-0.5cm}
\subfigure{\includegraphics[scale=0.35,angle=270]{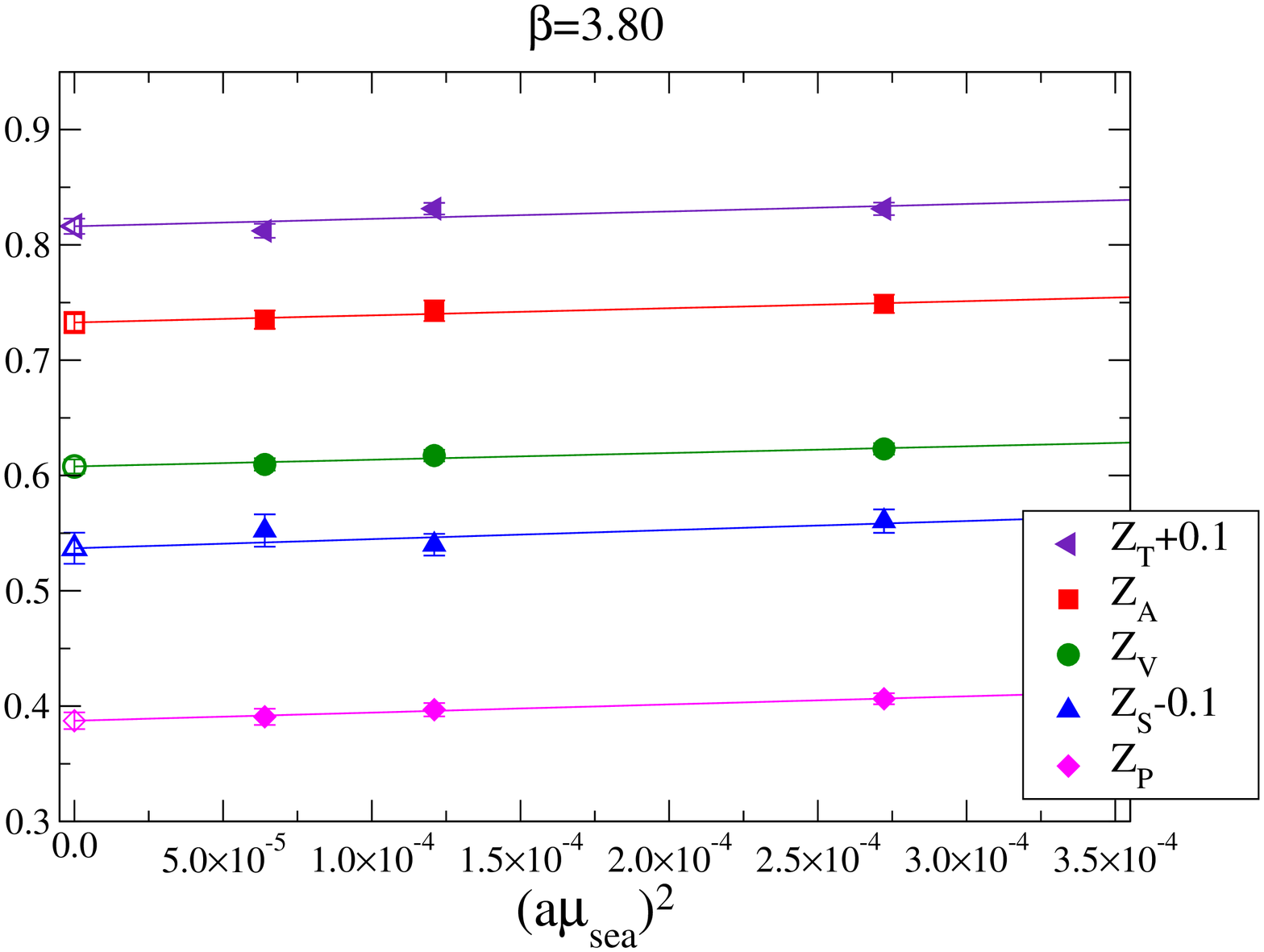}}
\subfigure{\includegraphics[scale=0.35,angle=270]{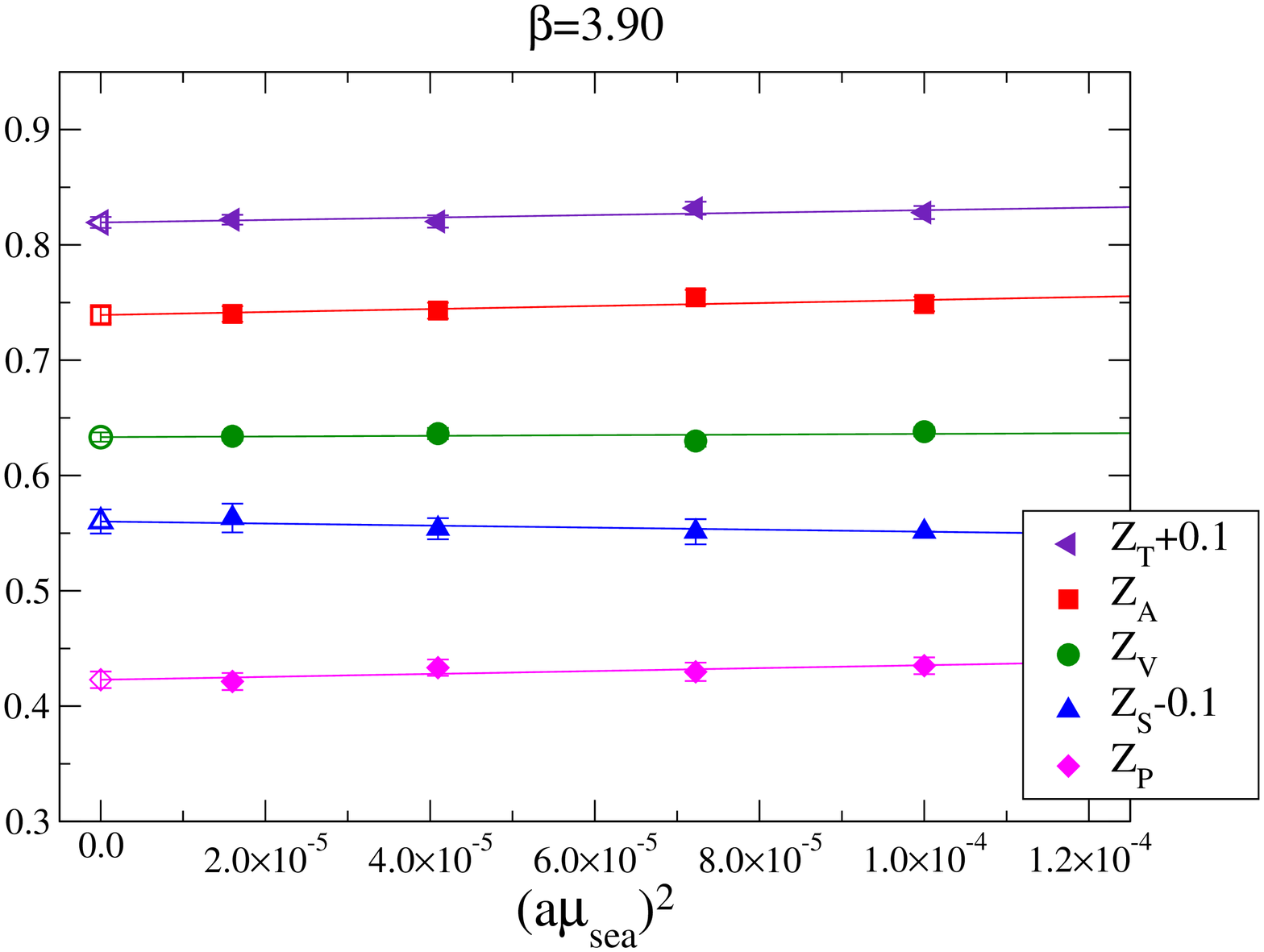}}
\subfigure{\includegraphics[scale=0.35,angle=270]{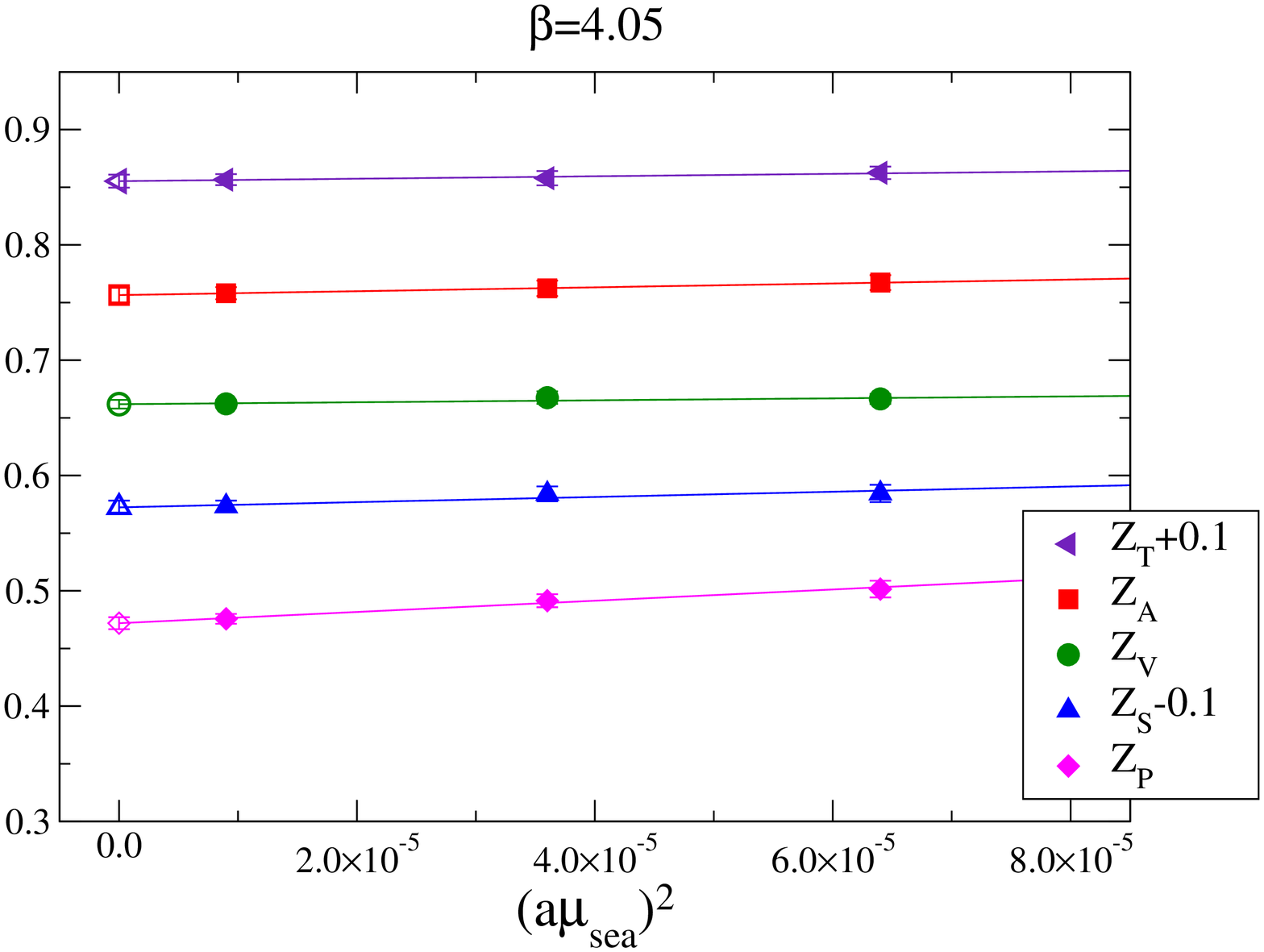}}
\end{center}
\vspace{-1.0cm}
\caption{{\sl Dependence of the RCs at $a^2 \tilde p^2\simeq 1$ on the sea quark
mass after valence chiral extrapolation. The solid lines are linear
extrapolations in $(a\mu_{sea})^2$ to the chiral limit.}}
\label{fig:zextras}
\end{figure}

In the tm formulation of lattice QCD with $N_f=2$ flavours of degenerate sea
quarks, the fermionic determinant is a function of the sea quark mass squared.
Therefore, we expect that the RCs, which by definition are insensitive to the
effects of spontaneous chiral symmetry breaking, will also exhibit the same
dependence. This is shown in Fig.~\ref{fig:zextras}. We find that all RCs depend
very weakly on the sea quark mass squared. We thus obtain our mass independent
RCs by performing the sea quark mass extrapolation to the chiral limit linearly
in $a^2 \mu_{sea}^2$. We also checked that chiral extrapolations based on a
first or second order polynomial fit in $a \mu_{sea}$ lead within errors to
practically equivalent results.

\subsubsection{Renormalization scale dependence and subtraction of the ${\cal O}
(g^2a^2)$ effects}
\label{sec:scaledep}
Once the RCs have been extrapolated to the chiral limit, we investigate their
dependence on the renormalization scale by evolving, at fixed coupling, the RCs
to a reference scale $\mu_0=1/a$. This is done using
\beq
\label{eq:zevol}
Z_\Gamma(a,\mu_0) = C_\Gamma(\mu_0,\mu) \,\, Z_\Gamma(a,\mu) \,,
\eeq
where the evolution function $C_\Gamma$ is expressed in terms of the beta
function $\beta(\alpha)$ and of the anomalous dimension of the relevant operator
$\gamma_\Gamma(\alpha)$ by
\beq
C_\Gamma(\mu_0,\mu)= \exp\left[ \int_{\alpha(\mu_0)}^
{\alpha(\mu)} \frac{\gamma_\Gamma(\alpha)}{\beta(\alpha)} d\alpha \right] ~.
\eeq 
This function is known, in the RI-MOM scheme, at the N$^2$LO for
$Z_T$~\cite{gracey} and at the N$^3$LO for $Z_S$ and $Z_P$~\cite{chetyr}. Since
$Z_V$, $Z_A$ and the ratio $Z_P/Z_S$ are scale independent, they have vanishing
anomalous dimensions; thus for these quantities we have $C_\Gamma(\mu_0,\mu)=1$.

In the non-perturbative calculation, the RCs evolved to the reference scale
$\mu_0=1/a$ maintain a dependence on the renormalization scale $p^2$ at which
they have been initially computed. We will keep track of this dependence, which
(at large enough $p^2$) is mostly due to discretization effects, by denoting
these RCs as $Z_\Gamma(1/a;a^2p^2)$, where the first variable indicates the
renormalization scale, $\mu_0=1/a$, and the second the dependence on the initial
momentum. 

\begin{figure}[p]
\begin{center}
\subfigure{\includegraphics[scale=0.27,angle=-90]{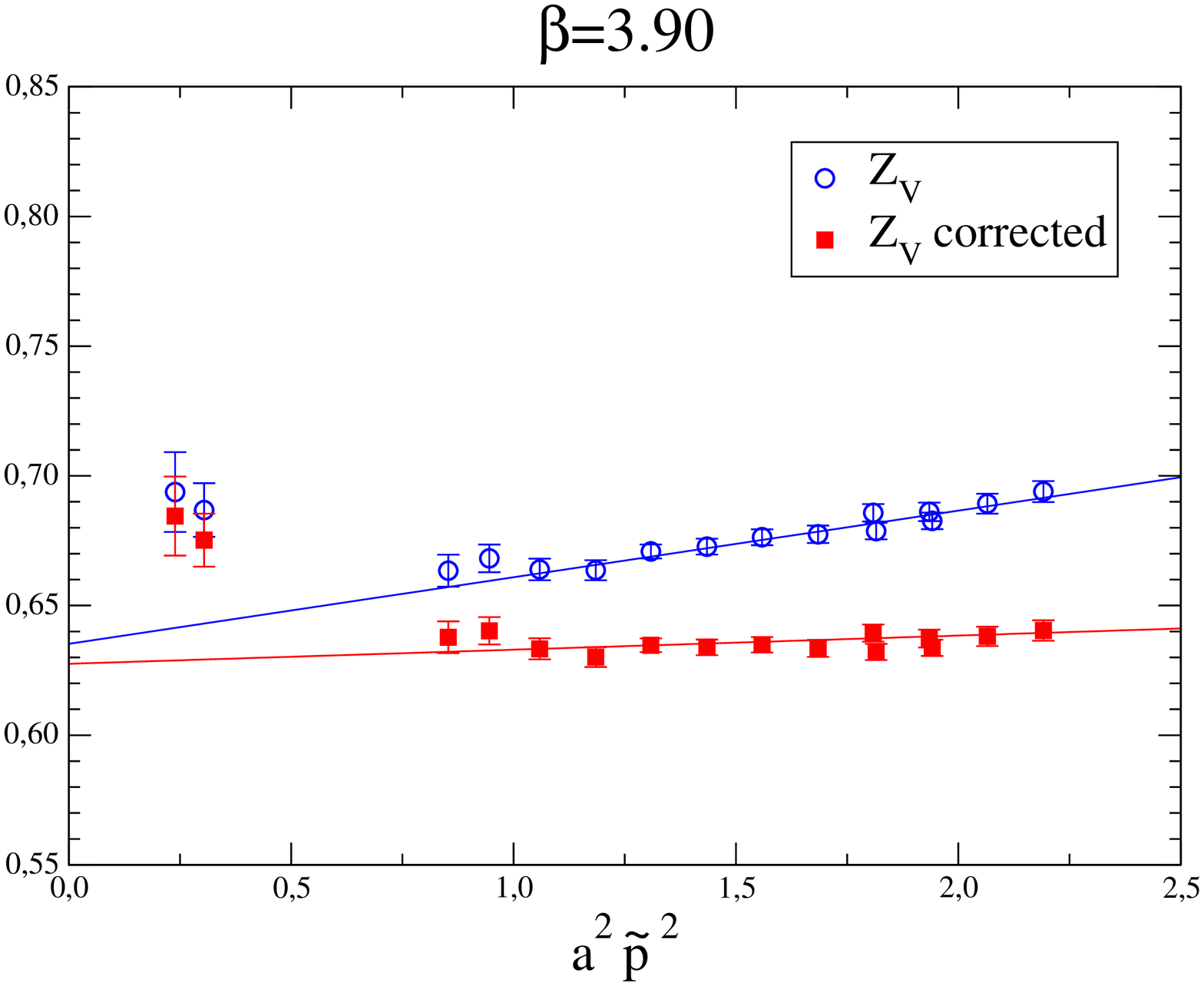}}
\subfigure{\includegraphics[scale=0.27,angle=-90]{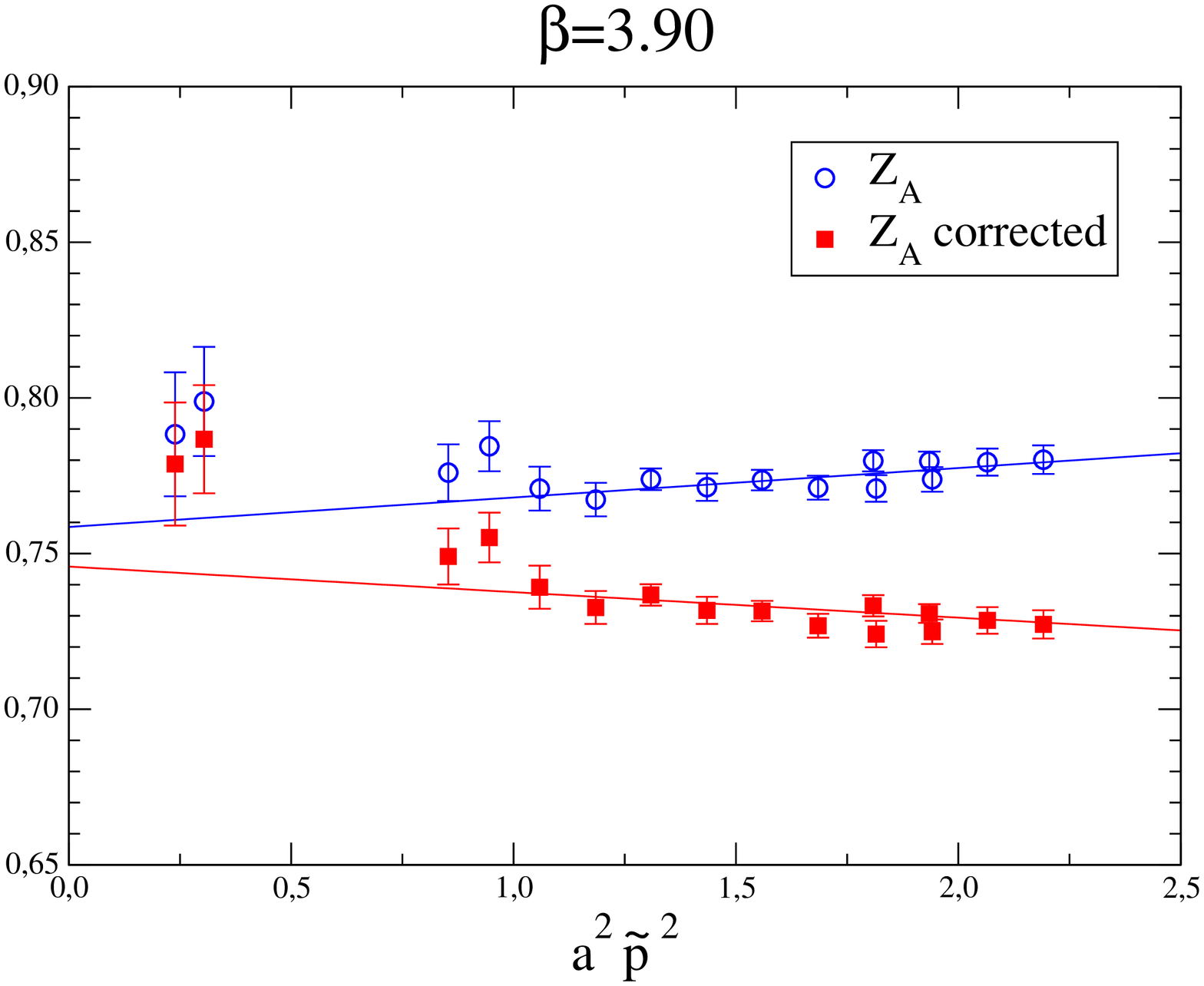}}
\subfigure{\includegraphics[scale=0.27,angle=-90]{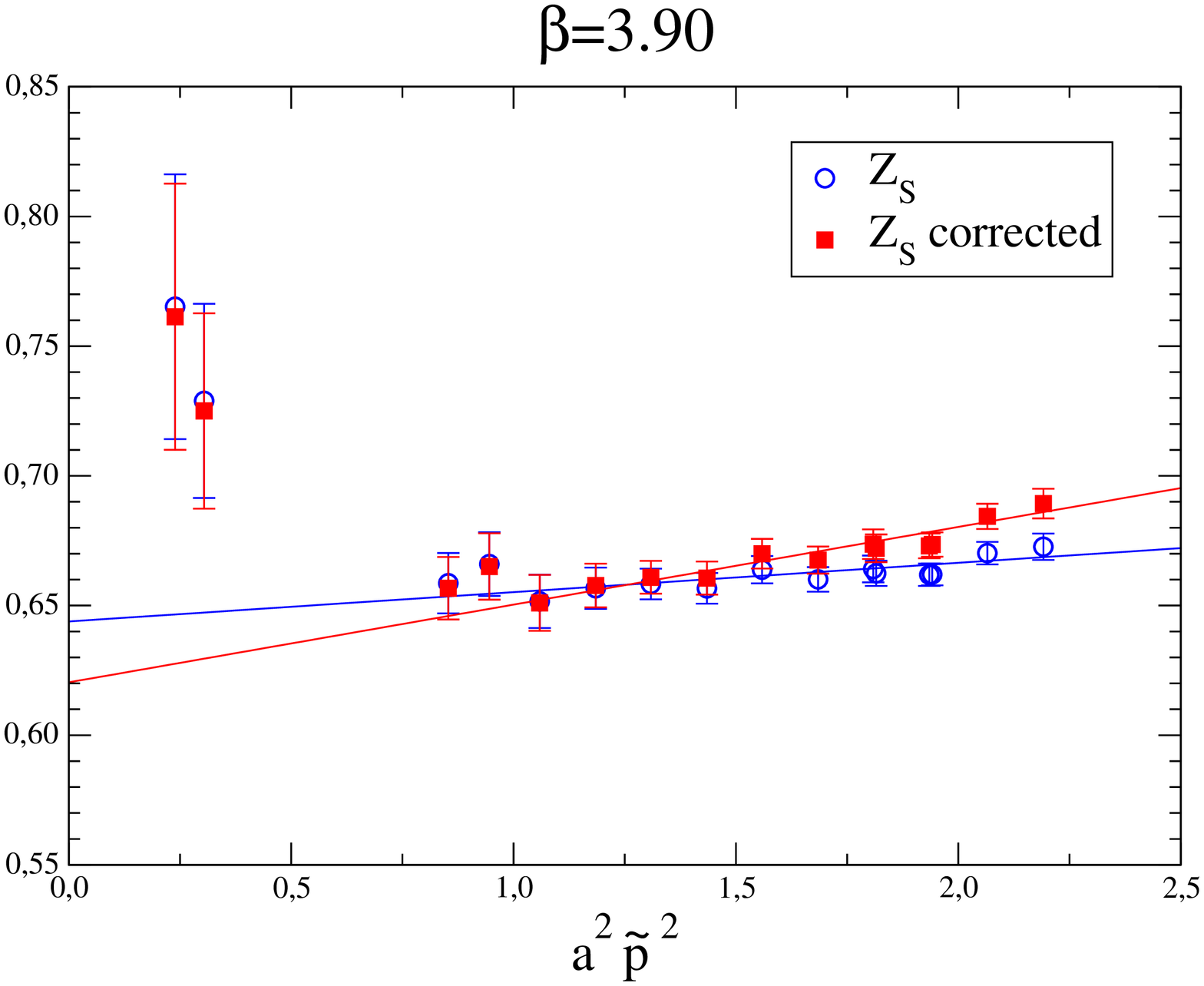}}
\subfigure{\includegraphics[scale=0.27,angle=-90]{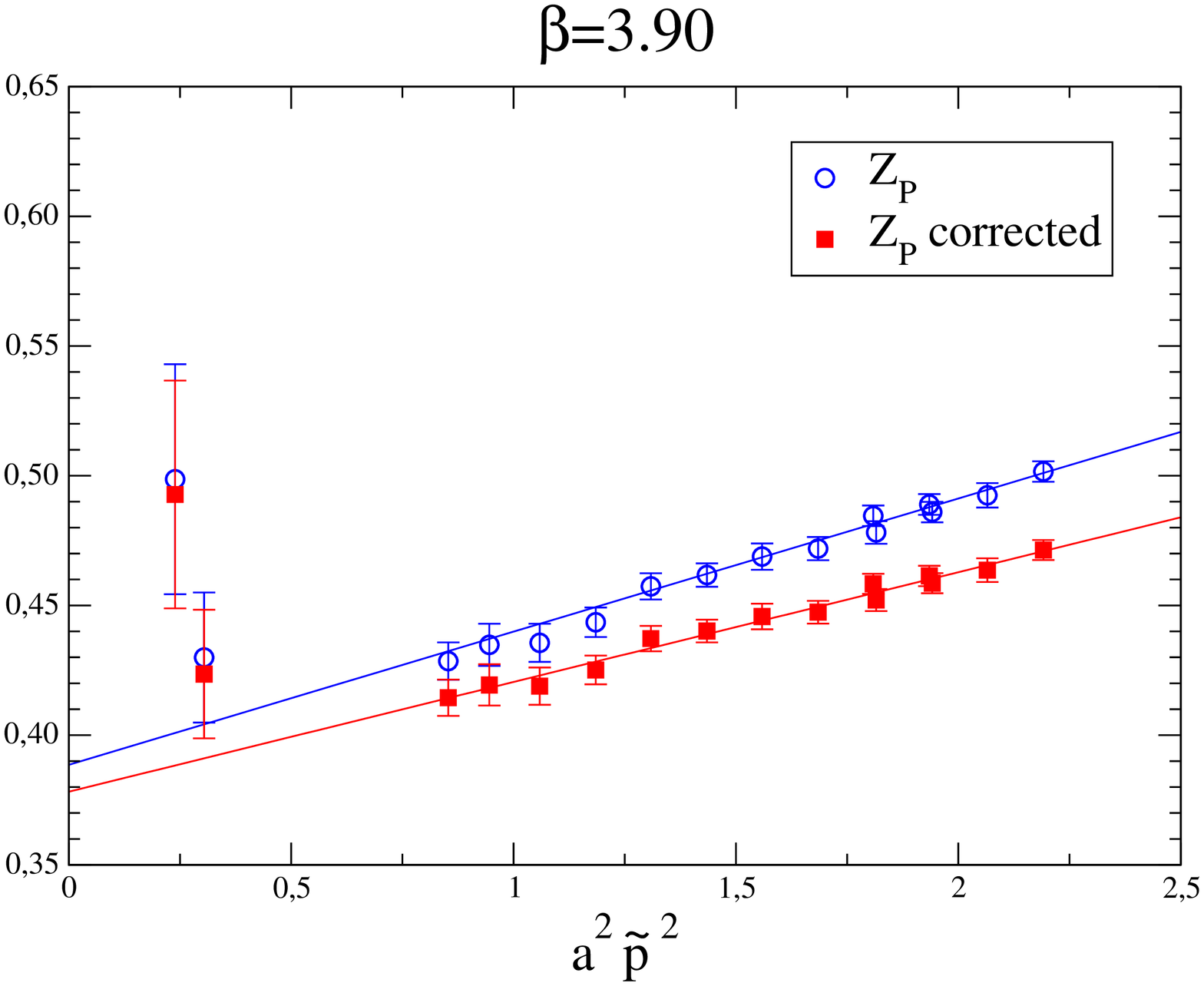}}
\subfigure{\includegraphics[scale=0.27,angle=-90]{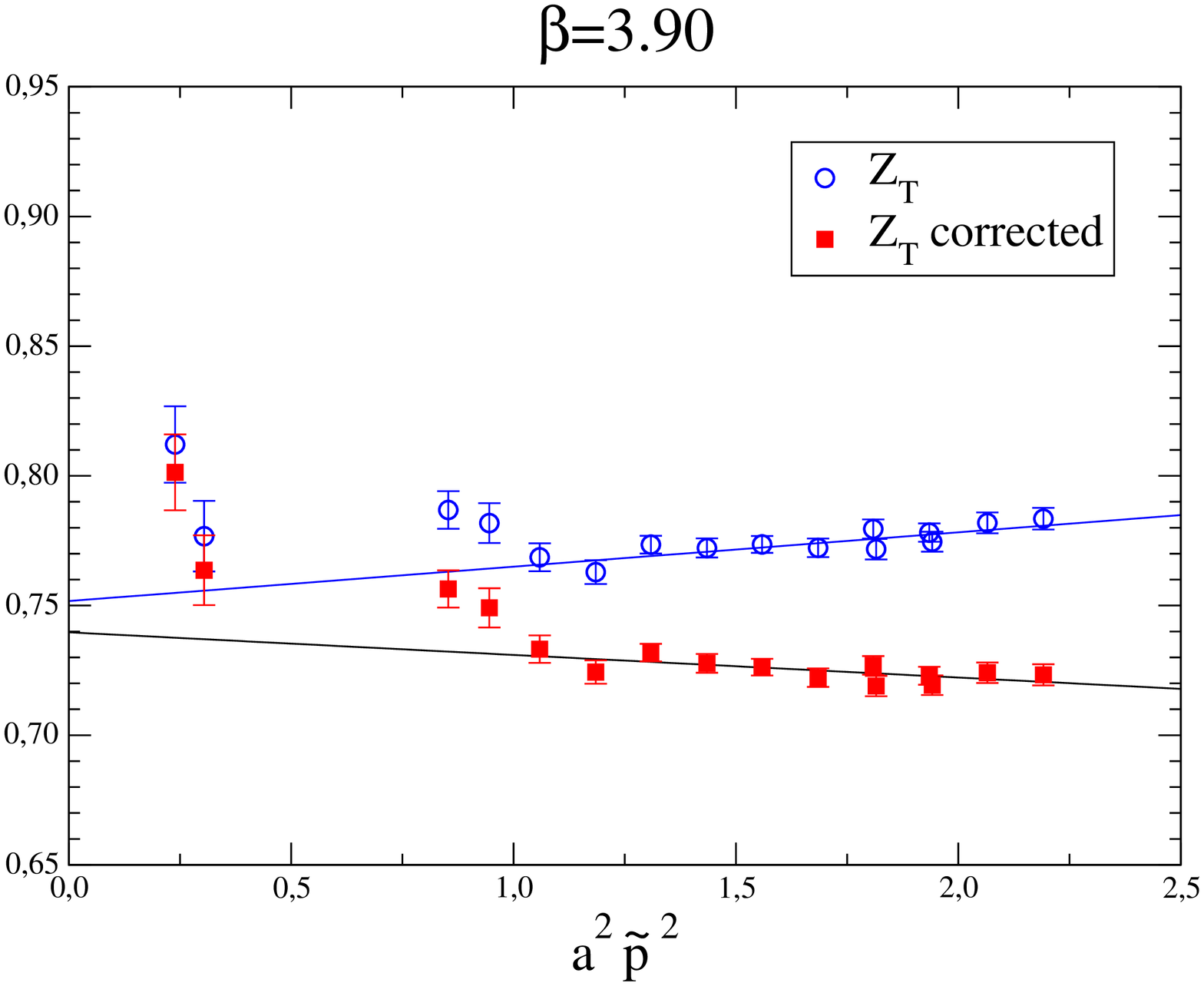}}
\subfigure{\includegraphics[scale=0.27,angle=-90]{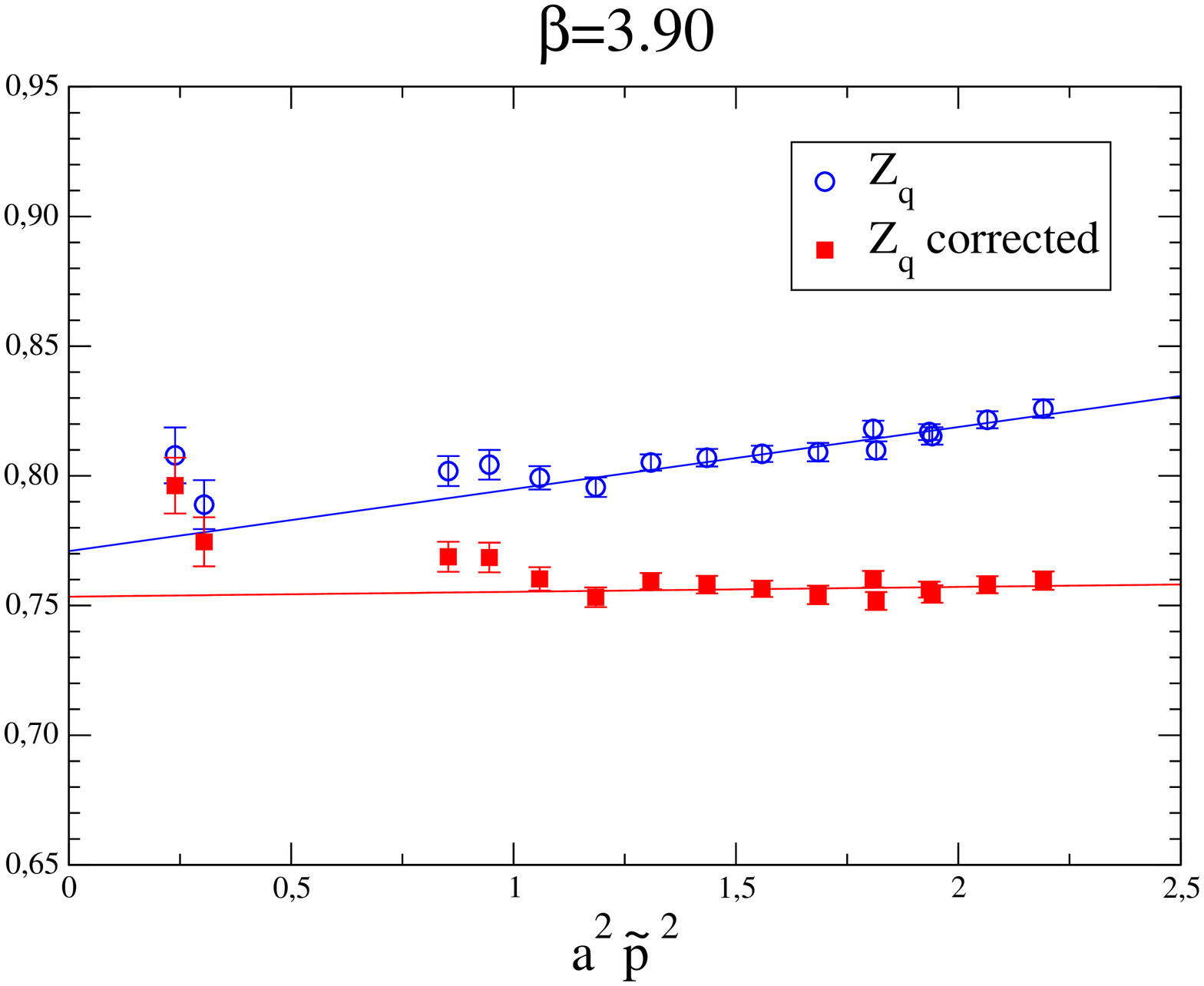}}
\vspace{-0.5cm}
\caption{{\sl The RCs $Z_\Gamma(1/a;a^2p^2)$ at $\beta=3.9$, evaluated at the
reference scale $\mu_0=1/a$, plotted against the initial renormalization scale
$a^2p^2$. Filled squares (empty circles) are results obtained with (without) the
subtraction of the ${\cal O}(g^2a^2)$ discretization effects computed in
perturbation theory, see Eq.~(\ref{eq:sub}). The solid lines are linear fits to
the data.}}
\label{fig:zcorr}
\end{center}
\end{figure}

In Fig.~\ref{fig:zcorr} we show the results for all $Z_\Gamma (1/a;a^2p^2)$ at
$\beta=3.9$ as a function of $a^2 \tilde p^2$ (empty symbols). The residual
$a^2p^2$--dependence which is observed in these results in the large momentum
region is practically linear, suggesting that leading discretization effects are
${\cal O}(a^2 p^2)$. As illustrated in the plots, this dependence is
particularly pronounced in the case of the pseudoscalar RC $Z_P$. 

In order to reduce the size of discretization errors, we analytically subtract
from the quark propagator and the amputated vertex functions the ${\cal O}(g^2
a^2)$ contributions, recently computed in lattice perturbation
theory~\cite{Og2a2}. Up to 1-loop and up to ${\cal O}(a^2)$, the amputated
projected Green functions ${\cal V}_\Gamma(p)$, defined in Eq.~(\ref{eq:rimom}),
have the simple and general expression
\bea
{\cal V}_\Gamma(p)^{\rm pert.} &=&
1 + \frac{g^2}{12\pi^2} \Big{\{} b_\Gamma^{(1)} + 
b_\Gamma^{(2)}\,\ln(a^2\,p^2) \nonumber \\
&+& a^2 \Big[ p^2 \big(c_\Gamma^{(1)} + c_\Gamma^{(2)}\,
\ln(a^2\,p^2)\big) + c_\Gamma^{(3)} \frac{\sum_\rho\, p_\rho^4}{p^2} \Bigr]
\Bigr{\}} + {\cal O}(a^4\,g^2,g^4) \ .
\label{eq:v1loop}
\eea
The values of the coefficients $b_\Gamma^{(i)}$ and $c_\Gamma^{(i)}$ are
collected in Table~\ref{tab:pt} for the lattice action used in the present
study, namely the tree-level Symanzik improved gluon action in the Landau gauge
and the tm fermionic action at maximal twist~\footnote{The same results are also
valid for the standard Wilson fermionic action, except for the exchange of the
values of the vector (V) and axial (A) coefficients.}.
\begin{table}
\begin{center}
\begin{tabular}{cr@{}lcr@{}lr@{}lr@{}l}
\hline \hline
\multicolumn{1}{c}{$\Gamma$}&
\multicolumn{2}{c}{$b_{\Gamma}^{(1)}$} &
\multicolumn{1}{c}{$b_{\Gamma}^{(2)}$} &
\multicolumn{2}{c}{$c_{\Gamma}^{(1)}$} &
\multicolumn{2}{c}{$c_{\Gamma}^{(2)}$} &
\multicolumn{2}{c}{$c_{\Gamma}^{(3)}$} \\
\hline \hline
V  &-0&.48369852(8)  &$\,$ 0    &1&.5240798(1)   &-1/&3      &-125/&288  \\
A  &3&.57961385(3)   &$\,$ 0    &0&.6999177(1)   &-1/&3      &-125/&288  \\
S  &0&.58345905(5)   &-3        &2&.3547298(2)   &-1/&4     &1/&2  \\
P  &8&.7100837(1)    &-3        &0&.70640549(6)  &-1/&4      &1/&2  \\
T  &0&.51501972(6)   &$\,$ 1    &0&.9724758(2)   &-13/&36    &-161/&216  \\
\hline \hline
\end{tabular}
\end{center}
\caption{\sl Values of the coefficients $b_{\Gamma}^{(i)}$ and
$c_{\Gamma}^{(i)}$ entering the 1-loop expression~(\ref{eq:v1loop}) of the
amputated projected Green functions ${\cal V}_\Gamma(p)$. Results are presented
for the case of the tree-level Symanzik improved gluon action in the Landau
gauge and the tm fermionic action at maximal twist.}\label{tab:pt}
\end{table}

A result similar to Eq.~(\ref{eq:v1loop}) also holds for the form factor 
$\Sigma_1(p)$ of the inverse quark propagator, from which the quark field RC 
$Z_q$ is evaluated:
\bea
\Sigma_1(p)^{\rm pert.} &=&
1 + \frac{g^2}{12\pi^2} \Big{\{} b_q^{(1)} +  b_q^{(2)}\,\ln(a^2\,p^2)
\nonumber \\
&+& a^2 \Big[ p^2 \big( c_q^{(1)} + c_q^{(2)}\, \ln(a^2\,p^2)\big) +  c_q^{(3)}
\frac{\sum_\rho\, p_\rho^4}{p^2} \Bigr] \Bigr{\}} + {\cal O}(a^4\,g^2,g^4) \ .
\label{eq:s1loop}
\eea
Our definition of this form factor on the lattice, which is equivalent to 
Eq.~(\ref{eq:zqri}) up to terms of ${\cal O}(a^2)$, is
\beq
\Sigma_1(p) \equiv \frac{-i}{12 N(p)}{\rm Tr}\left[\sum_{\rho}\phantom{}^\prime
\left(\gamma_\rho\, S(p)^{-1}\right)/\tilde p_\rho\, \right]~,
\eeq
where the sum $\sum_{\rho}^{\,\prime}$ only runs over the Lorentz indices for 
which $p_\rho$ is different from zero and $ N(p)=\sum_{\rho}^{\,\prime}\,1$.
With this definition, the coefficients of Eq.~(\ref{eq:s1loop}) take the values
\bea
&& b_q^{(1)}=-13.02327272(7)\, ; \quad b_q^{(2)} = 0 \,; 
\quad c_q^{(3)} =\frac{7}{240} \,;
\nonumber \\
&& c_q^{(1)}= 1.14716212(5) + \frac{2.07733285(2)}{N(p)} \,; \quad
c_q^{(2)}= -\frac{73}{360} -  \frac{157/180}{N(p)} ~.
\eea

Using Eqs.~(\ref{eq:v1loop}) and (\ref{eq:s1loop}), we then define the
{\em corrected} amputated Green functions and the {\em corrected} form factor
$\Sigma_1$ as
\bea
\label{eq:sub}
&& {\cal V}_\Gamma(p)^{\rm corr.} = {\cal V}_\Gamma(p) - \frac{g^2}{12\pi^2} 
a^2 \Big[\tilde p^2 \big(c_\Gamma^{(1)} +  c_\Gamma^{(2)}\,\ln(a^2\,\tilde
p^2)\big) +  c_\Gamma^{(3)} \frac{\sum_\rho\, \tilde p_\rho^4}{\tilde p^2}
\Bigr] \Bigr{\}}~, \nonumber \\
&& \Sigma_1(p)^{\rm corr.} = \Sigma_1(p) - \frac{g^2}{12\pi^2} 
a^2 \Big[\tilde p^2 \big(c_q^{(1)} +  c_q^{(2)}\,\ln(a^2\,\tilde p^2)\big) + 
c_q^{(3)} \frac{\sum_\rho\, \tilde p_\rho^4}{\tilde p^2} \Bigr] \Bigr{\}}~,
\eea
which are free of ${\cal O}(g^2 a^2)$ effects. Note that the ${\cal O}(a^2)$
terms depend not only on the magnitude, $\sum_\rho p^2_\rho$, but also on the
direction of the momentum, $p_\rho$, as manifested by the presence of $\sum_\rho
p^4_\rho$. As a consequence, different renormalization prescriptions, involving
different directions of the renormalization scale $p_\rho$, treat lattice
artifacts differently. In the numerical evaluation of the perturbative
correction in Eq.~(\ref{eq:sub}), we replaced $g^2$ with a simple boosted 
coupling~\cite{lm}, defined as $\tilde g^2=g_0^2/\langle P\rangle$, where the
average plaquette $\langle P\rangle$ is computed non-perturbatively.

The effect of the subtraction~(\ref{eq:sub}) is also illustrated in
Fig.~\ref{fig:zcorr}, which shows that the $a^2 p^2$--dependence of the RCs is
significantly reduced by the perturbative correction. By fitting the RCs as
\beq
\label{eq:oasub}
Z_\Gamma(1/a;a^2p^2) = Z_\Gamma(1/a) + \lambda_\Gamma\, a^2 \tilde p^2
\eeq
in the large momentum region $a^2 \tilde p^2 \simge 1$ (with $\lambda_\Gamma$
just constrained to depend smoothly on $g^2$ -- see Eq.~(\ref{eq:slopep2})), we
find that the slopes  $\lambda_\Gamma$ are reduced, after the perturbative
subtraction, at the level of $10^{-2}$ or smaller for $Z_V$, $Z_A$, $Z_T$ and
$Z_q$. For $Z_S$ we find that the slope, which is also about $10^{-2}$ before
implementing the correction, increases slightly after the subtraction. For
$Z_P$, on the other hand, the slope is rather large before the subtraction; i.e.
$\lambda_P \simeq 5 \cdot 10^{-2}$. As shown in Fig.~\ref{fig:zcorr}, the effect
of the subtraction is beneficial also in this case, but inadequate for
correcting the bulk of the observed $a^2p^2$-dependence. This is not completely
unexpected: when discretization effects are large, the subtraction of only the
leading ${\cal O}(g^2 a^2)$ terms may be not sufficient to reduce them to a
negligible level. Similar results, with approximately equal values of the slopes
$\lambda_\Gamma$, are obtained at all three values of the lattice spacing.

A useful way to investigate the size of lattice artifacts, which are left after
the perturbative subtraction, consists in comparing the renormalization scale
dependence of the RCs as observed at different lattice spacings. Specifically,
at each value of the lattice spacing we fix two common values of the
renormalization scale, $\mu$ and $\mu^\prime$, and compute the step scaling
functions
\beq
\Sigma_\Gamma(\mu,\mu^\prime;a) = 
\frac{Z_\Gamma(a,\mu)}{Z_\Gamma(a,\mu^\prime)} \, .
\eeq
If not for discretization effects, the step scaling functions should be
independent of the cut-off and equal to the evolution functions
$C_\Gamma(\mu,\mu^\prime)$ of Eq.~(\ref{eq:zevol}).

\begin{figure}[t]
\begin{center}
\includegraphics[scale=0.35,angle=270]{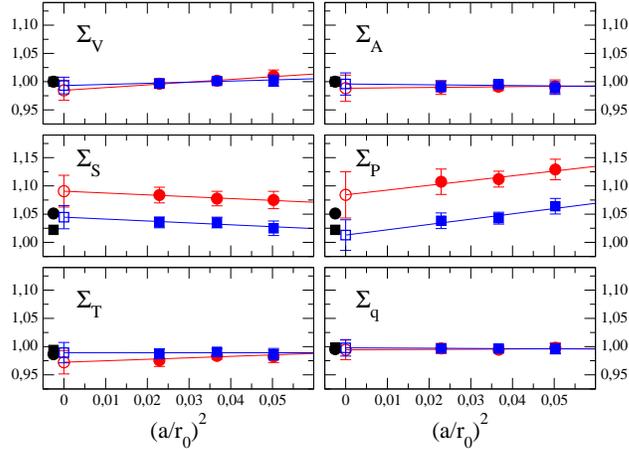}
\end{center}
\vspace{-0.5cm}
\caption{{\sl Values of the step scaling functions $\Sigma_\Gamma(\mu,\mu^\prime
;a)$ evaluated for $(\mu^2, \mu^{\prime\, 2}) = (8.5,6.0)\ \gev^2$ (red circles)
and $(\mu^2, \mu^{\prime\, 2}) = (8.5,7.25)\ \gev^2$ (blue squares) at the three
values of the lattice coupling. The solid lines and the empty markers show the
results of a linear extrapolation to the continuum limit. For comparison, the
corresponding perturbative estimates of the evolution functions
$C_\Gamma(\mu,\mu^\prime)$ are also shown in the plots (black circles and
squares).}}
\label{fig:zeta_ssf}
\end{figure}

In fig.~\ref{fig:zeta_ssf} we show the results obtained for the step scaling
functions of the quark bilinear operators and of the quark field RCs, as a
function of $(a/r_0)^2$, for two common pairs renormalization scales, $(\mu^2,
\mu^{\prime\, 2}) = (8.5,6.0)\ \gev^2$ and $(\mu^2, \mu^{\prime\, 2}) =
(8.5,7.25)\ \gev^2$, which both lie in the interval of momenta accessible at all
values of the lattice spacing.
We see from these plots that the dependence of the step scaling functions on the
lattice cut-off is tiny in most of the cases, being clearly visible only in the
case of $\Sigma_P$. Moreover, for the scale independent RCs $Z_V$ and $Z_A$, a
linear extrapolation of the results to the continuum limit leads to values which
are perfectly consistent with unity, within the statistical errors. For the
other RCs, the continuum limit of the step scaling functions leads to
non-perturbative results which are in good agreement with the perturbative
determinations of the evolution functions $C_\Gamma(\mu, \mu^\prime)$, computed
at the N$^2$LO for $Z_T$ and at the N$^3$LO for $Z_S$, $Z_P$ and $Z_q$. These
results provide evidence that higher order perturbative contributions to the
anomalous dimensions of the various operator are not relevant for describing the
renormalization scale dependence of the RCs in the region of momenta explored in
the present calculation.

In order to account for the residual discretization effects in the calculation
of the RCs, we follow two different approaches:
\begin{itemize}
\item[-] \underline{\em Extrapolation method ({\sf M1})}: after subtraction of
the ${\cal O}(g^2 a^2)$ terms according to Eq.~(\ref{eq:sub}), we extrapolate
the RCs linearly to $a^2p^2 \to 0$. Specifically, we fit $Z_\Gamma(1/a;a^2p^2)$
with Eq.~(\ref{eq:oasub}) in the large momentum region, $1.0\simle a^2 \tilde
p^2 \simle 2.2$. The slopes $\lambda_\Gamma$ exhibit only a very mild dependence
on the coupling. We parameterize this dependence by performing a simultaneous
extrapolation at the 3 values of the coupling and writing the slopes as
\beq
\label{eq:slopep2}
\lambda_\Gamma(g^2) = \lambda_\Gamma(g_0^2) + \lambda_\Gamma^\prime(g_0^2) 
\cdot (g^2-g_0^2)\ ,
\eeq
where $g_0$ is the coupling corresponding to the intermediate value $\beta=3.9$.
The intercept of the extrapolation as determined from the fit, and indicated
with $Z_\Gamma(1/a)$ in Eq.~(\ref{eq:oasub}), provides our final estimate of the
RC with the extrapolation method. The fit is illustrated, for all RCs, in
fig.~\ref{fig:z3beta}.
\begin{figure}[p]
\begin{center}
\subfigure{\includegraphics[scale=0.27,angle=-90]{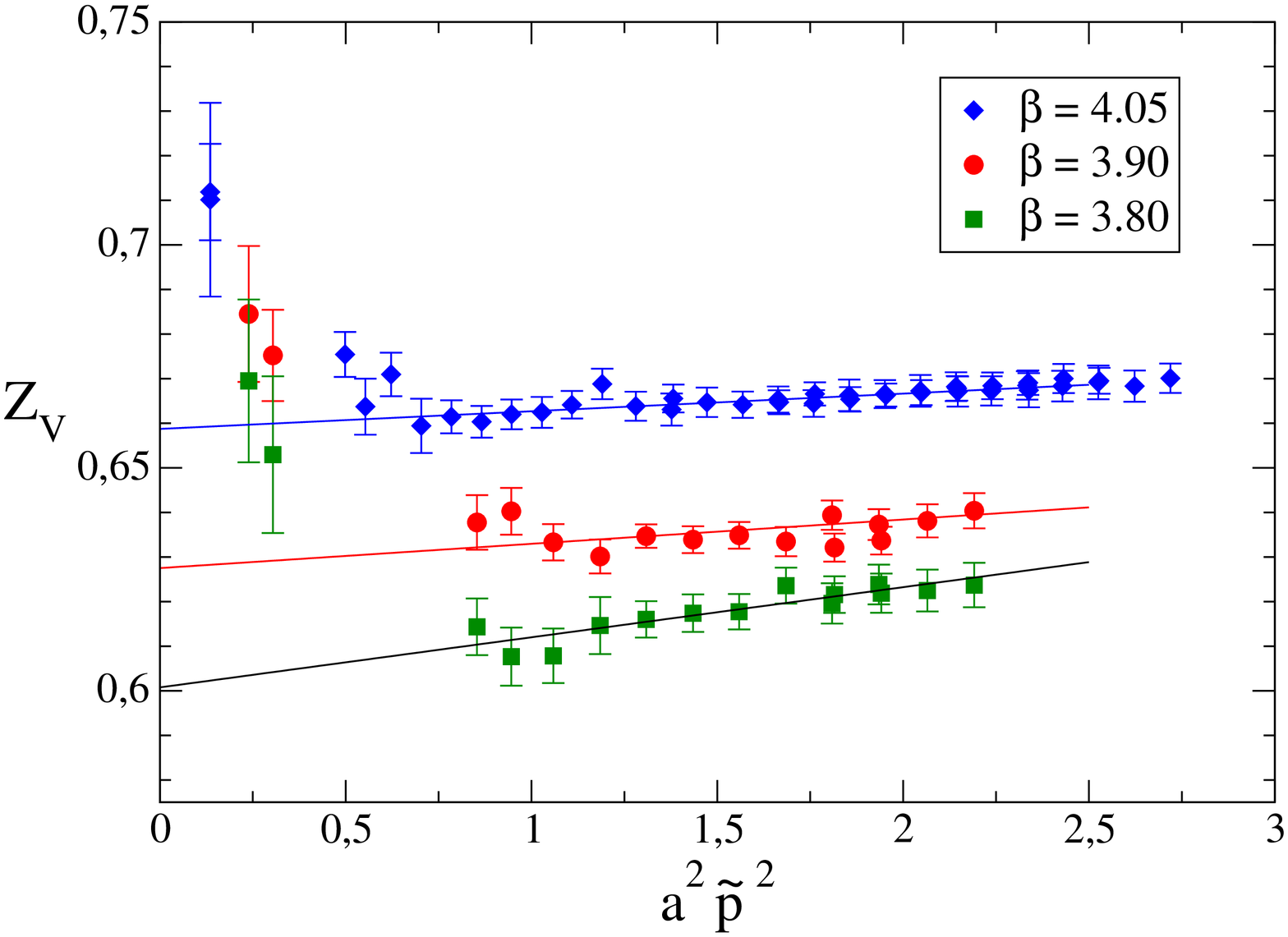}}
\subfigure{\includegraphics[scale=0.27,angle=-90]{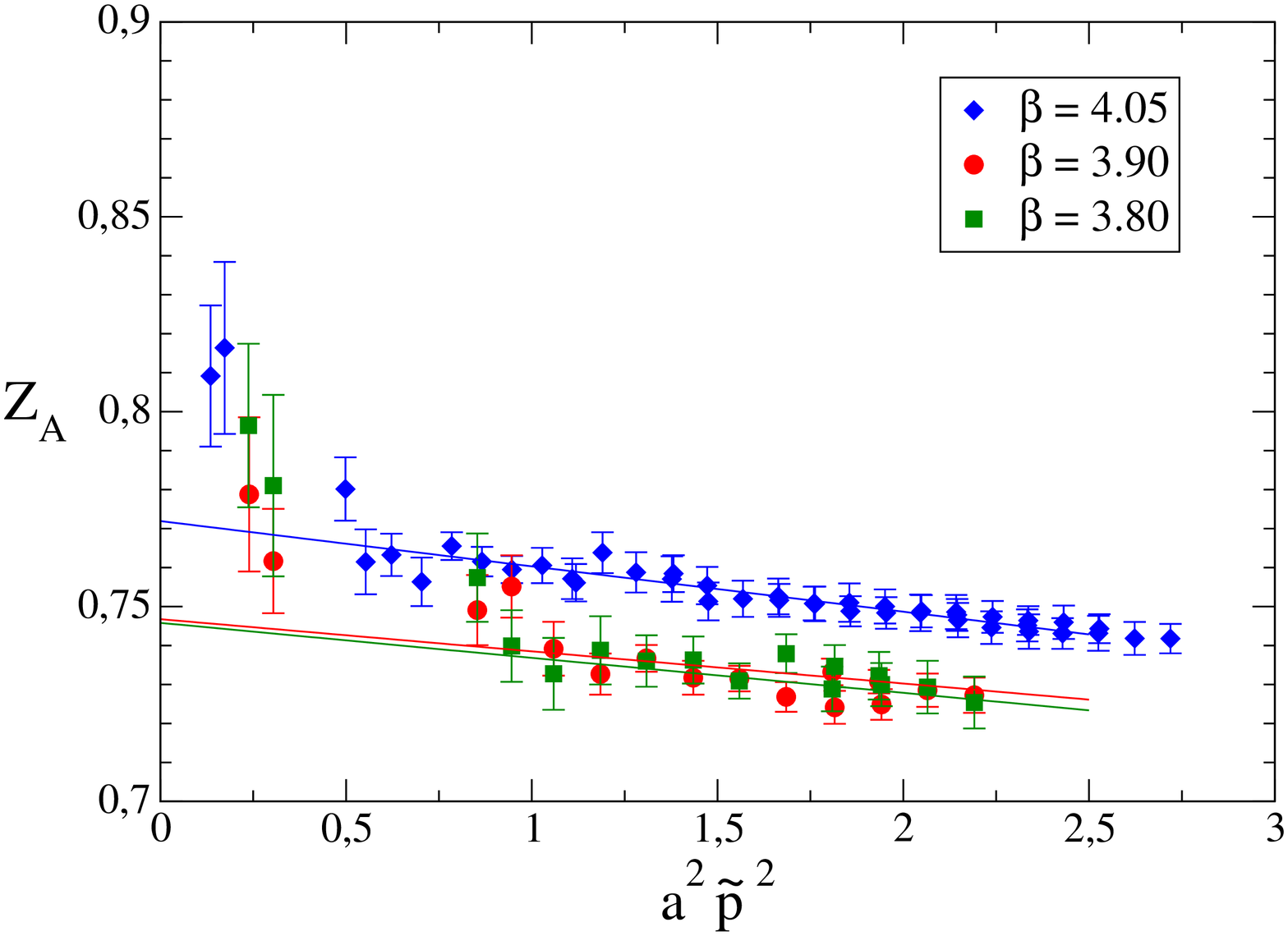}}
\subfigure{\includegraphics[scale=0.27,angle=-90]{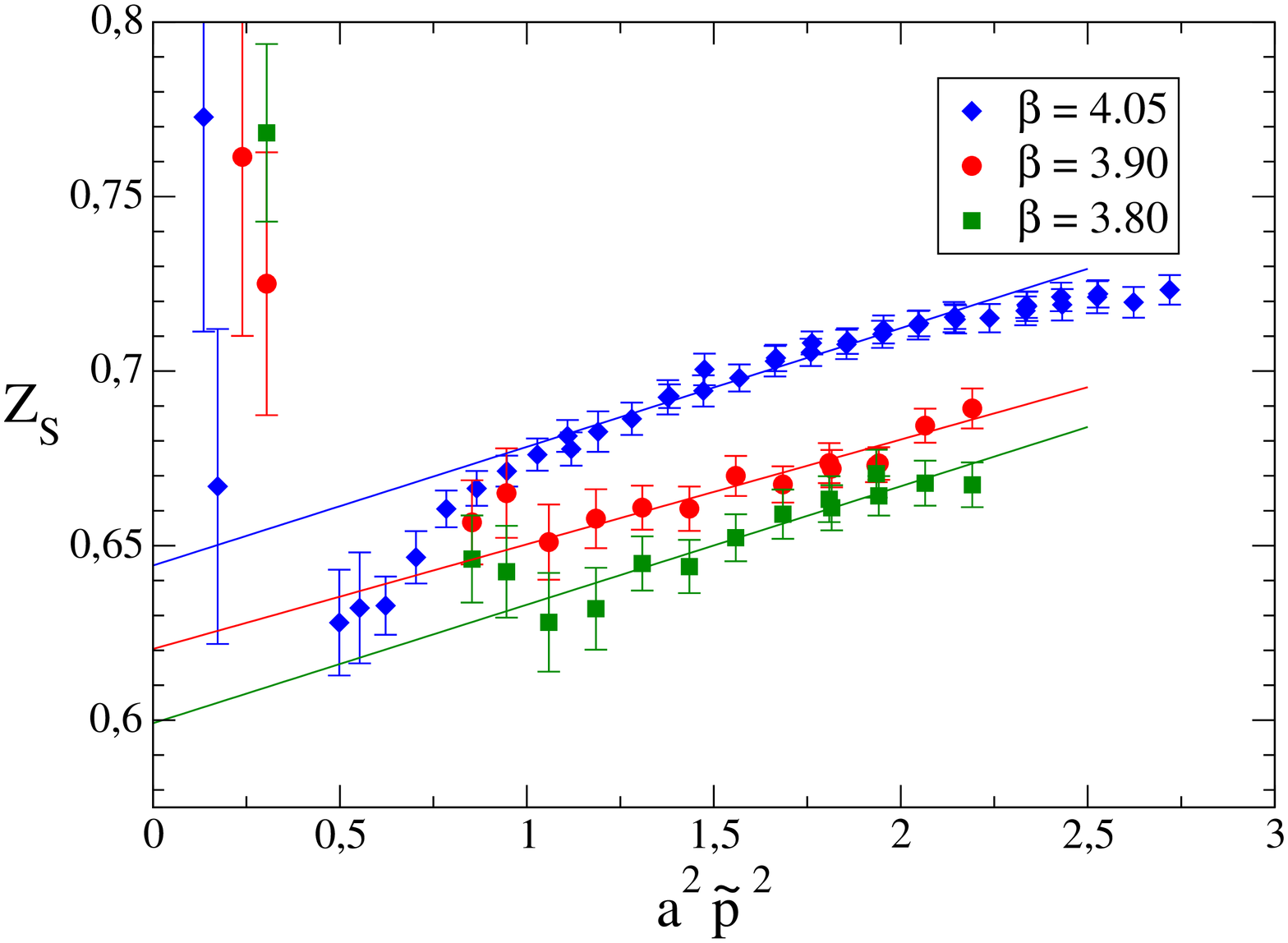}}
\subfigure{\includegraphics[scale=0.27,angle=-90]{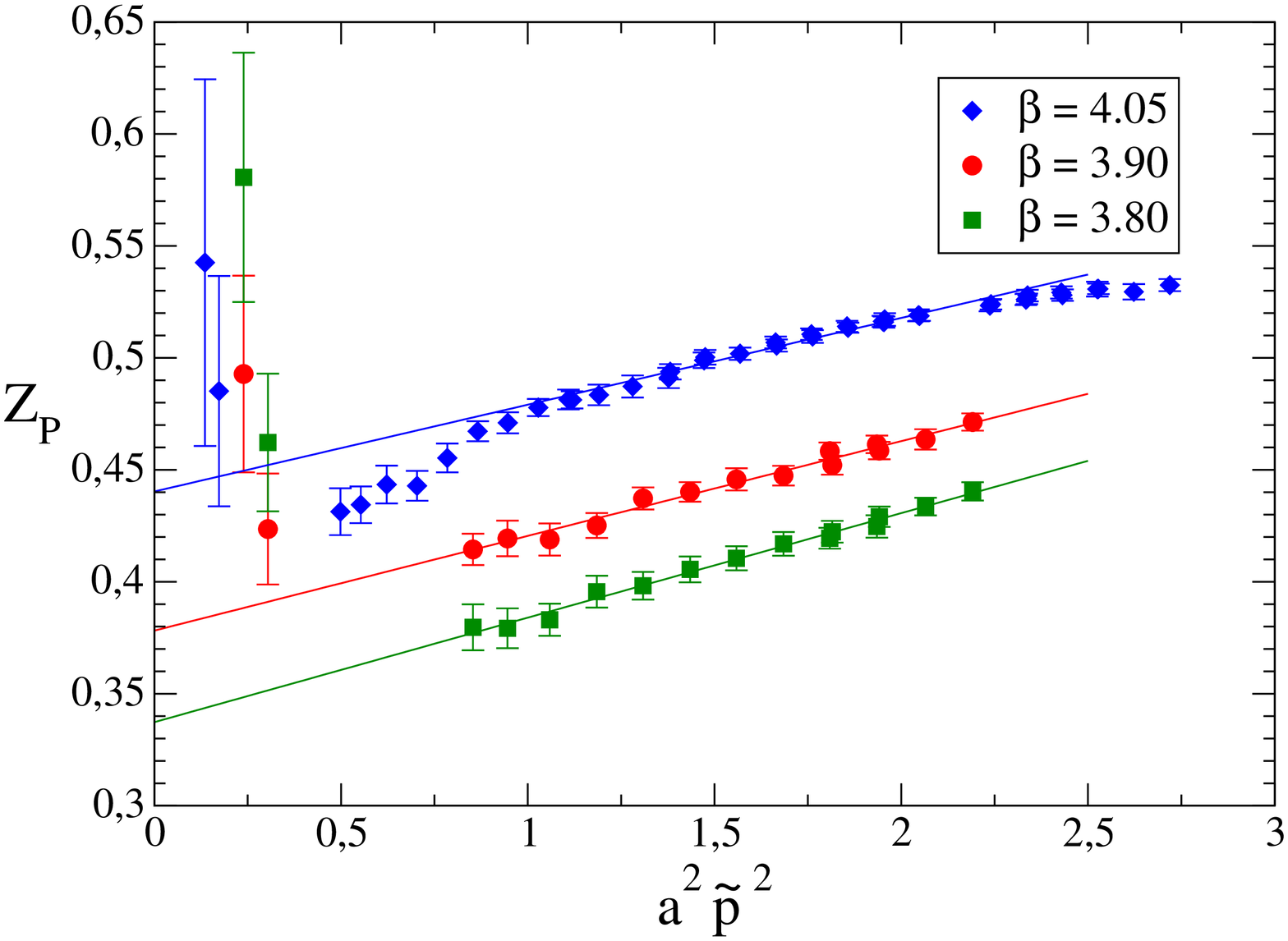}}
\subfigure{\includegraphics[scale=0.27,angle=-90]{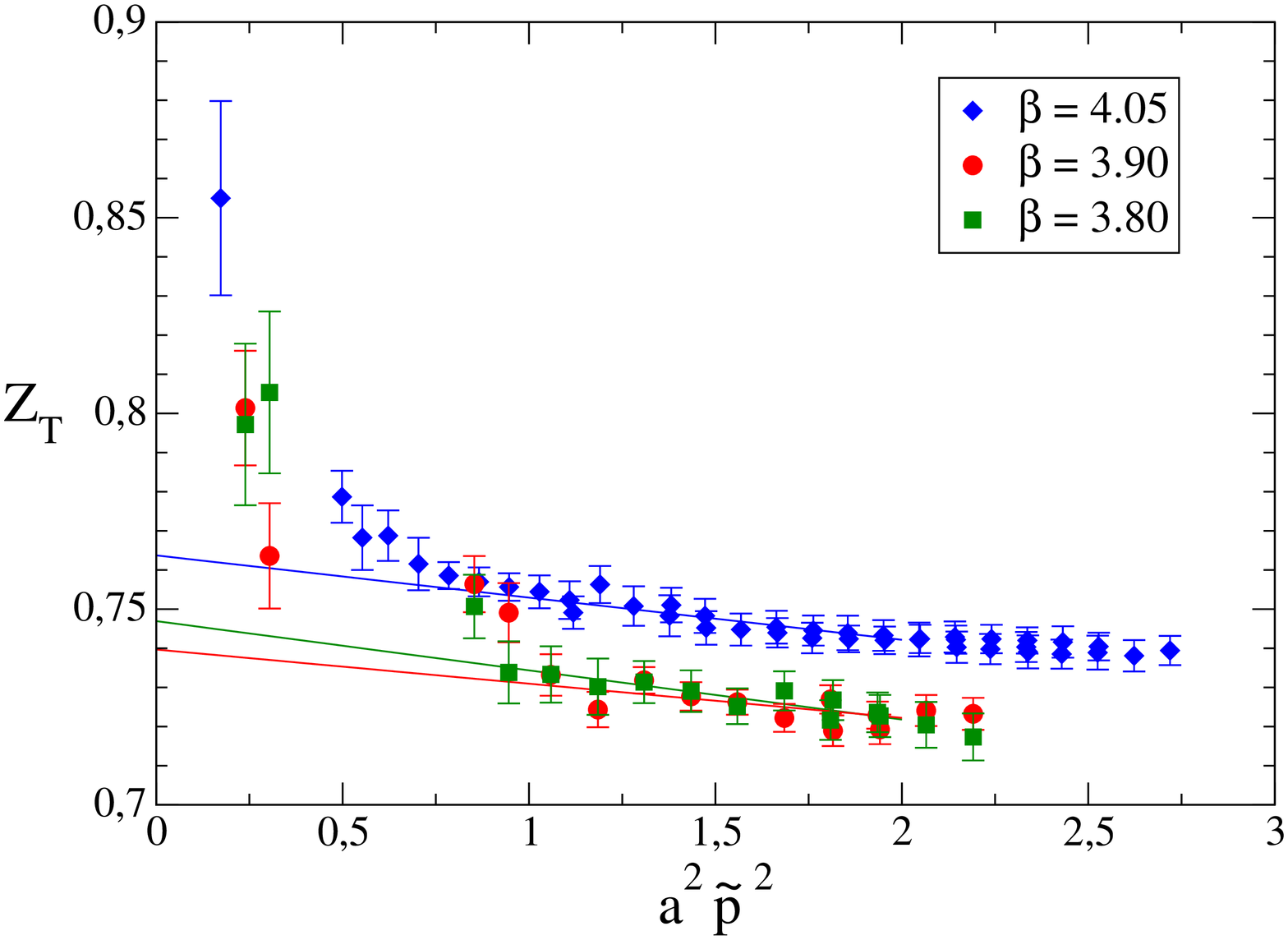}}
\subfigure{\includegraphics[scale=0.27,angle=-90]{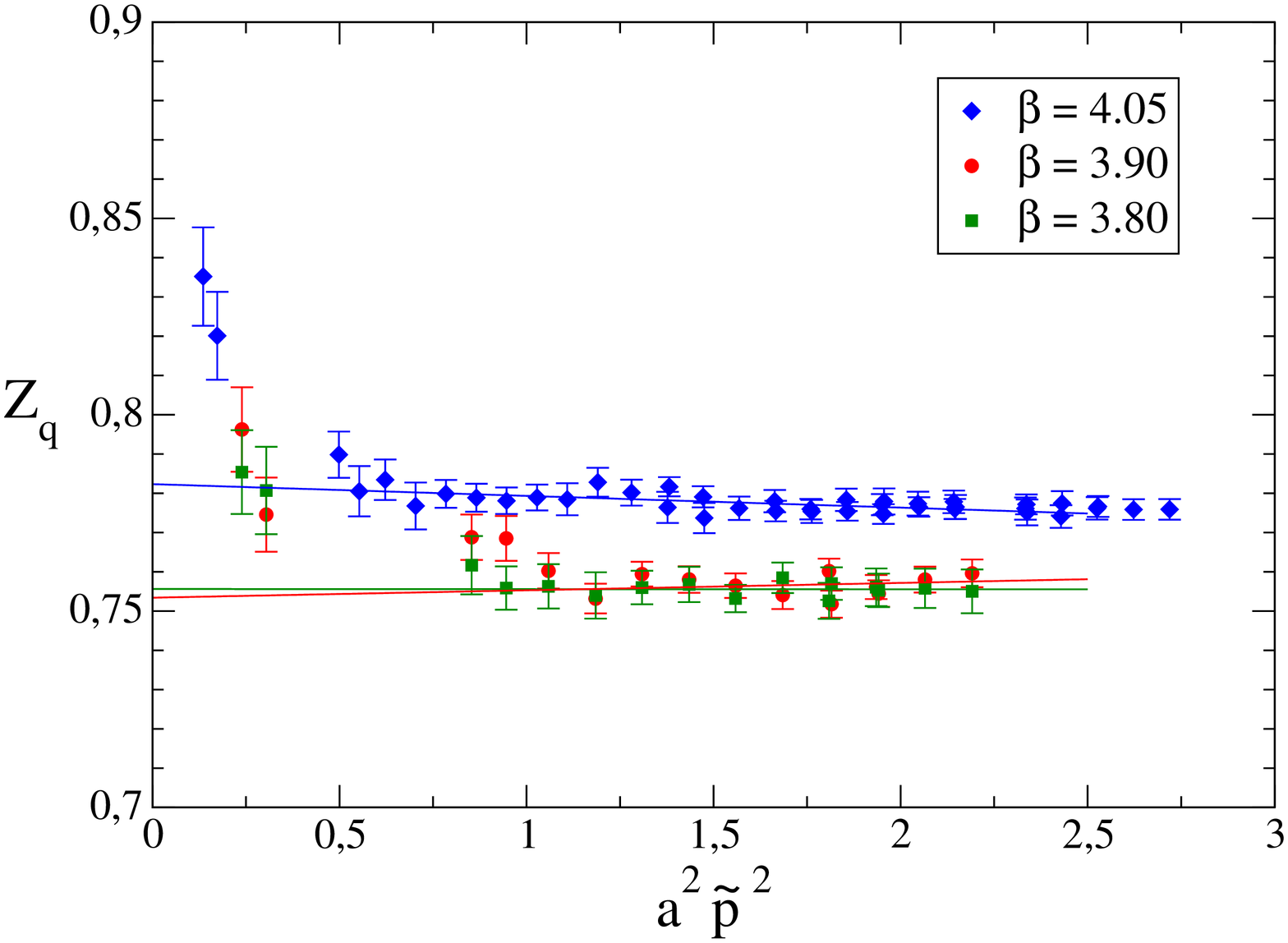}}
\vspace{-0.5cm}
\caption{{\sl The RCs $Z_\Gamma(1/a;a^2p^2)$, evolved to the reference scale
$\mu_0=1/a$, as a function of the initial renormalization scale $a^2 \tilde
p^2$. The solid lines are linear fits to the data.}}
\label{fig:z3beta}
\end{center}
\end{figure}

\item[-] \underline{\em $p^2$-window method ({\sf M2})}: in this approach we do
not attempt any additional subtraction of discretization effects, besides the
perturbative one. The final estimates of RCs are obtained by taking directly the
results of $Z_\Gamma(1/a;a^2p^2)$ at a fixed value of $p^2$ (in physical units).
In practice, this is done by fitting the RCs to a constant in the small momentum
interval $\tilde p^2\in[8.0,9.5]~\gev^2$. The idea behind this approach is that,
once RCs are combined with bare quantities, so as to construct the physical
observables of interest, the residual ${\cal O}(a^2)$ effects, which are present
in both RCs and bare quantities, will be extrapolated away in the continuum
limit.
\end{itemize}
The two methods are compared below, in a specific example. This consists in the
scaling of the pseudoscalar meson mass, composed by two degenerate quarks of
fixed (renormalized) mass. As expected, the two methods give different results
at fixed lattice cut-off, but the difference disappears in the continuum limit.

\subsection{Analysis of systematic errors}
Before presenting our final results for the RCs, we list and discuss in some
detail the main systematic uncertainties.

\begin{itemize}
\item[-] \underline{\em Finite size effects}: the RCs, determined in the  RI-MOM
scheme at large momenta, are short distance quantities and, as such, should not
be affected by significant finite size effects. In order to verify this
expectation, an additional RI-MOM calculation at $\beta=3.90$ has been
performed, on a $32^3 \times 64$ lattice. A comparison of these results with
those obtained on the smaller $24^3 \times 48$ lattice is illustrated in
Fig.~\ref{fig:z2vol}.
\begin{figure}[p]
\begin{center}
\subfigure{\includegraphics[scale=0.28,angle=270]{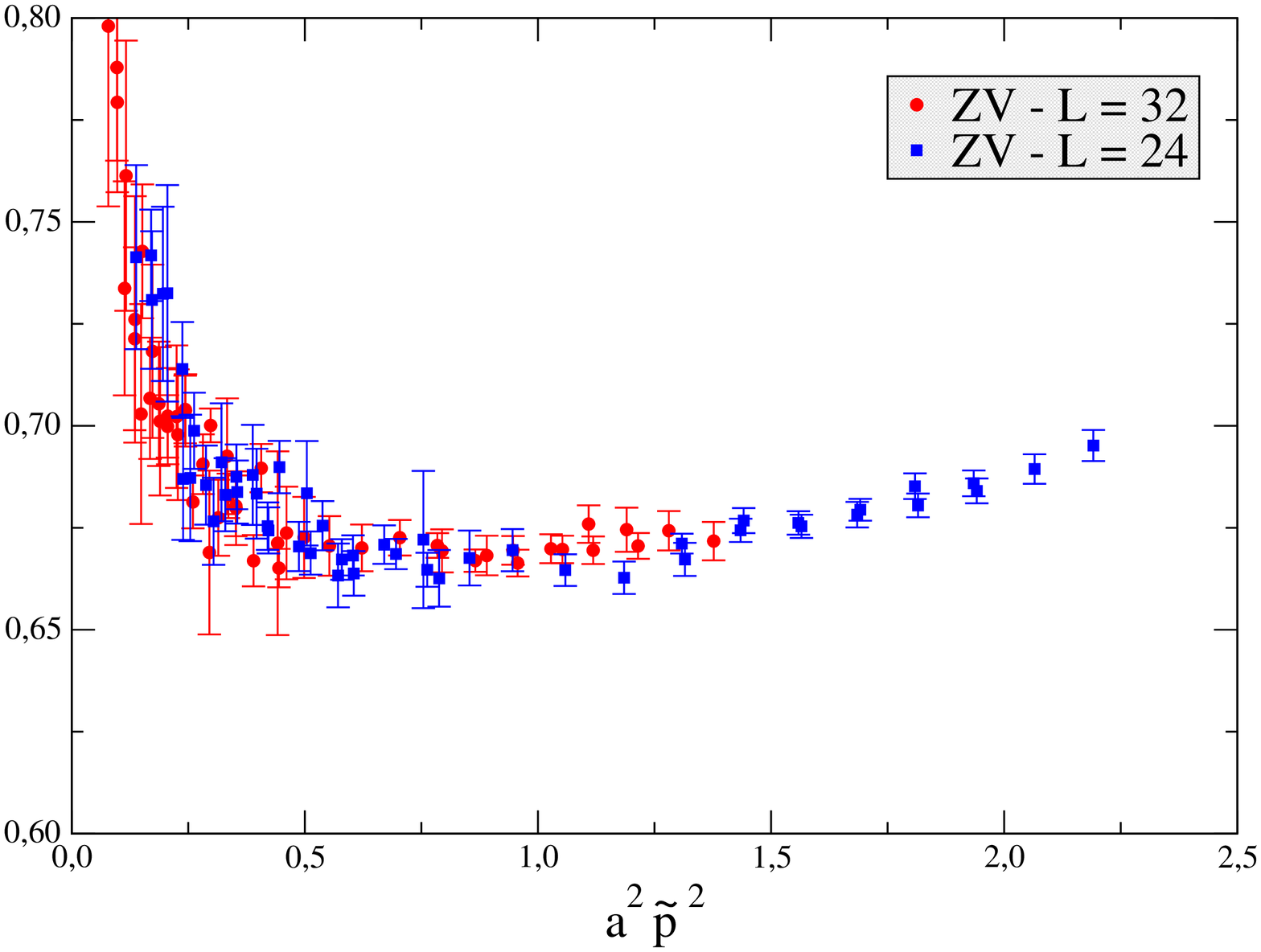}}
\subfigure{\includegraphics[scale=0.28,angle=270]{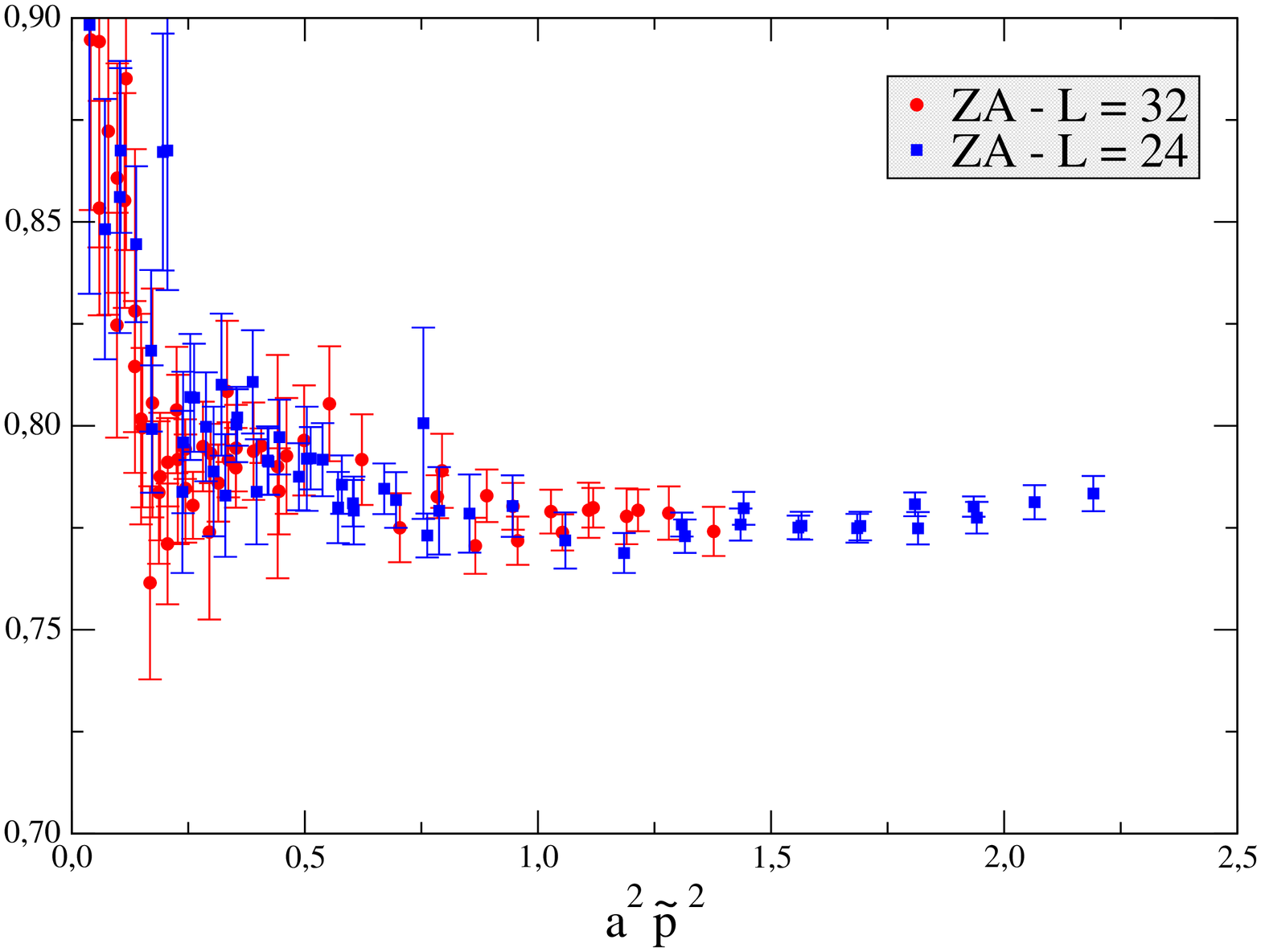}}
\subfigure{\includegraphics[scale=0.28,angle=270]{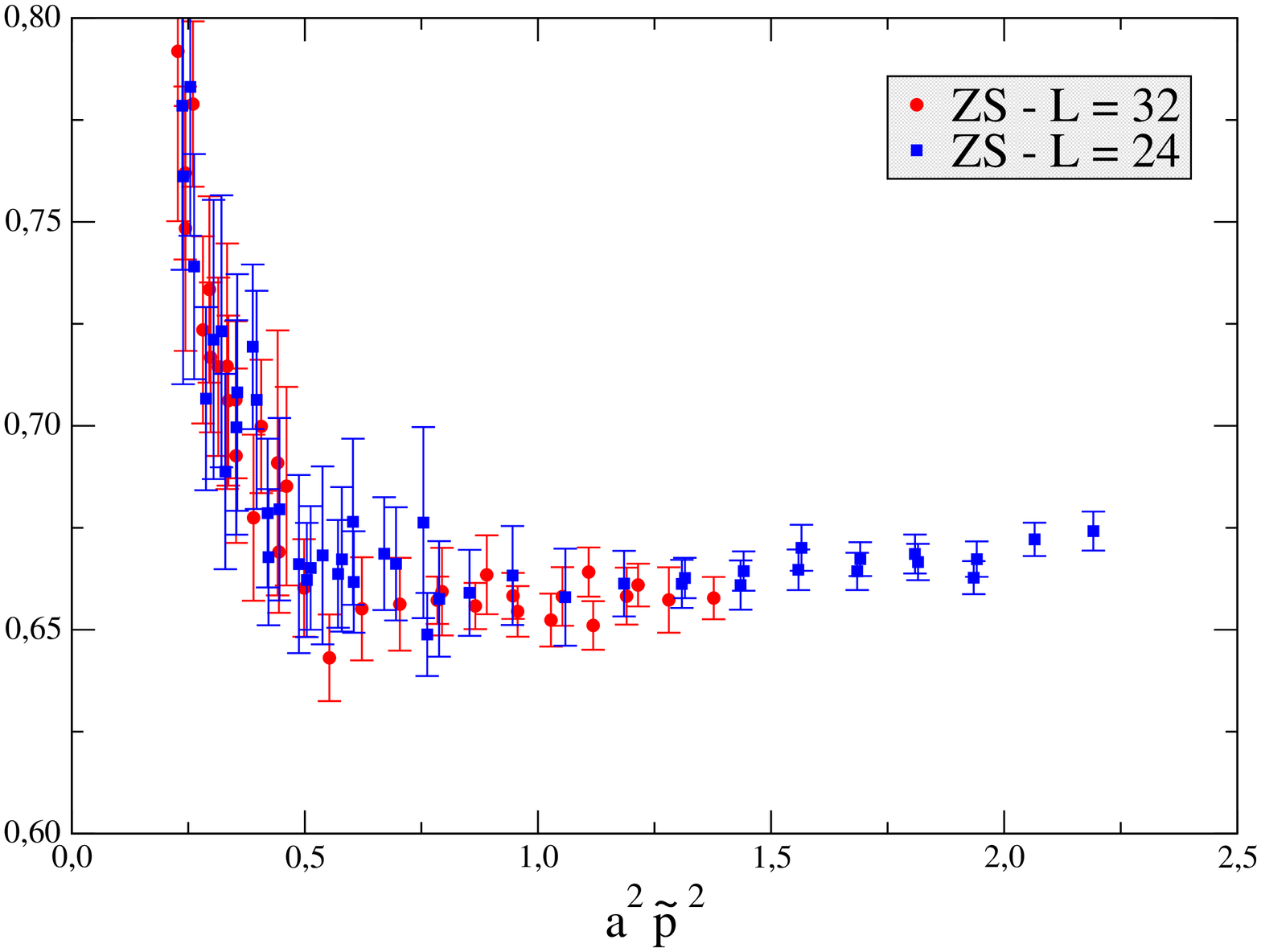}}
\subfigure{\includegraphics[scale=0.28,angle=270]{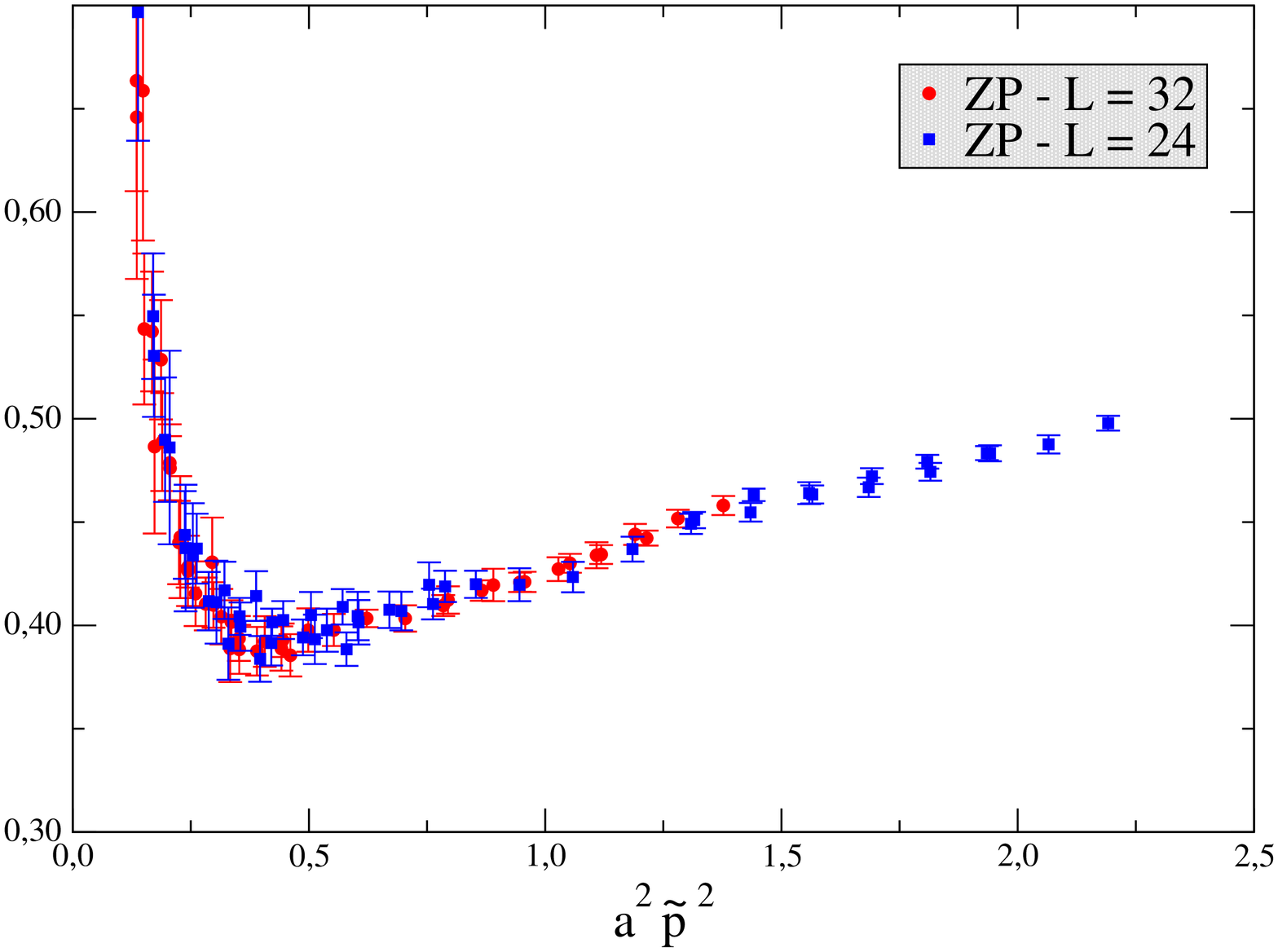}}
\subfigure{\includegraphics[scale=0.28,angle=270]{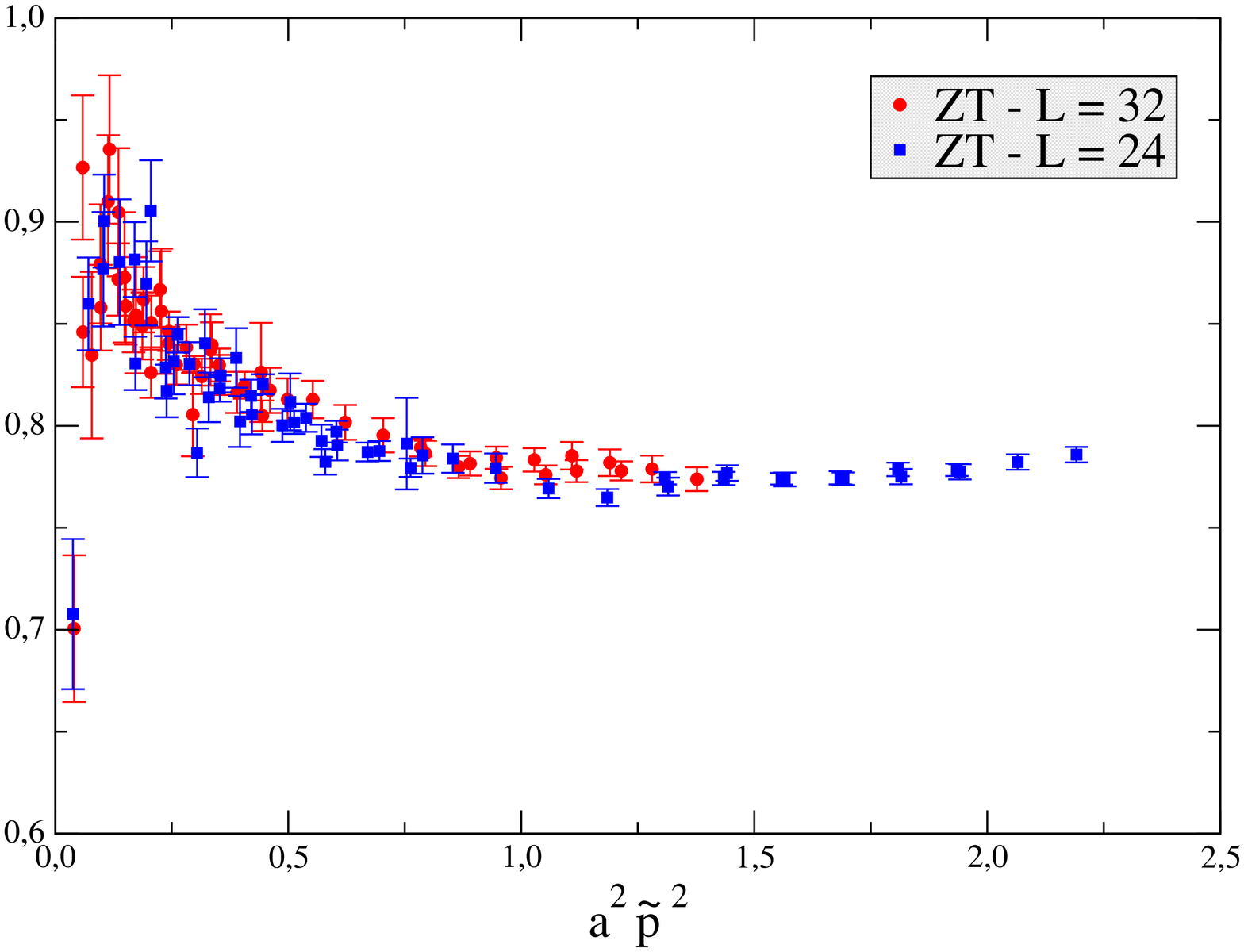}}
\subfigure{\includegraphics[scale=0.28,angle=270]{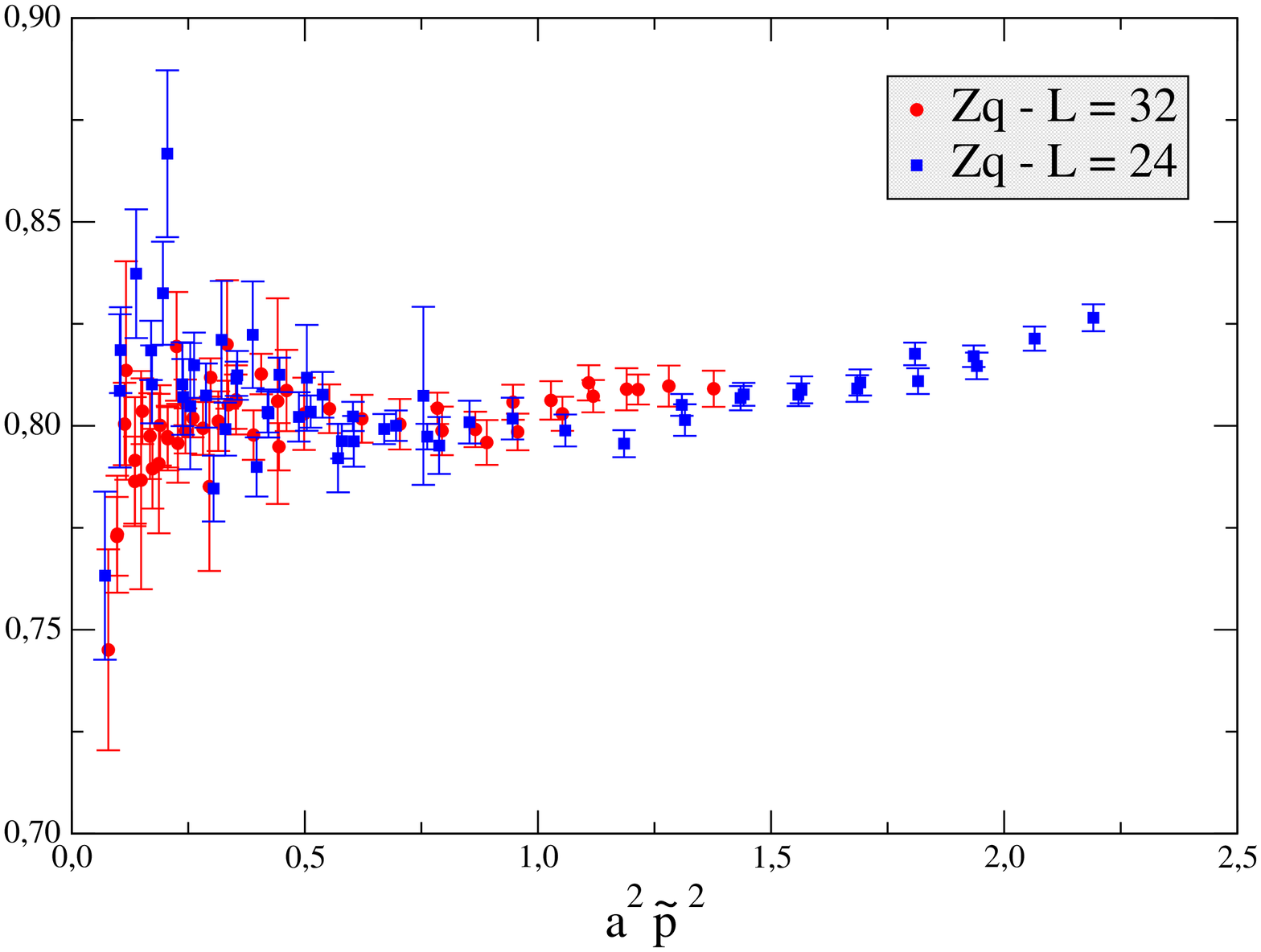}}
\end{center}
\vspace{-0.5cm}
\caption{\sl Comparison of the RCs obtained, at at $\beta=3.90$, on $24^3
\times 48$ and $32^3 \times 64$ lattices.}
\label{fig:z2vol}
\end{figure}
We find that, for the momentum range in which results on both volumes are
available, the differences are smaller than statistical errors.

\item[-] \underline{\em Subtraction of the Goldstone pole and chiral
extrapolations}: the chiral extrapolation of the RCs is rather delicate for
$Z_P$ (and to a lesser extent for $Z_S$), due to the presence of the Goldstone
pole which has to be subtracted. The uncertainty introduced by the subtraction
of the Goldstone pole contribution to $Z_S$ and $Z_P$ can be estimated by
comparing the results obtained following two different subtraction approaches:
the fit of Eq.~(\ref{eq:gpsfit}) or, alternatively, the procedure based on
Eq.~(\ref{eq:gv_pgb}). We find that differences are always negligible for $Z_S$,
whereas for $Z_P$ they are at the level of our statistical errors.

As illustrated in Figs.~\ref{fig:zextrav} and \ref{fig:zextras}, the chiral
extrapolations in the valence quark mass for the other RCs and in the sea quark
mass for all RCs are quite smooth. They are also well consistent with the
expected leading linear and purely quadratic dependence on the valence and sea
quark masses respectively. Therefore, we estimate that the uncertainties
associated with these chiral extrapolations are negligible with respect to the
statistical errors.

\item[-] \underline{\em Discretization effects}: the analysis of these effects
has been presented in section~\ref{sec:scaledep}. The leading ${\cal O}(g^2
a^2)$ discretization errors have been accounted for, using  the analytical
results of ref.~\cite{Og2a2}. With the method denoted as {\sf M1} in
section~\ref{sec:scaledep}, after the perturbative subtraction, residual ${\cal
O}(a^2p^2)$ effects are extrapolated away, with the Ansatz~(\ref{eq:oasub}).
These contributions are not subtracted, instead, in the approach denoted as {\sf
M2}. Discretization effects affecting both RCs and the bare matrix elements will
be corrected for, by extrapolating the physical (renormalized) quantity to the
continuum limit. 
 
\begin{figure}[t]
\begin{center}
\includegraphics[scale=0.35,angle=-90]{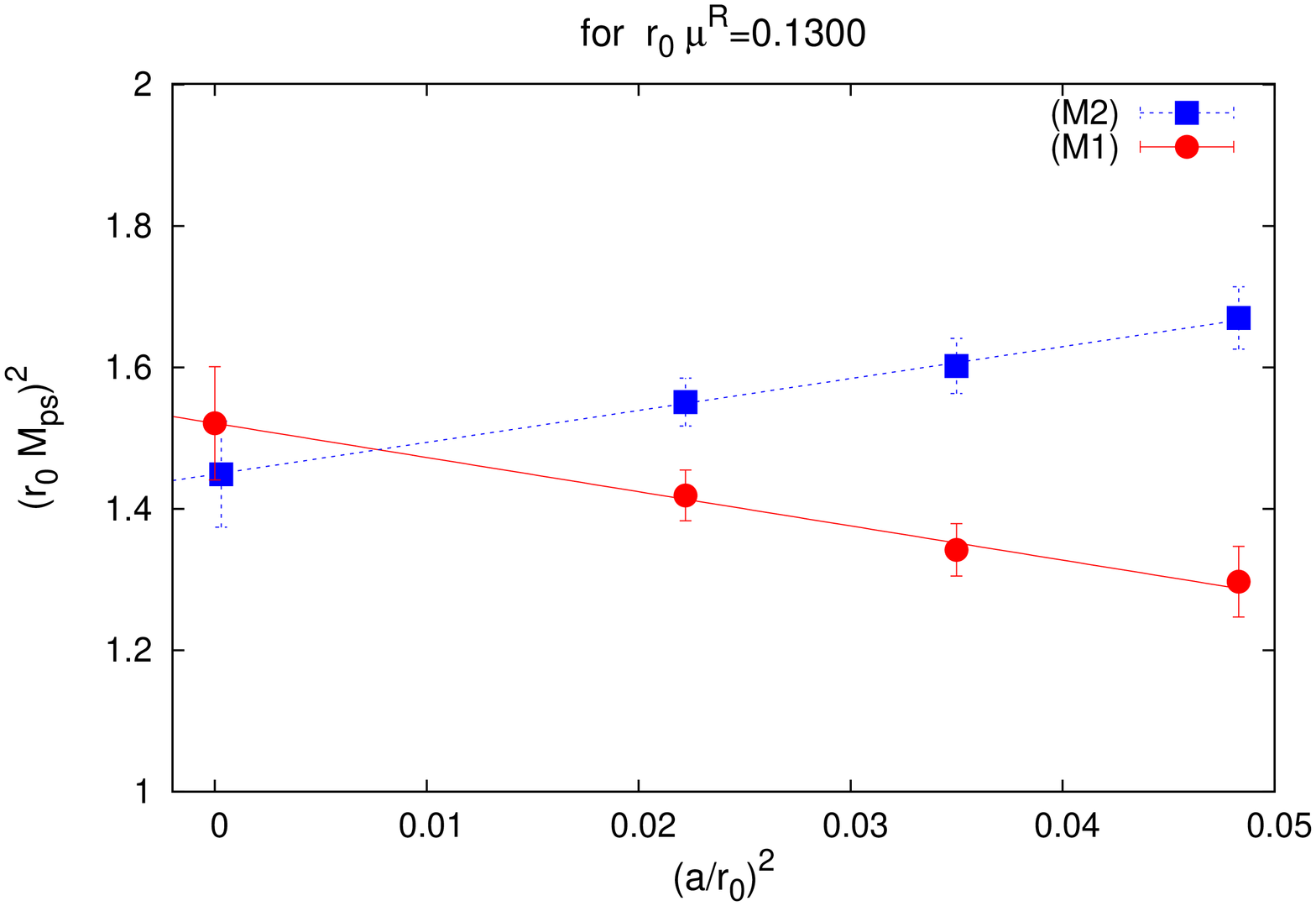}
\end{center}
\vspace{-0.5cm}
\caption{{\sl Scaling of the pseudoscalar meson mass squared, computed at a
fixed value of the quark mass. The {\sf M1} and {\sf M2} determinations for
$Z_{\mu}=Z_P^{-1}$ lead to compatible results in the continuum limit.}} 
\label{fig:Mps2_vs_a2}
\end{figure}
As an example of the results obtained by implementing the two approaches, we
show in Fig.~\ref{fig:Mps2_vs_a2} the continuum extrapolation of the
pseudoscalar mass squared, computed at a fixed value of the renormalized quark
mass $\mu_R$. The renormalization of the latter has been achieved using the
estimates {\sf M1} and {\sf M2} for $Z_P$ given in Tables~\ref{tab:final1} and
\ref{tab:final2}\footnote{Note that in tm lattice QCD we have
$Z_{\mu}=Z_P^{-1}$.}. Although the two $Z_P$ determinations differ by visible
discretization effects at finite lattice spacing, particularly at the two
coarsest lattice, the continuum limit results of the pseudoscalar mass squared
are consistent within the two determinations. This is evidence that the
discretization errors as well as other possible systematic effects in the
analysis of the RI-MOM RCs are well under control, at least within our current
statistical errors (say at the level of one standard deviation and possibly even
less).
\end{itemize}

\subsection{Final results and comparison with perturbation theory}
\begin{table}[p]
\begin{center}
\renewcommand{\arraystretch}{1.5}
\begin{tabular}{||c|c|c|c|c||}  \hline
$\beta$ &   Method         &    $Z_V$  &    $Z_A$  & $Z_P/Z_S$ \\ \hline \hline
      & RI-MOM ({\sf M1})  & 0.604(07) & 0.746(11) & 0.580(17) \\
3.80  & RI-MOM ({\sf M2})  & 0.623(05) & 0.727(07) & 0.653(07) \\
      &   Alt. methods     & 0.5816(02)& 0.747(22) & 0.498(22) \\
      &       BPT          & 0.61-0.70 & 0.71-0.77 & 0.72-0.80  \\ \hline
      & RI-MOM ({\sf M1})  & 0.624(04) & 0.746(06) & 0.626(13) \\
3.90  & RI-MOM ({\sf M2})  & 0.634(03) & 0.730(03) & 0.669(08) \\
      & Alt. methods & 0.6103(03)& 0.743(18) & 0.579(19) \\
      &   BPT        & 0.63-0.71 & 0.72-0.78 & 0.74-0.82   \\ \hline
      & RI-MOM ({\sf M1})  & 0.659(04) & 0.772(06) & 0.686(12) \\
4.05  & RI-MOM ({\sf M2})  & 0.662(03) & 0.758(04) & 0.710(07) \\
      & Alt. methods & 0.6451(03)& 0.746(18) & 0.661(21) \\
      &    BPT       & 0.65-0.73 & 0.74-0.80 & 0.76-0.83   \\ \hline
\multicolumn{5}{c}{}\\
\end{tabular}
\renewcommand{\arraystretch}{1.0}
\end{center}
\vspace{-1.0cm}
\caption{\sl Values of $Z_V, Z_A$ and $Z_P/Z_S$, obtained with the RI-MOM
methods {\sf M1} and {\sf M2} (see text for details), as well as with the
methods described in Sec.~\ref{sec:SI} (labelled as ``Alt. method''). The
predictions of one loop boosted perturbation theory (BPT) are also shown for
comparison.}
\label{tab:final1}
\vspace{0.4cm}
\begin{center}
\renewcommand{\arraystretch}{1.5}
\begin{tabular}{||c|c|c|c|c|c||}  \hline
$\beta$ & Method  & $Z_S^{RI}(1/a)$ & $Z_P^{RI}(1/a)$ & $Z_T^{RI}(1/a)$ &
$Z_q^{RI}(1/a)$ \\ \hline \hline
3.80 & RI-MOM ({\sf M1}) & 0.603(15) & 0.338(10) & 0.743(09) & 0.755(07) \\
     & RI-MOM ({\sf M2}) & 0.668(06) & 0.437(04) & 0.719(06) & 0.755(05) \\
     &  BPT        & 0.68-0.75 & 0.49-0.60 & 0.68-0.75 &  0.69-0.76 \\ \hline
3.90 & RI-MOM ({\sf M1}) & 0.615(09) & 0.377(06) & 0.743(05) & 0.758(04) \\
     & RI-MOM ({\sf M2}) & 0.669(05) & 0.447(05) & 0.724(03) & 0.755(03)  \\
     &  BPT        & 0.70-0.76 & 0.52-0.62 & 0.70-0.76 &  0.71-0.77 \\ \hline
4.05 & RI-MOM ({\sf M1}) & 0.645(06) & 0.440(06) & 0.763(06) & 0.782(05) \\
     & RI-MOM ({\sf M2}) & 0.678(04) & 0.480(04) & 0.752(04) & 0.778(03) \\
     &    BPT      & 0.72-0.78 & 0.55-0.65 & 0.72-0.78 &  0.73-0.79 \\ \hline
\end{tabular}
\renewcommand{\arraystretch}{1.0}
\end{center}
\caption{\sl Values of $Z_S, Z_P$ and $Z_T$, obtained with the RI-MOM methods 
{\sf M1} and {\sf M2} (see text for details). The predictions of one loop 
boosted perturbation theory (BPT) are also shown for comparison.}
\label{tab:final2}
\end{table}

\begin{table}[t]
\begin{center}
\renewcommand{\arraystretch}{1.5}
\begin{tabular}{||c|c|c|c|c|c|c|c||}  \hline
$\beta$ & $Z_V$ & $Z_A$ & $Z_S^{\msb}(2\ \gev)$& $Z_P^{\msb}(2\ \gev)$&
$Z_T^{\msb}(2\ \gev)$& $Z_q^{\msb}(2\ \gev)$ \\ \hline \hline
3.80 & \sf{0.604(07)} & \sf{0.746(11)} & \sf{0.734(18)} & \sf{0.411(12)} & \sf{0.733(09)} & \sf{0.745(07)} \\
3.90 & \sf{0.624(04)} & \sf{0.746(06)}& \sf{0.713(10)} & \sf{0.437(07)} & \sf{0.743(05)} & \sf{0.751(04)} \\
4.05 & \sf{0.659(03)} & \sf{0.772(06)}& \sf{0.699(06)} & \sf{0.477(06)} & \sf{0.777(06)} & \sf{0.780(05)} \\
\hline
\end{tabular}
\renewcommand{\arraystretch}{1.0}
\end{center}
\caption{\sl Final results for the RCs of bilinear quark operators and the
quark field, obtained with the RI-MOM method. The scheme dependent RCs are
given in the $\msb$ scheme at scale $\mu$=2 GeV.}
\label{tab:final3}
\end{table}
The final results for the bilinear quark operators and the quark field RCs,
obtained with the RI-MOM method, are collected in Tables~\ref{tab:final1} and
\ref{tab:final2} (labelled as RI-MOM ({\sf M1}) and RI-MOM ({\sf M2})). These
results are compared with those obtained in Sec.~\ref{sec:SI} for the scale
independent $Z_V$, $Z_A$ and $Z_P/Z_S$ (Alt. methods). In addition, we give in
Tables~\ref{tab:final1} and \ref{tab:final2} the predictions of one loop boosted
perturbation theory (BPT)~\cite{pt,Og2a2}, obtained with two definitions of the
boosted coupling: the first  is $\tilde g^2=g_0^2/\langle P\rangle$, also used
in the previous section, and the second is based on the one-loop
tadpole-improved expansion of $\log \langle P\rangle$~\cite{lm}.

From Tables~\ref{tab:final1} and \ref{tab:final2}, we see that the largest
deviations between the central values of the RI-MOM ({\sf M1}) determinations
and the alternative determinations of Section~\ref{sec:SI} are at the level of
3-4\% or smaller for $Z_A$ and $Z_V$. These differences are not always accounted
for by the errors quoted for each result. When statistically significant, these
differences provide an estimate of the systematic uncertainties affecting the
calculations. As discussed in the previous section, we expect these
uncertainties to be dominated by $O(a^2)$ discretization effects. Nevertheless,
we do not include these differences in the final errors. The reason is that such
estimates of discretization effects, though meaningful at finite lattice
spacing, are of no practical significance when the continuum limit of a physical
quantity is eventually computed.

The comparison of the $Z_P/Z_S$ results, obtained with the RI-MOM and the
alternative method, deserves a further comment. The difference between the two
results is about 15\% for the coarsest lattice and 4\% for the finest one. The
rather big difference noticed in the former case may be attributed not
only to discretization effects, but also to an imprecise tuning of the twist
angle to maximal twist at this coupling. This is expected to influence
substantially the OS pseudoscalar density.

As shown in Tables~\ref{tab:final1} and \ref{tab:final2}, the non-perturbative
results also lie in the range of one-loop BPT estimates, with the exception of
$Z_S$ and, more notably, of $Z_P$. We emphasise that the accuracy of $Z_P$ is
crucial in the determination of quark masses, since the bare quark mass,
computed with maximally twisted fermions, renormalizes with of $Z_P^{-1}$; see
for example ref.~\cite{Blossier:2007vv}.

In most phenomenological applications, the renormalization scheme of choice is
$\msb$ and the preferred reference scale is 2 GeV. For this reason, we provide
in Table~\ref{tab:final3} the values of the scheme dependent RCs $Z_S$, $Z_P$,
$Z_T$ and $Z_q$ in the $\msb$ scheme at the scale of 2 GeV. The conversion from
the RI-MOM scheme at $\mu=1/a$ to $\msb$ at $\mu=$2 GeV has been performed using
renormalization group-improved perturbation theory at the N$^2$LO for $Z_T$ and
at the N$^3$LO for $Z_S$, $Z_P$ and $Z_q$. For completeness, we also report in
the table the results for the scale independent $Z_V$ and $Z_A$. Note, however,
that the Ward identity determination of $Z_V$, given in Tables~\ref{tab:final1}
and \ref{tab:final2}, has a much better statistical accuracy.

It should be remarked that in Table~\ref{tab:final3} we only quote the results
obtained from RI-MOM renormalization conditions with the method {\sf M1}, since
they are in general affected by smaller discretization effects and should hence
be preferably used in computations at fixed gauge coupling without continuum
extrapolation~\footnote{Anyway, the values of the RCs in the $\msb$ scheme at
the scale of 2~GeV obtained with the method {\sf M2} can easily be obtained for
each $a=a(\beta)$ from their {\sf M1}-method counterparts and the RI-MOM results
in Tables~\ref{tab:final1} and \ref{tab:final2} with the help of the relation
$Z_\Gamma^{\msb}(2\ \gev)|_{\sf M2} / Z_\Gamma^{RI}(1/a)|_{\sf M2} =
Z_\Gamma^{\msb}(2\ \gev)|_{\sf M1} / Z_\Gamma^{RI}(1/a)|_{\sf M1} $.}. The case
of Fig. \ref{fig:Mps2_vs_a2} is an exception in this sense, because there an
accidental cancellation (enhancement) of the cutoff effects coming from the bare
pseudoscalar meson mass $M_{\rm ps}$ and from $Z_P$ computed with the method
{\sf M2} ({\sf M1}) leads to a similar scaling behaviour of the two sets of data
corresponding to the {\sf M2} and {\sf M1} determinations of $Z_P$. Moreover,
with respect to the method {\sf M1}, the method {\sf M2} tends to give results
that are affected by smaller statistical uncertainties, since the latter
procedure does not involve the extrapolation to $a^2p^2=0$ of
Eq.~(\ref{eq:oasub}).

\section{Conclusions}
In this paper we have presented a non-perturbative determination of the RCs of
bilinear quark operators for the tree-level Symanzik improved gauge action and
the $N_f=2$ twisted mass fermion action at maximal twist, which guarantees
automatic ${\cal O}(a)$--improvement. The values of these constants do not
depend on whether tm, OS or Wilson fermions are used. The RCs of the five
bilinear quark operators and the quark field have been determined with the
RI-MOM method. We have also obtained an independent estimate of $Z_V$, based on
a Ward identity, and have introduced a new method for $Z_A$ and $Z_S/Z_P$, based
on a combined use of standard tm and OS fermions. The main results for the RCs
are collected in Tables~\ref{tab:final1}, \ref{tab:final2} and \ref{tab:final3},
where they are also compared with the predictions of one-loop perturbation
theory. We find that the differences between perturbative and non-perturbative
determinations are large in the cases of $Z_S$ and $Z_P$. This result is of
particular relevance to the lattice determinations of quark masses and the
chiral condensate.
 
\section*{Appendix: ${\mathbf{\cal O}(a)}$-improvement of RI-MOM
re\-nor\-ma\-li\-za\-tion constants}

In this Appendix we prove that:
\begin{enumerate}
\item at maximal twist the RI-MOM ``form factors'' from which RCs are
extracted~\cite{rimom} are automatically O($a$) improved at all $p^2$ values;
\item at generic values of the twist angle, form factor improvement is
guaranteed only at ``large'' $p^2$ values; i.e.\ where spontaneous chiral
breaking effects can be neglected. 
\end{enumerate}
In both cases the resulting renormalization constants will be improved. In the
first case this is obvious because they are extracted from ratios of
automatically improved form factors. In the second case the desired conclusion
follows from the observation that RCs are short-distance quantities
(insensitive to infrared effects), evaluated from large $p^2$ correlators. In
the ensuing sections we give a detailed proof of these two statements.

\subsection*{Improvement at maximal twist}
In the RI-MOM scheme the RC $Z_\Gamma^{fh}$ of the bilinear operator
\beq
O_\Gamma^{fh}=\bar q^f\Gamma q^h,
\eeq
is defined by the condition
\beq
\lim_{\mu_{{\rm val}}^{f,h},\,\mu_{\rm sea} \to 0} \;
Z_q^{-1} Z_\Gamma^{fh} \, {\rm Tr} [\Lambda_\Gamma^{fh} (p)\, 
P_\Gamma]\Big{|}_{p^2=\mu^2}=1 \, ,
\label{ZDEF}\eeq
where $f,h=1,2$ are flavour indices, and $P_\Gamma$ is the projector onto the
Dirac structure of the bilinear. The other quantities entering Eq.~(\ref{ZDEF})
are defined by
\beq
\lim_{\mu_{{\rm val}}^{f},\,\mu_{\rm sea} \to 0} \;
Z_q \,\frac{-i}{12}\,\left.\frac{{\rm Tr}[\slash p\, S_q^f(p)^{-1}]}
{p^2}\right\vert_{p^2=\mu^2}=1 
\label{ZQDEF}
\eeq
with 
\beqn
&&\Lambda_\Gamma^{fh}(p) = S_q^f(p)^{-1} \,\, G_\Gamma^{fh}(p) \,\, S_q^h(p)^{-1}\, ,
\label{GAMDEF}\\ 
&& S_q^f(p) = a^4 \sum_x \, e^{-i p\cdot x} \, \langle  q^f(x) \bar
 q^f(0)\rangle\, , 
\label{PROP}\\
&&G_\Gamma^{fh}(p) = a^8 \sum_{x,y} \, e^{-i p\cdot (x-y)} \, \langle  q^f(x)
O_\Gamma^{fh}(0)\bar  q^h(y) \rangle \, .
\label{GFH}
\eeqn
Two useful observations are: (i)  the ``form factors'' ${\cal{V}}_q(p)={\rm
Tr}[\slash p\, S^f(p)^{-1}]$ and ${\cal{V}}^{fh}_\Gamma(p)={\rm
Tr}[\Lambda_\Gamma^{fh}(p)\, P_\Gamma]$ are H(4) invariant by construction. In
the continuum limit they approach quantities which are invariant under both O(4)
and parity. (ii) $Z_q$ carries no flavour index $f$ since, owing to its
definition~(\ref{ZQDEF}), it depends quadratically on $r_q$.

The proof that our estimators of the RI-MOM RCs are O($a$)-improved at all quark
masses and external momenta in maximally twisted QCD follows closely the
argument of ref.~\cite{FR} (subsequently refined in ref.~\cite{FMPR}) for
automatic improvement of correlation functions. It is based on the exact
invariance of the maximally twisted QCD action under the transformation
${\cal{P}}\times {\cal D}_d \times (\mu_q\to -\mu_q)$, where $\mu_q$ is the
twisted quark mass. For completeness we recall the definition of the discrete
transformations ${\cal{P}}$ (continuum parity) and $ {\cal D}_d$ in the physical
quark basis 
\beq
{\cal{P}}:\left \{\begin{array}{ll} 
&\hspace{-.3cm} U_0(x)\rightarrow U_0(x_{\cal{P}})\, ,\qquad U_k(x)\rightarrow 
U_k^{\dagger}(x_{\cal{P}}-a\hat{k})\, ,\qquad k=1,2,3\\
&\hspace{-.3cm} q^f(x)\rightarrow \gamma_0  q^f(x_{\cal{P}})\\ 
&\hspace{-.3cm}\bar{ q}^f(x) 
\rightarrow\bar{ q}^f(x_{\cal{P}})\gamma_0 
\end{array}\right . \label{PAROP}  
\eeq
\beq 
\hspace{-7.cm}{\cal{D}}_d : \left \{\begin{array}{lll}     
U_\mu(x)&\rightarrow U_\mu^\dagger(-x-a\hat\mu) \\
 q^f(x)&\rightarrow e^{3i\pi/2}  q^f(-x)  \\
\bar{ q}^f(x)&\rightarrow e^{3i\pi/2} \bar{ q}^f(-x)  
\end{array}\right . \label{FIELDT} 
\eeq
where $x_{\cal{P}}=(-{\bf{x}},x_0)$ and $\hat\mu$ is the unit vector in the 
$\mu$-direction. Actually the proof is also valid in the more general ``mixed
action'' setting, where Osterwalder-Seiler fermions are used as valence quarks.
This is because the corresponding action is similarly invariant under
${\cal{P}}\times {\cal D}_d \times (M_q\to -M_q)$, where $M_q$ denotes the set
of all quark (sea and valence) twisted mass parameters.

\subsubsection*{The proof}
The proof can be given in three steps:

$\bullet$ Due to H(4) symmetry, the lattice form factors ${\cal{V}}_q(p)$ and
${\cal{V}}^{fh}_\Gamma(p)$ have no terms of the form $p_0^k+p_1^k+p_2^k+p_3^k$
and $(p_0p_1p_2p_3)^k$, with odd $k$. Thus they are invariant under parity
transformation; i.e.\ they satisfy the relations 
\beq
{\cal{V}}_q(p) = {\cal{V}}_q(p_{\cal{P}}) \, ,\qquad
{\cal{V}}^{fh}_\Gamma(p)={\cal{V}}^{fh}_\Gamma(p_{\cal{P}})\, ,\qquad
p_{\cal{P}}=(-{\bf{p}},p_0)\, .\label{PAFF}
\eeq
\begin{itemize}
\item As has been proved in~\cite{FMPR}, the invariance of maximally twisted QCD
under ${\cal{P}}\times {\cal D}_d \times (\mu_q\to -\mu_q)$ implies that
O($a^j$) cut-off artefacts of lattice correlators arise, in their Symanzik
expansion, from the insertion of Symanzik operators of parity $(-1)^j$.
\item Subsequently, O($a^j$) terms with odd $j$ must then be absent in the
Symanzik expansion of parity-even form factors, like ${\cal{V}}_q(p)$ and
${\cal{V}}^{fh}_{\Gamma}(p)$. 
\end{itemize}
The Symanzik expansion of the form factors above is straightforwardly obtained
from that of the correlators in terms of which they are defined.

\subsection*{Improvement at a generic value of the twist angle}
\label{sec:IGT}
Since RCs are UV quantities, they should be independent of the QCD infrared
structure and properties; in particular they should be unaffected by spontaneous
chiral symmetry breaking. Thus, if we compute RCs at zero quark mass and at high
momentum scale $p^2$ such that chiral breaking effects are negligible, we can
fully exploit the ${\cal{R}}_5$-symmetry of the continuum theory. This will be
useful because in turn the lattice action is invariant under ${\cal{R}}_5\times
{\cal D}_d$, where
\beq
{\cal{R}}_5=\prod_f {\cal{R}}^f_5\, , \qquad {\cal{R}}^f_5 : \left 
\{\begin{array}{ll}     
 q^f&\rightarrow\gamma_5  q^f  \\
\bar{ q}^f&\rightarrow -\bar{ q}^f\gamma_5     
\end{array}\right . 
\label{PSIBAR} 
\eeq
is a transformation belonging to the SU$(N_f)_L\times$SU$(N_f)_R$ chiral group
in the unitary case (or the graded SU$(N_f^{sea}+N_f^{val}|N_f^{val})_L \times
$SU$(N_f^{sea}+N_f^{val}|N_f^{val})_R$ chiral group in the mixed action case).

\subsubsection*{The proof}
The proof of the O($a$) improvement of RCs in lattice QCD at generic values of
the twist angle, e.g.\ of the bare twist angle  $\omega_0 = {\rm arctan}( \mu_q
/(m_0 - M_{cr} ))$, runs on lines similar to those of the previous section, with
${\cal{R}}_5$ replacing ${\cal{P}}$. One can, in fact, argue as follows.

$\bullet$ RCs constants can be extracted from the large-$p^2$ behaviour of the
${\cal{V}}_q(p)$ and ${\cal{V}}^{fh}_\Gamma(p)$ form factors in the chiral
limit. In this regime chiral breaking effects can be ignored and
${\cal{V}}_q(p)$ and ${\cal{V}}^{fh}_\Gamma(p)$ are invariant under
${\cal{R}}_5$.

$\bullet$ The invariance of the theory under ${\cal{R}}_5\times {\cal D}_d$
implies that the ${\cal{R}}_5$-parity of an operator equals the parity of its
naive dimension. Thus, O($a^j$) lattice artefacts of correlators or form factors
arise in the Symanzik expansion from the insertion of operators of
${\cal{R}}_5$-parity equal to $(-1)^j$.

$\bullet$ Therefore O($a^j$) terms with odd $j$ cannot be present in the
Symanzik expansion of ${\cal{R}}_5$-even form factors, like ${\cal{V}}_q(p)$ and
${\cal{V}}^{fh}_\Gamma(p)$. In fact, the continuum target theory is invariant
under the chiral group (including ${\cal{R}}_5$), while O($a^j$) terms would be
odd under ${\cal{R}}_5$.

\subsection*{A final comment}
The ${\cal O}(a)$ improvement of RI-MOM RCs at any value (including zero) of the
twist angle $\omega_0$ hardly comes as a surprise because the RC-estimators
computed at generic $\omega_0$ become $\omega_0$-independent in the chiral
limit, provided they are analytic in the complex mass parameter $(m_0 - M_{cr} )
+ i\mu_q$, which is certainly the case at sufficiently large $p^2$. As a
consequence, a unique set of RI-MOM RCs, all free from ${\cal O}(a^j)$ ($j$ odd)
artifacts, exist for lattice QCD with Wilson quarks and a given pure gauge
lattice action. These RCs can in principle be extracted from simulations at
arbitrary (and not necessarily all equal) values of the twist angle. Computing
the RI-MOM RCs at maximal twist ($|\omega_0| = \pi/2$), as we do in the present
study, is just a simple and technically convenient choice. In our case this
choice was a quite natural one, as we could then evaluate the necessary
correlators on the $N_f=2$ gauge ensembles which were produced for studying
large volume physics.


\begin{thebibliography}{9}

\bibitem{rimom}
G.~Martinelli, C.~Pittori, C.~T.~Sachrajda, M.~Testa and A.~Vladikas,
Nucl.\ Phys.\ B {\bf 445} (1995) 81
[arXiv:hep-lat/9411010].

\bibitem{Dimopoulos:2007fn}
  P.~Dimopoulos, R.~Frezzotti, G.~Herdoiza, A.~Vladikas, V.~Lubicz, S.~Simula
and M.~Papinutto,
  PoS {\bf LAT2007} (2007) 241
  [arXiv:0710.0975 [hep-lat]].

\bibitem{plb}
  Ph.~Boucaud {\it et al.}  [ETM Collaboration],
  Phys.\ Lett.\  B {\bf 650} (2007) 304
  [arXiv:hep-lat/0701012].

\bibitem{ETMC-long} 
  Ph.~Boucaud {\it et al.}  [ETM collaboration],
  Comput.\ Phys.\ Commun.\  {\bf 179} (2008) 695
  [arXiv:0803.0224 [hep-lat]].

\bibitem{urbach}
  C.~Urbach  [European Twisted Mass Collaboration],
  PoS {\bf LAT2007} (2007) 022
  [arXiv:0710.1517 [hep-lat]].

\bibitem{Baron:2009wt}
  R.~Baron {\it et al.},
  arXiv:0911.5061 [hep-lat].

\bibitem{pt}
  S.~Aoki, K.~i.~Nagai, Y.~Taniguchi and A.~Ukawa,
  Phys.\ Rev.\  D {\bf 58} (1998) 074505
  [arXiv:hep-lat/9802034].

\bibitem{Og2a2}
  M.~Constantinou, V.~Lubicz, H.~Panagopoulos and F.~Stylianou,
  JHEP {\bf 0910} (2009) 064
  [arXiv:0907.0381 [hep-lat]].

\bibitem{Gockeler:2010yr}
  M.~Gockeler {\it et al.},
  arXiv:1003.5756 [hep-lat].

\bibitem{xlf_1} 
  K.~Jansen, M.~Papinutto, A.~Shindler, C.~Urbach and I.~Wetzorke  [XLF
  Collaboration],
  JHEP {\bf 0509} (2005) 071
  [arXiv:hep-lat/0507010].

\bibitem{becirevic} D.~Becirevic {\it et al.},
  Phys.\ Rev.\  D {\bf 74} (2006) 034501
  [arXiv:hep-lat/0605006].

\bibitem{OS} K.~Osterwalder and E.~Seiler,  Annals Phys.
{\bf 110} (1978) 440; 
\\
R.~Frezzotti and G.~C.~Rossi,
  JHEP {\bf 0410} (2004) 070
  [arXiv:hep-lat/0407002].

\bibitem{FR}
  R.~Frezzotti and G.~C.~Rossi,
  JHEP {\bf 0408} (2004) 007
  [arXiv:hep-lat/0306014].

\bibitem{frezzotti} R.~Frezzotti {\it et al.},  [Alpha collaboration],
  JHEP {\bf 0108} (2001) 058
  [arXiv:hep-lat/0101001].

\bibitem{dsv} P.~Dimopoulos, H.~Simma and A.~Vladikas,
  JHEP {\bf 0907} (2009) 007
  [arXiv:0902.1074 [hep-lat]].

\bibitem{chris} M.~Foster and C.~Michael  [UKQCD Collaboration],
  Phys.\ Rev.\  D {\bf 59} (1999) 074503
  [arXiv:hep-lat/9810021].

\bibitem{Frezzotti:2008dr}
  R.~Frezzotti, V.~Lubicz and S.~Simula,
  Phys.\ Rev.\  D {\bf 79} (2009) 074506
  [arXiv:0812.4042 [hep-lat]].

\bibitem{zeta}
D.~Becirevic, V.~Gimenez, V.~Lubicz, G.~Martinelli, M.~Papinutto and J.~Reyes,
JHEP {\bf 0408} (2004) 022
[arXiv:hep-lat/0401033].

\bibitem{qprop}
  D.~Becirevic, V.~Gimenez, V.~Lubicz and G.~Martinelli,
  Phys.\ Rev.\  D {\bf 61} (2000) 114507
  [arXiv:hep-lat/9909082].

\bibitem{alain1}
J.~R.~Cudell, A.~Le Yaouanc and C.~Pittori,
Phys.\ Lett.\ B {\bf 454} (1999) 105
[arXiv:hep-lat/9810058].

\bibitem{alain2}
J.~R.~Cudell, A.~Le Yaouanc and C.~Pittori,
Phys.\ Lett.\ B {\bf 516} (2001) 92
[arXiv:hep-lat/0101009].

\bibitem{gv}
L.~Giusti and A.~Vladikas,
Phys.\ Lett.\ B {\bf 488} (2000) 303
[arXiv:hep-lat/0005026].

\bibitem{gracey}
J.~A.~Gracey,
Nucl.\ Phys.\ B {\bf 662} (2003) 247
[arXiv:hep-ph/0304113].

\bibitem{chetyr}
K.~G.~Chetyrkin and A.~Retey,
Nucl.\ Phys.\ B {\bf 583} (2000) 3
[arXiv:hep-ph/9910332].

\bibitem{lm}
  G.~P.~Lepage and P.~B.~Mackenzie,
  Phys.\ Rev.\  D {\bf 48} (1993) 2250
  [arXiv:hep-lat/9209022].

\bibitem{Blossier:2007vv}
  B.~Blossier {\it et al.}  [European Twisted Mass Collaboration],
  JHEP {\bf 0804} (2008) 020
  [arXiv:0709.4574 [hep-lat]].

\bibitem{FMPR}
  R. Frezzotti, G. Martinelli, M. Papinutto and G.C. Rossi, 
  JHEP {\bf 0604} (2006) 038 
  [arXiv:hep-lat/0503034] and PoS LAT2005:285,2006 [arXiv:hep-lat/0509168].

\bibitem{Frezzotti:2005gi}
  R.~Frezzotti, G.~Martinelli, M.~Papinutto and G.~C.~Rossi,
  JHEP {\bf 0604} (2006) 038
  [arXiv:hep-lat/0503034]
and
  PoS {\bf LAT2005} (2006) 285
  [arXiv:hep-lat/0509168].


\end{thebibliography}
\end{document}